\documentclass[aps,10pt,prd,notitlepage,showpacs,nofootinbib,superscriptaddress, compressed]{revtex4-1}
\usepackage{graphicx}
\usepackage[utf8]{inputenc} 
\usepackage{adjustbox}
\usepackage{amsmath}
\usepackage{amssymb}
\usepackage{rotating}
\usepackage{float}
\usepackage{comment}
\usepackage{multirow}
\usepackage{slashed}
\usepackage[normalem]{ulem}
\usepackage{braket}
\usepackage{bigstrut}
\usepackage{gensymb}
\usepackage{caption}
\usepackage{subcaption}
\usepackage[usenames,dvipsnames]{color}
\usepackage[colorinlistoftodos]{todonotes}
\usepackage[colorlinks=true,citecolor=darkred,urlcolor=darkred, pdfborder={0 0 0}]{hyperref}
\definecolor{darkred}{rgb}{0.6,0,0}
\usepackage[colorinlistoftodos]{todonotes}
\definecolor{linkcolor}{rgb}{0,0,0.5}

\newcommand {\ignore}[1]{}

\newcommand{\bea}{\begin{eqnarray}}
\newcommand{\eea}{\end{eqnarray}}

\def\gsim{\raise0.3ex\hbox{$\;>$\kern-0.75em\raise-1.1ex\hbox{$\sim\;$}}}
\def\lsim{\raise0.3ex\hbox{$\;<$\kern-0.75em\raise-1.1ex\hbox{$\sim\;$}}}

\usepackage{soul}

\definecolor{mightnightblue}{RGB}{25,25,112}

\definecolor{brown}{rgb}{0.59, 0.29, 0.0}

\newcommand{\UNISA}{\affiliation{Dipartimento di Fisica ``E.R Caianiello'', Universit\`a degli Studi di Salerno,\\ Via Giovanni Paolo II, 132 - 84084 Fisciano (SA), Italy}}
\newcommand{\INFN}{\affiliation{Istituto Nazionale di Fisica Nucleare - Gruppo Collegato di Salerno - Sezione di Napoli,\\ Via Giovanni Paolo II, 132 - 84084 Fisciano (SA), Italy.}}

\def\21{\mathrm{$SU(2)_L \otimes U(1)_Y$}}

\usepackage{fontawesome5}
\usepackage{hyperref}

\makeatletter
\newcommand{\github}[1]{%
	\href{#1}{\faGithubSquare}%
}
\makeatother

\begin{document}
\bibliographystyle{unsrt}   
\title{Probing chiral and flavored $Z^\prime$ from cosmic bursts through neutrino interactions}
\author{ShivaSankar K.A.}\email{a-shiva@particle.sci.hokudai.ac.jp}
\affiliation{Department of Physics, Hokkaido University, Sapporo 060-0810, Japan}
\author{Arindam Das}
\email{adas@particle.sci.hokudai.ac.jp}
\affiliation{Institute for the Advancement of Higher Education, Hokkaido University, Sapporo 060-0817, Japan}
\affiliation{Department of Physics, Hokkaido University, Sapporo 060-0810, Japan} 
\author{Gaetano Lambiase}\email{lambiase@sa.infn.it}
\UNISA\INFN
\author{Takaaki Nomura}\email{nomura@scu.edu.cn}
\affiliation{College of Physics, Sichuan University, Chengdu 610065, China}
\author{Yuta Orikasa}\email{yuta.orikasa@utef.cvut.cz}
\affiliation{Institute of Experimental and Applied Physics, Czech Technical University in Prague, Husova 240/5, 110 00 Prague 1, Czech Republic}
\date{\today}
\begin{abstract}
The origin of tiny neutrino mass is an unsolved puzzle leading to a variety of phenomenological aspects beyond the Standard Model (BSM). We consider $U(1)$ gauge extension of the Standard Model (SM) where so-called seesaw mechanism is incarnated with the help of thee generations of Majorana type right-handed neutrinos followed by the breaking of $U(1)$ and electroweak gauge symmetries providing anomaly free structure. In this framework, a neutral BSM gauge boson $Z^\prime$ is evolved. To explore the properties of its interactions we consider chiral (flavored) frameworks where $Z^\prime$ interactions depend on the handedness (generations) of the fermions. In this paper we focus on $Z^\prime-$neutrino interactions which could be probed from cosmic explosions. We consider $\nu \overline{\nu} \to e^+ e^-$ process which can energize gamma-ray burst (GRB221009A, so far the highest energy) through energy deposition. Hence estimating these rates we constrain $U(1)$ gauge coupling $(g_X)$ and $Z^\prime$ mass $(M_{Z^\prime})$ under Schwarzchild (Sc) and Hartle-Thorne (HT) scenarios. We also study $\nu-$DM scattering through $Z^\prime$ to constrain $g_X-M_{Z^\prime}$ plane using IceCube data considering high energy neutrinos from cosmic blazar (TXS0506+056), active galaxy (NGC1068), the Cosmic Microwave Background (CMB) and the Lyman-$\alpha$  data, respectively. Finally highlighting complementarity we compare our results with current and prospective bounds on $g_X-M_{Z^\prime}$ plane from scattering, beam-dump and $g-2$ experiments. \github{https://github.com/darth-photon/PICSHEP.git}
\end{abstract}
\maketitle
\section{Introduction}
Experimental observations of tiny neutrino mass and flavor mixing \cite{ParticleDataGroup:2020ssz} allow the Standard Model (SM) of particle physics to step beyond. Additionally, from the studies of bullet cluster, large-scale cosmological data and galaxy rotation curve we come to know that roughly one-fourth of the energy budget of the Universe has been apportioned to some non-luminous objects called dark matter (DM), which strongly suggest going beyond the SM (BSM) \cite{Planck:2018vyg, Fermi-LAT:2011vow}. A simple but interesting way to explain the origin of tiny neutrino mass is the seesaw mechanism \cite{Minkowski:1977sc,Yanagida:1979as,Gell-Mann:1979vob,Mohapatra:1979ia,Schechter:1980gr} where SM is extended by adding  singlet heavy Right Handed Neutrinos (RHNs). The latter is an appropriate realization of the idea of a dimension five operator within the SM framework \cite{Weinberg:1979sa}, where a heavy mass scale can be integrated out, followed by the violation of the lepton number by two units. 

Among a variety of BSM scenarios explaining the origin of tiny neutrino mass and flavor mixing, we study the $U(1)$ gauge extension of the SM that can also be a suitable choice to explain the origin of tiny neutrino mass at the tree level and flavor mixing since SM-singlet RHNs are naturally required to cancel anomalies associated with the general U(1) gauge symmetry. Under particular set-up such scenarios can also accommodate potential DM candidates. These scenarios give rise to a massive, neutral BSM gauge boson $Z^\prime$ after the breaking of $U(1)$ symmetry. For example, we consider a pioneering scenario called B$-$L (baryon minus lepton) \cite{Davidson:1978pm,Marshak:1979fm}, where three generations of SM-singlet RHNs are introduced to achieve an anomaly free scenario. The SM-singlet scalar field in this scenario can acquire vacuum expectation value (VEV) through the B$-$L symmetry breaking. Hence the Majorana mass terms for the RHNs can be generated which could be responsible for the light neutrinos to achieve tiny masses through the seesaw mechanism and flavor mixing. 

We consider an aspect of general $U(1)_X$ extension of the SM where three generations of RHNs can be introduced to cancel gauge and mixed gauge-gravity anomalies. In this scenario, a SM-singlet scalar field is considered, which acquires a VEV following the breaking of the general $U(1)_X$ symmetry \cite{Das:2016zue,Das:2017flq} . This leads to the generation of Majorana masses for the heavy neutrinos, giving rise to the seesaw mechanism which is responsible for generating tiny neutrino masses and flavor mixing. After solving the anomaly cancellation conditions we find that left and right-handed fermions interact differently with $Z^\prime$ manifesting the chiral nature of the model which could be probed in high energy \cite{Das:2021esm} and low energy experiments involving neutrino-electron \cite{Chakraborty:2021apc}  and neutrino-nucleon scattering \cite{Asai:2023xxl} and different beam dump experiments including FASER, ILC-beam dump and DUNE \cite{Asai:2022zxw}. Additionally we consider two more varieties of general $U(1)$ extension of the SM as $U(1)_{xq-\tau_R^3}$ and $U(1)_{q+xu}$ scenarios \cite{Appelquist:2002mw,Carena:2004xs,Hashimoto:2014ela}, respectively. In the first case we fix general $U(1)$ charges for the RHNs as well as the SM-singlet BSM scalar field and in the second case we fix general $U(1)$ charge for the left-handed fermion doublets of the SM, respectively. In these scenarios $Z^\prime$ interactions depend on the handedness (left or right) of the fermions irrespective of the generation or flavor. 

We also consider some well-known flavored scenarios where $Z^\prime$ interactions depend on the generation or flavor of the fermions irrespective of its handedness. We first consider  $L_i-L_j$ scenarios where two particular flavors of unlike leptons $(i \neq j)$ are charged under the $U(1)$ gauge group but the third generation is not. If the $i$th generation is positively charged, the $j$th generation will be negatively charged under the prescribed $U(1)$ extension \cite{Foot:1990mn,He:1990pn,He:1991qd} whereas the remaining fields are uncharged under this new gauge group. There is another alternative flavored scenario which is known as B$-3 L_i$ \cite{Lee:2010hf,Chang:2000xy,Bauer:2020itv}. Here quark charges are their respective baryon numbers $(\frac{1}{3})$ where the $i$th generation leptonic charge is $-3$ under the flavored $U(1)$ gauge groups while the remaining two generations of the leptons are uncharged. The fermions charged under the flavored gauge group only interact with the $Z^\prime$ boson. These scenarios also contain SM-singlet scalar which acquires VEV and gives rise to the Majorana mass term for the RHNs to generate the tiny neutrino mass through the seesaw mechanism. In these models $Z^\prime$ interacts with the charged leptons and neutrinos, depending on the choice of gauge structure. 

The above scenarios have interesting motivations in the context of $Z^\prime-$neutrino interactions which could be probed with different cosmic phenomena. We consider the neutrino-antineutrino annihilation into electron-positron pair production considering a $Z^\prime$ mediated process in addition to $Z$ and $W$ mediated processes from the SM. It has been pointed out in \cite{Eichler:1989ve} that electron neutrinos evolve from the accreting primary member and the anti-neutrinos evolve from the disrupted secondary member of a Neutron Star (NS) binary system accelerating towards its collapse and emitting huge energy in the form of Gamma-Ray Burst (GRB). These neutrinos and antineutrinos could annihilate into electrons and positrons out of the orbital plane causing electron-positron pair production, which is thought to energize the GRB  $\mathcal{O}(\geq 10^{51} \rm{erg})$ above the neutrino-sphere of a type-II supernova \cite{Bethe:1985sox,Co87,Goodman:1986we} by analyzing energy deposition rate. The strong gravitational field effects were investigated in \cite{Salmonson:1999es,Salmonson:2001tz} and results showed that the efficiency of the neutrino-antineutrino annihilation process, compared to the Newtonian calculations, may enhance up to a factor of $30$ in case of a collapsing NS. The energy deposition rate for an isothermal accretion disk in a Schwarzchild or Kerr metric was considered in \cite{Asano:2000ib,Asano:2000dq}. Time-dependent models in which black hole (BH) accretion disks evolving as a remnant of NS-mergers have been studied in \cite{1999A&A...344..573R}, while other models include pair-annihilation during the evolution \cite{Just:2015dba,2020ApJ...902L..27F}. These works suggest that neutrino-antineutrino annihilation in General Relativity (GR) may not be efficient enough to power GRBs. In this respect, the  Blandford-Znajek process \cite{10.1093/mnras/179.3.433} could be a promising mechanism for launching jets from spinning supermassive BH powering the accreting supermassive BH.  In this context, we mention that recently an energetic GRB (GRB221009A) has been observed \cite{Burns:2023oxn,LIGOScientific:2017ync-1,Murase:2022vqf} having isotropic energy of $E_{\rm iso} \simeq 1.2 \times 10^{55}$ erg \cite{Burns:2023oxn} which can be applied in order to constrain the effect of neutrino energy deposition involving $Z^\prime$ mediated process in the framework of GR involving Newtonian, Schwarzchild (Sc) and Hartle-Thorne (HT) metrics.

A remarkable concurrence of neutrino emission with gamma rays was observed from the blazar TXS0506+056 by the IceCube \cite{ICECUBE:2018dnn}, where a neutrino was observed with an energy around 290 TeV (IC 0922A event), with jets pointing towards the Earth. High energy neutrinos from distant astrophysical sources pass through a dense spike of DM \cite{Gondolo:1999ef,Shapiro:2022prq} neighboring the central BH. As a result, there is a high chance of neutrino-DM scattering which is challenging because observing neutrino is tough. Therefore we consider high-energy neutrinos from astrophysical sources. These high-energy neutrinos could boost DM particles if they are light, so that they can be detected in the ground-based experiments \cite{Fayet:2006sa,Boehm:2003hm,Arguelles:2017atb,Olivares-DelCampo:2017feq,Wang:2021jic}. General $U(1)$ extensions of the SM can contain potential DM candidates under proper gauge structure to ensure their stability. Such DM candidates can be charged under the $U(1)_X$ gauge group allowing to interact with $Z^\prime$. As the lepton doublets are charged under the $U(1)$ gauge group, the neutrino-DM interaction can be possible in $t-$channel which further might be constrained using IceCube data. Neutrino-DM scattering is under scanner for a long period of time through setting limits on the interactions considering the high energy neutrinos that pass through DM staring from blazar before reaching the Earth \cite{Choi:2019ixb,Blennow:2019fhy,Cline:2022qld,Ferrer:2022kei}. In this paper we perform model-dependent analyses of neutrino-DM scattering in the context of chiral and flavored extension of the SM through $Z^\prime$ in the $t-$channel where couplings of the neutrino and potential DM candidate depend on the corresponding $U(1)$ charges.

IceCube collaboration recently observed a point-like steady-state source of high energy neutrinos from a nearby active galaxy NGC1068 \cite{ICECUBE:2022der}, a radio galaxy in nature and jets pointing about $90^{\degree}$ away from line of sight \cite{Crenshaw:2000pf,Jaffe2004}, posting a landmark direct evidence of TeV neutrino emission. As a result, the Earth is exposed to equatorial emissions perpendicular to the jets. IceCube observed $79^{+22}_{-20}$ neutrinos from NGC1068 at a significance of 4.2$\sigma$ having energies between 1 TeV to 15 TeV. At the galactic center there is a supermassive BH which could be surrounded by a dense DM spike obstructing the emitted neutrinos, and resulting in neutrino-DM interaction further affecting neutrino emissions. This mechanism depends on the region closer or away from the supermassive BH allowing DM spike to play a very crucial role where both neutrinos and DM candidates are considered to be weakly coupled particles \cite{Mangano:2006mp,Boehm:2013jpa,Bertoni:2014mva,McMullen:2021ikf,Murase:2019xqi,Carpio:2022sml,Akita:2023yga}. Neutrino-DM interaction can be probed utilizing the observed high-energy neutrinos emitted from active galactic nuclei which has been studied in \cite{Cline:2023tkp} using a vector interaction under B$-$L framework. The DM-neutrino interaction also affects both the Cosmic Microwave Background (CMB) power spectrum and the late-time matter power spectrum such as the Lyman-$\alpha$ flux spectrum \cite{Mosbech:2020ahp,Hooper:2021rjc}. In this paper, we explore neutrino-DM interaction under chiral and flavor extensions. We consider three types of potential DM candidates, for example, complex scalar \cite{McDonald:1993ex,Barger:2008jx}, Majorana \cite{Okada:2020cue} and Dirac fermions \cite{FileviezPerez:2018toq,Rizzo:2021pxo}. 

We estimate bounds on the general $U(1)$ coupling $(g_X)$ with respect to $Z^\prime$ mass $(M_{Z^\prime})$ studying GRB, and comparing the enhancement due to the involvement of $Z^\prime$ using GRB 221009A. In addition to that we study neutrino-DM scattering using cosmic blazar TXS0506+056 and active galaxy NGC1068 data from IceCube to estimate bounds on $g_X-M_{Z^\prime}$ plane solving cascade equations \cite{Lipari:1993hd,Fedynitch:2015zma,Arguelles:2016zvm,Vincent:2017svp} for chiral and flavored scenarios. We obtain the bound from the Lyman-$\alpha$ data on the same plane. The bounds obtained from the cosmic bursts are compared with the bounds from existing direct experiments from $\nu_e/\bar{\nu_e}-e^-$ scattering in BOREXINO \cite{Bellini:2011rx}, TEXONO \cite{TEXONO:2009knm}, $\nu_{\mu}, \bar{\nu_{\mu}}-e^-$ scattering in CHARM-II \cite{CHARM-II:1994dzw}, coherent elastic $\nu$-nucleus scattering from COHERENT experiment \cite{COHERENT:2020ybo}, neutrino magnetic moment in GEMMA \cite{Beda:2010hk}, proton beam-dump in CHARM \cite{CHARM:1985anb}, Nomad \cite{NOMAD:2001xxt}, $\nu-$cal \cite{Blumlein:2013cua} and $e^\pm$ beam dump in Orsay \cite{Davier:1989wz}, KEK \cite{Beer:1986qr}, E141\cite{Riordan:1987aw}, E137 \cite{Bjorken:1988as},  NA64 \cite{NA64:2019auh}, NA64$_\mu$\cite{Andreev:2024sgn}, E774 \cite{Bross:1989mp} and $\nu-$decay in PS191 \cite{Bernardi:1987ek}, respectively. We consider limits on $g_X-M_{Z^\prime}$  plane from the existing long-lived particle search results \cite{LHCb:2018roe} and dark photon search \cite{LHCb:2019vmc} in LHCb, visible \cite{BaBar:2014zli} and invisible \cite{BaBar:2017tiz} decay of dark photon in BaBar and prompt production of GeV scale dark resonance search in CMS \cite{CMS:2023slr} experiments. We also compare our results with the existing bounds on $g_X-M_{Z^\prime}$ plane from light vector boson decay into $\mu^+\mu^-$ and $\pi^{\pm}$ in KLOE experiment \cite{KLOE-2:2016ydq}, light gauge boson decay into $e^- e^+$ from MAMI detector by A1 collaboration \cite{Merkel:2014avp} and the decay of dark photons into $e^- e^+$, $\mu^-\mu^+$ in association with photon from Initial State Radiation (ISR) respectively. We compare our results with the existing bounds from the dark photon decay into $\pi^0$ in NA48/2 \cite{NA482:2015wmo} experiment and dark photon search using $e^-e^+$ final state from a radiated dark gauge boson in $e^- -$nucleus scattering in APEX experiment \cite{APEX:2011dww}. We also find bounds from CCFR  \cite{CCFR:1991lpl}, BaBar 4$\mu$ \cite{Bauer:2018onh} and $4\mu$ search from CMS \cite{CMS:2018yxg} and ATLAS \cite{ATLAS:2023vxg}, respectively for $L_e-L_{\mu}$, $L_{\mu}-L_{\tau}$ and $B-3L_{\mu}$ scenarios. Finally we estimate bounds on $g_X-M_{Z^\prime}$ plane from di-lepton and dijet searches in LEP-II \cite{ALEPH:2013dgf} and dilepton searches from heavy resonance from CMS \cite{CMS:2021ctt,CMS:2016xbv} and ATALS \cite{ATLAS:2019erb,ATLAS:2017eiz} experiments in LHC respectively. We compare our results with the bounds on $g_X-M_{Z^\prime}$ plane using $g-2$ data \cite{Muong-2:2023cdq} and prospective bounds from the beam-dump scenarios in DUNE, FASER and ILC \cite{Asai:2022zxw} for heavy $Z^\prime$ in order to show complementarity.  

We arrange our paper in the following way. We describe the BSM scenarios involving the flavored and chiral models in Sec.~\ref{models}. The aspects of neutrino heating have been described in Sec.~\ref{neut-heat}. Neutrino-DM scattering has been discussed using cosmic blazar TXS0506+056 and active galaxy NGC1068 data from IceCube in Sec.~\ref{BLAGN}. We discuss the results in Sec.~\ref{dis} and finally conclude the paper in Sec.~\ref{conc}.

\section{Beyond the standard model scenarios}
\label{models}
We consider two aspects of $U(1)$ extension of the SM: in one case, left and right handed fermions are differently charged under $U(1)$ gauge group being independent of generation, and, on the other hand, we consider a flavored scenario where leptons are differently charges under $U(1)$ gauge group irrespective of the handedness. We describe the scenarios in the following:
\subsection{General U$(1)$ extensions }
General $U(1)$ extension of SM involves three generations of SM-singlet RHNs and BSM scalar. Three generation of RHNs are introduced to cancel gauge and mixed gauge-gravity anomalies. The $U(1)$ symmetry is broken by the VEV of the SM-singlet scalar helping the Majorana mass of the RHNs to be generated following the generation of light neutrino mass satisfying neutrino oscillation data and flavor mixing. Due to the breaking of $U(1)$ gauge symmetry mass of a neutral BSM gauge boson $Z^\prime$ is generated. The particle contents of general $U(1)$ extensions of the SM are given in Tab.~\ref{tab:charges}, where general charges are related by gauge and mixed gauge-gravity anomaly cancellation conditions as: 
\begin{table}[h]
	\begin{center}\label{tab:tab1}
		\begin{tabular}{| c| c || c | c |c||c|}
			\hline
			\hspace{0.5cm}Fields \hspace{0.5cm}   & \hspace{0.5cm} $SU(3)_c\otimes SU(2)_L\otimes U(1)_Y$ \hspace{0.5cm} & \hspace{0.5cm} $U(1)_X$ \hspace{0.5cm} &\hspace{0.5cm} $U(1)_{xq-\tau_R^3}$ \hspace{0.5cm}&\hspace{0.5cm} $U(1)_{q+x u}$ \hspace{0.5cm}\\
			\hline \hline
			$q_L^i$ \ \             & $(3, 2, \frac{1}{6})$\ \      & $x_q= \frac{1}{6}x_H + \frac{1}{3} x_{\Phi}$ & $x$ & $\frac{1}{3}$ \\[0.1cm]
            $u_R^i$ \ \             & $(3, 1,  \frac{2}{3})$\ \      & $x_u=\frac{2}{3}x_H + \frac{1}{3}x_{\Phi} $& $-1+4 x$ & $\frac{x}{3}$ \\[0.1cm]
            $d_R^i$ \ \             & $(3, 1, -\frac{1}{3})$\ \      & $x_d=-\frac{1}{3} x_H + \frac{1}{3} x_{\Phi} $ & $1-2x $ & $\frac{2-x}{3} $ \\[0.1cm]
            \hline \hline
			$\ell_L^i$ \ \             & $(1, 2, -\frac{1}{2})$\ \         &$x_\ell=-\frac{1}{2} x_H - x_{\Phi}$ &$-3x$ &$-1$ \\[0.1cm]
			$ e_R^i$ \ \        & $(1, 1, -1)$\ \                  & $x_e=-x_H - x_{\Phi}$ &$1-6x$ &$-(\frac{2+x}{3})$ \\[0.1cm]
            \hline \hline
			$H$       \ \  & $(1, 2, -\frac{1}{2})$\ \               & $-\frac{1}{2} x_H$ & $1-3x$ & $\frac{1-x}{3}$ \\[0.1cm]
			\hline \hline 
			$ N^j$ \ \           & $(1,1,0)$ \ \ 					    &$x_N=-x_{\Phi}$ &$-1$ &$\frac{-4+x}{3}$ \\[0.1cm]
			$\Phi$			\ \ &$(1,1,0)$ 				  \ \	    &$2\;x_{\Phi}$ &$2$ &$-2 (\frac{-4+x}{3})$ \\
			\hline \hline
		\end{tabular}
	\end{center}
	\caption{Fields of minimal $U(1)$ extensions of the SM where $i, j=1,~2,~3$ with $x_H$, $x_\Phi$ and $x$ being free parameters.}
	\label{tab:charges}
\end{table} 
\begin{align}
\left[ {\rm SU}(3)_C \right]^2 \otimes {\rm U}(1)_X&\ :&
			2x_q - x_u - x_d &\ =\  0~, \nonumber \\
 \left[ {\rm SU}(2)_L \right]^2\otimes {\rm U}(1)_X &\ :&
			3x_q + x_\ell &\ =\  0~, \nonumber \\
 \left[ {\rm U}(1)_Y \right]^2 \otimes {\rm U}(1)_X &\ :&
			x_q - 8x_u - 2x_d + 3x_\ell - 6x_e &\ =\  0~, \nonumber \\
\left[ {\rm U}(1)_X \right]^2 \otimes {\rm U}(1)_Y &\ :&
			{x_q}^2 - {2x_u}^2 + {x_d}^2 - {x_\ell}^2 + {x_e}^2 &\ =\  0~, \nonumber \\
\left[ {\rm U}(1)_X \right]^3&\ :&
			{6x_q}^3 - {3x_u}^3 - {3x_d}^3 + {2x_\ell}^3 - {x_N}^3 - {x_e}^3 &\ =\  0~, \nonumber \\
\left[ {\rm grav.} \right]^2\otimes {\rm U}(1)_X&\ :&
			6x_q - 3x_u - 3x_d + 2x_\ell - x_N - x_e &\ =\  0~, 
\label{anom-f}
\end{align}
respectively. The Yukawa interactions in this scenario can be written as
\begin{equation}
{\cal L}^{\rm Yukawa} = - Y_u^{\alpha \beta} \overline{q_L^\alpha} H u_R^\beta
                                - Y_d^{\alpha \beta} \overline{q_L^\alpha} \tilde{H} d_R^\beta
				 - Y_e^{\alpha \beta} \overline{\ell_L^\alpha} \tilde{H} e_R^\beta
				- Y_\nu^{\alpha \beta} \overline{\ell_L^\alpha} H N_R^\beta- \frac{1}{2}Y_N^\alpha \Phi \overline{(N_R^\alpha)^c} N_R^\alpha + {\rm h.c.}~
\label{LYk}   
\end{equation}
where $H$ is the SM Higgs doublet, $\tilde{H} = i \tau^2 H^*$ with $\tau^2$ being the second Pauli matrix and $\Phi$ is the SM-singlet BSM scalar respectively and from Eq.~(\ref{LYk}) we write the following conditions
\begin{eqnarray}
\frac{1}{2} x_H^{}= - x_q + x_u \ =\  x_q - x_d \ =\  x_\ell - x_e=\  - x_\ell + x_\nu~;~~~~
-2 x_\Phi^{}	&=&  2 x_N~. 
\label{Yuk}
\end{eqnarray} 
Solving Eqs.~(\ref{anom-f}) and (\ref{Yuk}) we express the $U(1)_X$ charges in terms of $x_H^{}$ and $x_\Phi^{}$ so that the $U(1)_X^{}$ charges of the SM charged fermions can be expressed as linear combination of  $U(1)_Y$ and B$-$L charges. It implies that the left and right handed fermions interact differently with $Z^\prime$. Fixing $x_\Phi^{}=1$ with $x_H^{}=-2$, the $U(1)_X$ charges of $\ell_L^i$ and $q_L^i$ reduce to zero converting the model into $U(1)_{\rm R}$ scenario. With $x_\Phi=1$ and vanishing $x_H$ the $U(1)_X$ charge assignment reduces into B$-$L scenario. With $x_\Phi=1$ taking $x_H=-1$, $-0.5$ and $1$, $U(1)_X$ charge of $e_R^i$, $u_R^i$ and $d_R^i$ will be zero. 

Considering minimal $U(1)_{xq-\tau^3_R}$ scenario and using $x_q= x$, $x_\nu= -1$ and  $x_\Phi= 2$ in Eqs.~(\ref{anom-f}) and (\ref{LYk}) we find the charges of other SM particles in Tab.~\ref{tab:charges}. It rescales $U(1)$ charges to $6xY-\tau_R^3$. Taking $x=0$ and $x=\frac{1}{3}$ we reproduce $U(1)_R$ and $U(1)_{\rm B-L}$ cases. Using $x=\frac{1}{4}$, $\frac{1}{2}$ and $x=\frac{1}{6}$ we find that $U(1)_{xq-\tau^3_R}$ charges of $u_R^i$, $d_R^i$ and $e_R^i$ become zero, respectively. Solving the anomaly cancellation conditions we find $x_u=1-2x$ and $x_d=4x-1$ as an alternative solution. However, this solution is not mentioned in our paper because it demands an extension of the scalar sector of using doublets bypassing the minimal nature of models \cite{Hashimoto:2014ela}. 


There is another scenario $U(1)_{q+xu}$  \cite{Appelquist:2002mw,Carena:2004xs} where we fix $x_q =\frac{1}{3}$, $x_\ell= -1$. Using Eqs.~(\ref{anom-f}) and (\ref{LYk}), we derive U(1)$_{q+xu}$ charges of remaining fermion and scalar sectors of the model stated in Tab.~\ref{tab:charges}. Considering $x=1$ we reproduce the B$-$L charge assignments. However, we cannot reproduce the $U(1)_{\rm R}$ scenario varying $x$. Putting $x=0$, $2$ and $-2$ the charge of $u_R^i$, $d_R^i$ and $e_R^i$ under $U(1)_{q+xu}$ gauge group will be zero. We comment that due to $x=-1$, $U(1)_\chi$ scenario is followed from $SO(10)$ grand unification. 

The renormalizable scalar potential in a singlet scalar extended scenario under a general $U(1)$ extension of SM is
\begin{align}
  V \ = \ m_h^2(H^\dag H) + \lambda_H^{} (H^\dag H)^2 + m_\Phi^2 (\Phi^\dag \Phi) + \lambda_\Phi^{} (\Phi^\dag \Phi)^2 + \lambda^\prime (H^\dag H)(\Phi^\dag \Phi).
  \label{pot1x}
\end{align}
We assume $\lambda^\prime$ to be very small for simplicity. After the breaking of $U(1)_X^{}$ gauge and electroweak symmetries, the scalar fields $H$ and $\Phi$ develop their vacuum expectation values (VEVs) as 
\begin{align}
  \braket{H} \ = \ \frac{1}{\sqrt{2}}\begin{pmatrix} v+h\\0 
  \end{pmatrix}~, \quad {\rm and}\quad 
  \braket{\Phi} \ =\  \frac{v_\Phi^{}+\phi}{\sqrt{2}}~
  \label{scalar-1}
\end{align}
where at the potential minimum the electroweak scale is demarcated as $v=246$ GeV and $v_\Phi^{}$ is taken to be a free parameter. The breaking of the general $U(1)$ generates the mass of $Z^\prime$ which is given by $M_{Z^\prime}^{}=   g_X^{} \sqrt{ 4 x_\Phi^2 v_\Phi^{2}+ x_H^2 v^2}$ which reduces to $M_{Z^\prime} \simeq 2 g_X x_\Phi v_\Phi$ under the consideration of $v_\Phi \gg v$. Without the loss of generality we consider $x_\Phi^{}=1$. The quantity $M_{Z^\prime}$ is a free parameter and $g_X^{}$ is the general $U(1)$ coupling considered for the cases shown in Tab.~\ref{tab:charges}. From the Yukawa interactions given in Eq.~(\ref{LYk}), we find that the RHNs interact with the SM-singlet scalar field $\Phi$ generating the Majorana mass for the RHNs after the breaking of general $U(1)$ symmetry. After the electroweak symmetry breaking, the Dirac mass term is generated from the interaction among $H$, $\ell_L$ and $N_R$. These two mass terms allow the seesaw mechanism to originate the tiny neutrino masses and flavor mixing. The corresponding mass terms are $m_{N_R^\alpha}^{} \ = \ \frac{Y^\alpha_{N}}{\sqrt{2}} v_\Phi^{}$ and $m_{D}^{\alpha \beta} \  =  \ \frac{Y_{\nu}^{\alpha \beta}}{\sqrt{2}} v$. Hence we obtain the neutrino mass matrix as 
\begin{equation}
   m_\nu= \begin{pmatrix} 0&m_D^{}\\ m_D^T&m_N^{} \end{pmatrix}~
\label{num-1}
\end{equation}
and diagonalizing Eq.~(\ref{num-1}), we obtain the light neutrino mass eigenvalue as $-m_D^{} m_N^{-1} m_D^T$. Then the light neutrino flavor eigenstate $\nu^i_L$, from $\ell_L^i$, is written by light and heavy mass eigenstates $\nu_\alpha$ and $N_\alpha$ as $\nu^i_L \simeq U_{i\alpha} \nu_\alpha + V_{i \alpha} N_\alpha$ where $U_{i \alpha} \simeq (U_{\rm PMNS})_{i \alpha}$ and $V_{i \alpha} \simeq (m_D m_N^{-1})_{i \alpha}$ with $U_{\rm PMNS}$ being PMNS matrix. The RHNs interact with the SM gauge bosons through mixing. The mixing among light and heavy neutrinos are very small as we consider $m_D/m_N \ll 1$ and we ignore it when we calculate $Z^\prime$ decay below. The $Z^\prime$ interaction with the SM leptons under general $U(1)$ scenarios, where left and right handed SM fermions interact differently with $Z^\prime$ and the interaction Lagrangian, can be written as
\bea
\mathcal{L}^{f} = -g_X (\overline{f_L}\gamma^\mu q_{f_{L}^{}}^{}  f_L+ \overline{f_R}\gamma^\mu q_{f_{R}^{}}^{}  f_R) Z_\mu^\prime.
\label{Lag1}
\eea
where $q_{f_{L}^{}}^{}$ and $q_{f_{R}^{}}^{}$ are general $U(1)$ charges of the SM left and right handed fermions from Tab.~\ref{tab:charges}, manifesting chiral scenario. We calculate the partial decay widths of $Z^\prime$ into different SM charged fermions for a single generation as
\begin{align}
\label{eq:width-ll}
\Gamma(Z^\prime \to \bar{f} f)
= N_C^{} \frac{M_{Z^\prime}^{} g_{X}^2}{24 \pi} \left[ \left( q_{f_L^{}}^2 + q_{f_R^{}}^2 \right) \left( 1 - \frac{m_f^2}{M_{Z^\prime}^2} \right) + 6 q_{f_L^{}}^{} q_{f_R^{}}^{} \frac{m_f^2}{M_{Z^\prime}^2} \right]
\sqrt{1-\frac{4 m_f^2}{M_{Z^\prime}^2}}~,
\end{align}    
where $m_f$ is the SM fermion mass and $N_C^{}=1(3)$ for the SM leptons(quarks) being the color factor. 

The partial decay width of $Z^\prime$ into a pair of single generation light neutrinos is given by 
\begin{align}   
\label{eq:width-nunu}
    \Gamma(Z^\prime \to \nu \nu)
    =  \frac{M_{Z^\prime}^{} g_{X}^2}{24 \pi} q_{f_L^{}}^2~
\end{align} 
where tiny neutrino mass has been neglected. Here $q_{f_L^{}}^{}$ stands for general $U(1)$ charge of $\ell_L$ from Tab.~\ref{tab:charges}. In general $U(1)$ extended SM scenarios, $Z^\prime$ decays into a pair of heavy Majorana neutrinos following 
\bea
\mathcal{L}_N= -\frac{1}{2}g_X q_{N_R} \overline{N} \gamma_\mu \gamma_5 N Z_{\mu}^\prime.
\label{neut}
\eea
We calculate the corresponding partial decay width of $Z^\prime$ into a pair of single generation of the heavy neutrino as
\begin{align}
\label{eq:width-NN}
    \Gamma(Z^\prime \to N_R^\alpha N_R^\alpha)
    = \frac{M_{Z^\prime}^{} g_{X}^2}{24 \pi} q_{N_R^{}}^2 \left( 1 - \frac{4 M_N^2}{M_{Z^\prime}^2} \right)^{\frac{3}{2}}~
\end{align}
with $q_{N_R^{}}^{}$ is the general $U(1)$ charge of the RHNs from Tab.~\ref{tab:charges} and $M_N^{}$ is the RHN mass\footnote{In this analysis we considered the active-sterile mixing to be negligibly small. After the light-heavy mixing, the flavor eigenstate will be expressed in terms of mass eigenstates following $\nu_L^i \simeq U_{i \alpha} \nu_\alpha+ V_{i \alpha} N_\alpha$. In our analysis we considered that $|U_{i\alpha}|^2 \simeq 1$ under the effect of non-unitarity, lepton flavor violation as studied in \cite{Das:2012ze,Das:2017nvm,Das:2019fee,Chiang:2019ajm} reproducing neutrino oscillation data. Therefore Eq.~(\ref{eq:width-nunu}) will not be greatly affected. However, from these studies we find that limit on the active-sterile mixing can be less that $\mathcal{O}(10^{-5})$ providing a strong bound. Due to active-sterile mixing, a new mode $Z^\prime \to N \nu$ could appear and the corresponding partial decay-width will be proportional to $g_X^2 |V_{i\alpha}|^2$ producing an effect of less than $g_X^2 \mathcal{O}(10^{-10})$. As a result we did not take this aspect in consideration.}.  

\begin{figure}[h]
\includegraphics[width=0.49\textwidth,angle=0]{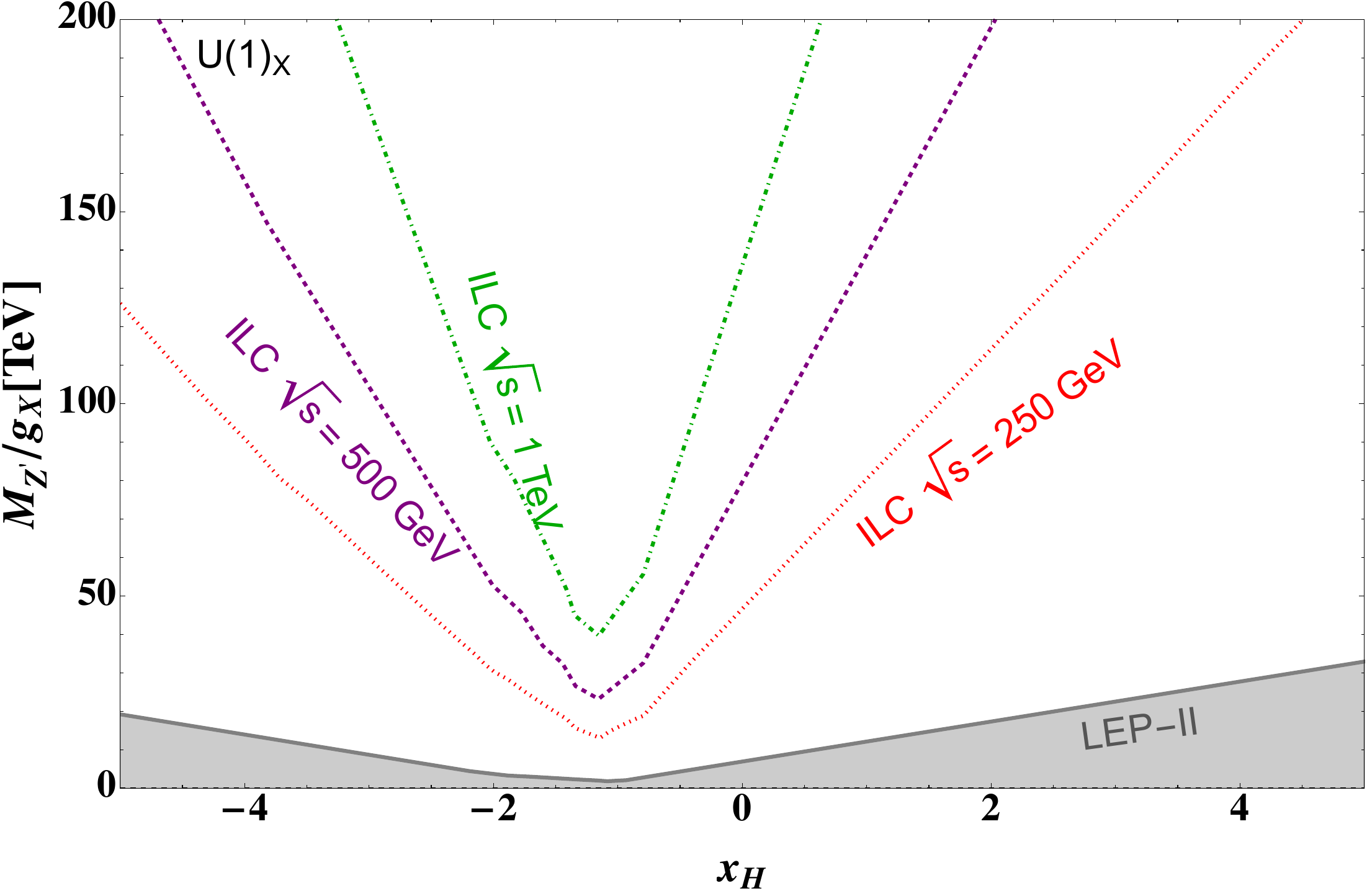}
\includegraphics[width=0.495\textwidth,angle=0]{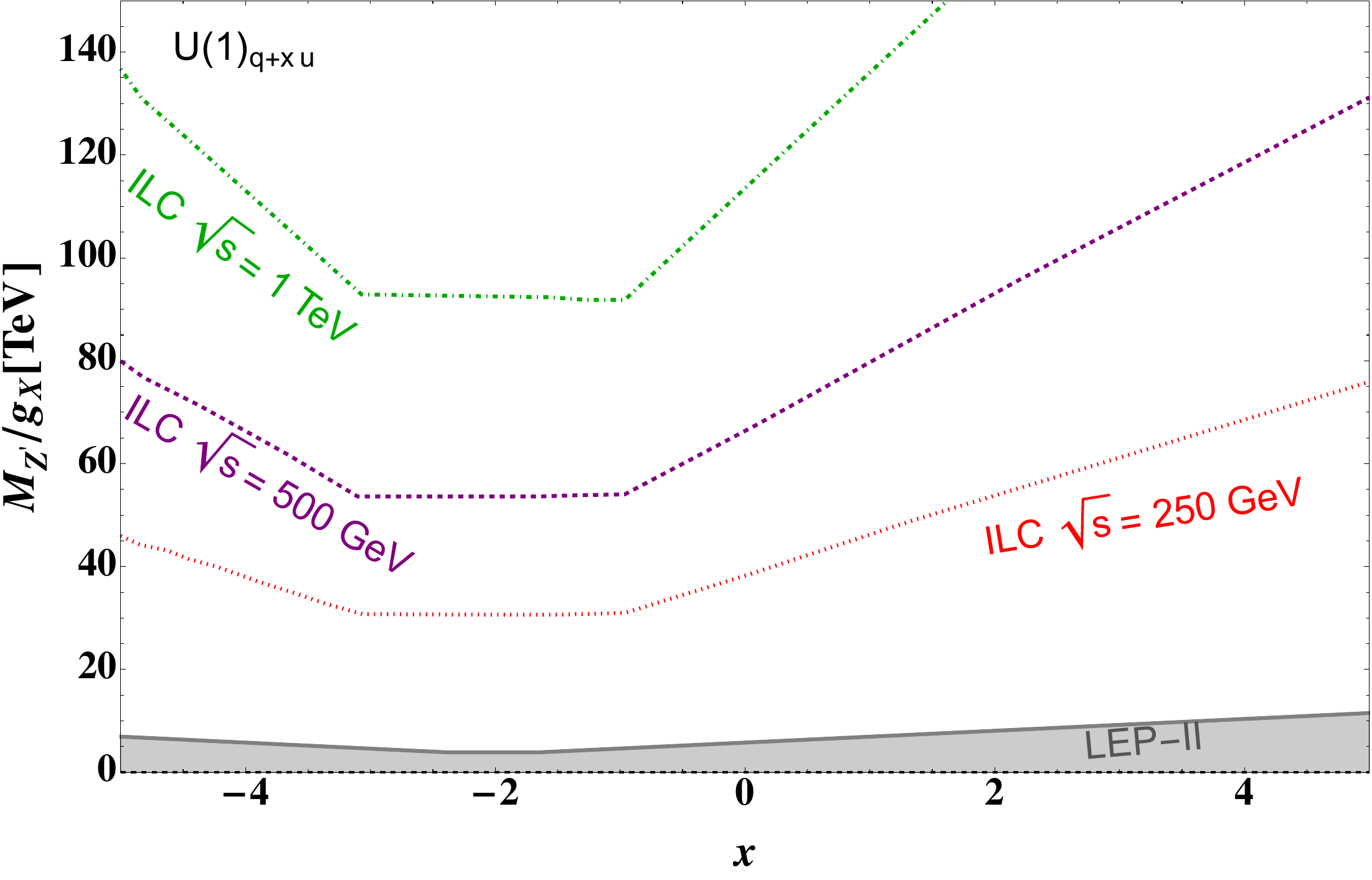}
\includegraphics[width=0.49\textwidth,angle=0]{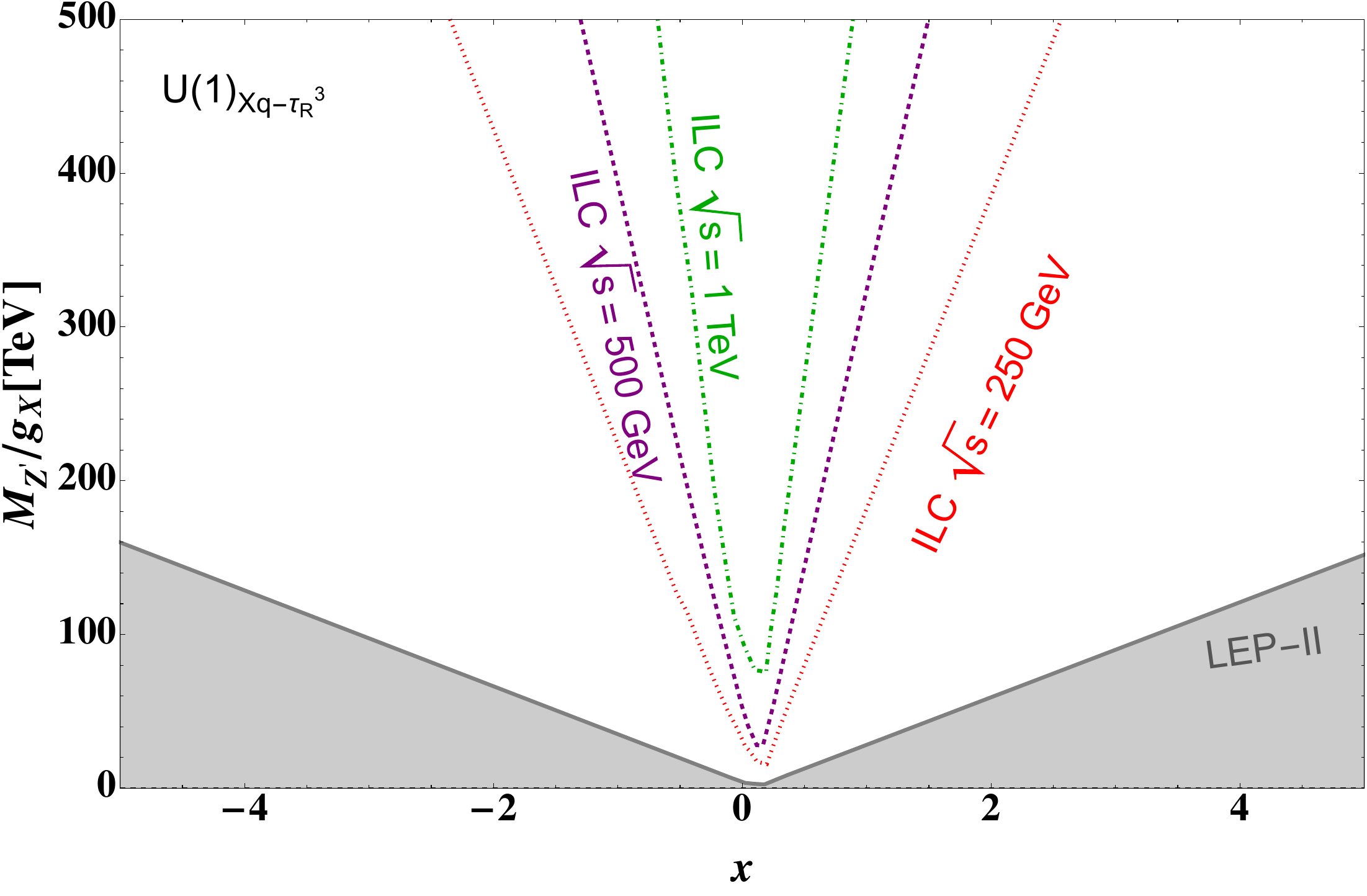}
\caption{Current LEP-II bounds (grey-shaded, ruled out) and prospective limits from ILC on $M_{Z^\prime}/g_X$ at 95$\%$ C.L.}
\label{gX1}
\end{figure}

The fermionic interactions in the general $U(1)$ scenario can be tested at $e^-e^+$ colliders being mediated by $Z^\prime$ which is much heavier than the center of mass energy $(\sqrt{s})$ of the collider. We can parametrize the $e^+e^-\to f\bar{f}$ process through contact interaction described by the effective Lagrangian \cite{Eichten:1983hw,LEP:2003aa,Schael:2013ita}  
\bea
{\cal L}_{\rm eff} \ = \ \frac{g_X^2}{(1+\delta_{ef}) (\Lambda_{AB}^{f\pm})^2} \sum_{A,B=L,R}\eta_{AB}(\overline{e} \gamma^\mu P_A e)(\overline{f} \gamma_\mu P_B f) \, 
\label{eq1}
\eea
where $g^2_X/4\pi$ is taken to be $1$ by convention, $\delta_{ef}=1\ (0)$ for $f=e$ ($f\neq e$),  $\eta_{AB}=\pm 1$ or 0, and $\Lambda_{AB}^{f\pm}$ is the scale of the contact interaction, having either constructive ($+$) or destructive ($-$) interference with the SM processes of fermion pair production \cite{Kroha:1991mn,Carena:2004xs}. We give the analytical expression of $Z^\prime$ exchange matrix element involved in $e^-e^+ \to f \overline{f}$ process as 
\bea
\frac{g_X^2}{M_{Z^\prime}^2-s} [\overline{e} \gamma^\mu (x_\ell P_L+ x_e P_R) e] [\overline{f} \gamma_\mu (x_{f_L} P_L+x_{f_R} P_R) f] \, 
\label{eq2}
\eea
where $x_{\ell (e)}$ and $x_{f_{L(R)}}$ are general $U(1)$ charges of $e_{L (R)}$ and $f_{L(R)}$ from Tab.~\ref{tab:charges}. Matching Eqs.~(\ref{eq1}) and (\ref{eq2}) under the limit $M_{Z^\prime} \gg \sqrt{s}$, where $\sqrt{s}_{\rm LEP}=209$ GeV we find the bound on $M_{Z^\prime}$ as
\bea
M_{Z^\prime}^2  \ \gtrsim \ \frac{{g_X}^2}{4\pi} |{x_{e_A}} x_{f_B}| (\Lambda_{AB}^{f\pm})^2. 
\label{Lim}
\eea
Bounds on effective scales $(\Lambda_{AB}^{f\pm})$ for different fermions with all possible chirality structures $AB=LL, RR, LR, RL, VV$ and $AA$ at $95\%$ C. L. from LEP-II \cite{Schael:2013ita} where $f$ stands for different leptons and quarks. Hence we estimate strongest limits on $M_{Z^\prime}/ g_X$ combining the results  obtained for different leptons and quarks in the final state using Eq.~(\ref{Lim}). This quantity further defines the VEV of general $U(1)$ theories. Similarly we calculate prospective bounds of $M_{Z^\prime}/g_X$ at the ILC using $\sqrt{s}=$250 GeV, 500 GeV and 1 TeV for different $x_H$ using the limits on $\Lambda_{AB}^{f\pm}$ where $AB=LL, RR, VV$ and $AA$ obtained at $95\%$ C. L. from \cite{LCCPhysicsWorkingGroup:2019fvj}. The limits are shown in Fig.~\ref{gX1} at 95$\%$ CL and listed in Tab.~\ref{tab3} for different charges showing similarities between $U(1)_X$ and $U(1)_{x q-\tau_R^3}$ scenarios while $U(1)_{q+xu}$ scenario is different from the rest due to $U(1)$ charges.  
\begin{table}[h]
\begin{center}
\begin{tabular}{|c|c|c|c|c|c|c|c|c|}
\hline
\multirow{2}{*}{Machine} & \multirow{2}{*}{$\sqrt s$} & \multicolumn{7}{|c|}{95\% CL lower limit on the $M_{Z^\prime}/g_X$ of U$(1)_X$ scenario (in TeV)} \\ \cline{3-9}
& & $x_H=-2$ &  $x_H=-1$ & $x_H=-0.5$ & $x_H=0$ & $x_H=0.5$ & $x_H=1$ & $x_H=2$ \\ \hline
LEP-II & 209 GeV & 5.0 & 2.2& 4.4& 7.0 &10.3&11.1 &18.0 \\ \hline
\multirow{3}{*}{ILC} & 250 GeV & 31.6 &16.3 &29.5 & 48.2 &64.3 &79.0 &113.7 \\ \cline{2-9}
& 500 GeV & 54.4 &26.3 & 50.1& 81.6 &110.2 &139.1 &199.7 \\ \cline{2-9}
 & 1 TeV & 88.6 &47.7 & 84.8& 137.2 &185.8 & 238.2&339.2 \\ 
\hline
\hline
\multirow{2}{*} & \multirow{2}{*} & \multicolumn{7}{|c|}{95\% CL lower limit on the $M_{Z^\prime}/g_X$ of U$(1)_{xq-\tau_R^3}$ scenario (in TeV)} \\ \cline{3-9}
& & $x=-2$ &  $x=0$ & $x=\frac{1}{3}$ & $x=0.25$ & $x=0.5$ & $x=\frac{1}{6}$ & $x=2$ \\ \hline
LEP-II & 209 GeV & 60.3 & 5.0& 7.0& 4.4 &11.1&2.2 &56.6 \\ \hline
\multirow{3}{*}{ILC} & 250 GeV & 415.9 &31.6 &48.2 & 29.5 &ds79.0 &16.3 &378.0 \\ \cline{2-9}
& 500 GeV & 728.7 &54.4 & 81.6& 50.1 &139.1 &26.3 &673.1 \\ \cline{2-9}
 & 1 TeV & 1272.6 &88.6 & 137.2& 84.8 &238.2 & 47.7&1163.4 \\ 
\hline
\hline
\multirow{2}{*} & \multirow{2}{*} & \multicolumn{7}{|c|}{95\% CL lower limit on the VEV of U$(1)_{q+x u}$ scenario (in TeV)} \\ \cline{3-9}
& & $x=-2$ &  $x=-1$ & $x=-0.5$ & $x=0$ & $x=0.5$ & $x=1$ & $x=2$ \\ \hline
LEP-II & 209 GeV & 3.2 & 4.2& 4.65& 5.4 &5.9&7.0 &7.4 \\ \hline
\multirow{3}{*}{ILC} & 250 GeV & 30.3 &30.2 &33.8 &37.5 &41.5 &48.2 &53.5 \\ \cline{2-9}
& 500 GeV & 53.1 &53.4 & 59.8& 66.5 &72.1 &81.6 &93.2 \\ \cline{2-9}
 & 1 TeV & 92.6 &91.1 & 101.9& 114.0 &124.3 & 137.2&158.0 \\ 
\hline
\end{tabular}
\end{center}
\caption{Lower limits on $M_{Z^\prime}/g_X$ at $95\%$ CL in general $U(1)$ extensions from $e^+e^-\to f\bar{f}$ processes \cite{Schael:2013ita, LCCPhysicsWorkingGroup:2019fvj}.}
\label{tab3}
\end{table}
\subsection{Flavored scenarios}
{\it \bf  $L_i-L_j$ scenarios:} We find anomaly free, leptonic flavor dependent gauged $L_e-L_\mu$, $L_e-L_\tau$ and $L_\mu-L_\tau$ scenarios where  $Z^\prime$ mass is originated after the spontaneously broken $U(1)_{L_i-L_j}$  symmetry. Under this gauge group, $i(j)$th flavor of the lepton is positively(negatively) charged whereas the remaining fields are uncharged under $U(1)_{L_i-L_j}$ gauge group. The particle content of the scenarios are given in Tab.~\ref{tab:charges2}.
\begin{table}[h]
	\begin{center}
 \label{tab:tab2}
		\begin{tabular}{| c| c || c | c | c| c|}
			\hline
			\hspace{0.5cm}Fields \hspace{0.5cm}   & \hspace{0.5cm}$SU(3)_c\otimes SU(2)_L\otimes U(1)_Y$\hspace{0.5cm} & \hspace{0.5cm} $U(1)_{L_{e}-L_{\mu}}$ \hspace{0.5cm}&$U(1)_{L_{\tau}-L_{e}}$ \hspace{0.5cm} &$U(1)_{L_{\mu}-L_{\tau}}$ \hspace{0.5cm} \\
			\hline \hline
			$q_L^i$ \ \             & $(3, 2, \frac{1}{6})$\ \      &0 &0&0 \\[0.1cm]
            $u_R^i$ \ \             & $(3, 1,  \frac{2}{3})$\ \      &0 &0&0 \\[0.1cm]
            $d_R^i$ \ \             & $(3, 1, -\frac{1}{3})$\ \      &0 &0&0 \\[0.1cm]
            \hline \hline
			$\ell_L^i$ \ \             & $(1, 2, -\frac{1}{2})$\ \         &$\{1,-1,0\}$ & $\{-1,0,1\}$ &$\{0,1,-1\}$ \\[0.1cm]
			$ e_R^i$ \ \        & $(1, 1, -1)$\ \                  &$\{1,-1,0\}$ &$\{-1,0,1\}$ &$\{0,1,-1\}$ \\[0.1cm]
           \hline \hline
			$H$       \ \  & $(1, 2, -\frac{1}{2})$\ \               & 0&0&0  \\[0.1cm]
			\hline \hline 
			$ N^j$ \ \           & $(1,1,0)$ \ \ 					    & $\{1,-1,0\}$ &$\{-1,0,1\}$ &$\{0,1,-1\}$ \\[0.1cm]
			$\Phi$			\ \ &$(1,1,0)$ 				  \ \	    &$-2$& $-2$ &$-2$ \\
			\hline \hline
		\end{tabular}
	\end{center}
	\caption{$L_i-L_j$ extensions of the SM where $i$th and $j$th fermions being oppositely charged under $L_i-L_j$.}
	\label{tab:charges2}
\end{table} 
The Yukawa interactions in this scenario can be written in a general form as 
\bea
{\cal L}^{\rm Yukawa} &=& - Y_u^{\alpha \beta} \overline{q_L^\alpha} H u_R^\beta
                                - Y_d^{\alpha \beta} \overline{q_L^\alpha} \tilde{H} d_R^\beta
				 - Y_e^{\alpha \beta} \overline{\ell_L^\alpha} \tilde{H} e_R^\beta- Y_\nu^{i} \overline{\ell_L^i} H N_R^i \nonumber
                - Y_\nu^{j} \overline{\ell_L^j} H N_R^j 
                \\
				&&-\frac{1}{2}Y_N^i \Phi \overline{(N_R^i)^c} N_R^i-\frac{1}{2}Y_N^j \Phi^* \overline{(N_R^j)^c} N_R^j 
				-Y_\nu^{k} \overline{\ell_L^k} H N_R^k+ {\rm h.c.}- \frac{1}{2}M_N^k \overline{{N_R^k}^c} N_R^k
                -\frac{1}{2} M_N^{ij} \overline{{N_R^i}^c} N_R^j
\label{LYk-Fl-1}   
\eea
where $\alpha, \beta$ denote three generations, $i (j)$ denote the  $i(j)$th generations of leptons with $k\equiv 6-i-j$. The renormalizable scalar potential in this scenario is the same as in Eq.~(\ref{pot1x}) and the VEVs of scalar fields are also given in the same way. 
Since the SM Higgs doublet does not have $U(1)_{L_i-L_j}^{}$ charge the $Z^\prime$ mass is written as $ M_{Z^\prime}^{}=  2 g_X^{} v_\Phi^{}$ where $g_X^{}$ is flavor conserving $U(1)_{L_i-L_j}$ gauge coupling. Following Eq.~(\ref{LYk-Fl-1}) Majorana and Dirac masses of RHNs are generated after the breaking of $U(1)_{L_i-L_j}$ and electroweak symmetries $(\tilde{m}_D^{ii, jj, kk}= \frac{Y_\nu^{i,j, k}}{\sqrt{2}}v, \tilde{m}_N^{i(j)}= \frac{Y_N^{i (j)}}{\sqrt{2}} v_\Phi)$. There are also a direct Majorana mass terms with $M^k_N$ and $M^{ij}_N$ for RHNs. From Dirac and Majorana masses, seesaw mechanism  is evolved to generate tiny neutrino mass and flavor mixing as we discussed below Eq.~(\ref{num-1}). Note that the structure of the Dirac and Majorana mass matrices is constrained in this case due to flavor dependent $U(1)_{L_i - L_j}$ charge. In fact the structure of the Majorana mass matrix of RHNs changes by the charge of $\Phi$ where we fixed it as $-2$ for simplicity. Since the neutrino mass is not our focus, in this work we do not discuss the realization of neutrino masses and mixings in this paper. The interaction Lagrangian between $Z^\prime$ and SM leptons can be written as 
\bea
\mathcal{L}^{f} = -g_X (\overline{f_L^i}\gamma^\mu   f_L^i+ \overline{f_R^i}\gamma^\mu   f_R^i- \overline{f_L^j}\gamma^\mu   f_L^j- \overline{f_R^j}\gamma^\mu   f_R^j) Z_\mu^\prime~,
\eea
where $f^{i(j)}$ denotes $i(j)$th $i\neq j$ generation of SM leptons. Hence partial decay widths of $Z^\prime$ into corresponding charged leptons, neutrinos and heavy neutrinos can be written as
\bea
\label{eq:width-ll-1}
    \Gamma(Z^\prime \to \bar{\ell^{i (j)}} \ell^{i (j)})&=& \frac{M_{Z^\prime}^{} g_{X}^2}{12 \pi} \left( 1 + 2 \frac{m_f^2}{M_{Z^\prime}^2} \right) \sqrt{1-\frac{4 m_f^2}{M^2_{Z^\prime}}},  \nonumber \\
    \Gamma(Z^\prime \to \nu^{i (j)} \nu^{i (j)})&=&  \frac{M_{Z^\prime}^{} g_{X}^2}{24 \pi},  \Gamma(Z^\prime \to N_R^{i(j)} N_R^{i(j)})= \frac{M_{Z^\prime}^{} g_{X}^2}{24 \pi}  \left( 1 - \frac{4 M_N^2}{M_{Z^\prime}^2} \right)^{\frac{3}{2}}    
\eea 
respectively where $m_f$, $M_{N}^{}$ being SM lepton and heavy neutrino masses. LEP-II constraint on $L_e-L_{\mu, \tau}$ scenario is $M_{Z^\prime}/g_X > 7$ TeV comparing the effective scales using $e^-e^+ \to f \overline{f}$ process following Eqs.~(\ref{eq1})-(\ref{Lim}). The prospective limits at 250 GeV, 500 GeV and 1 TeV ILC will be $M_{Z^\prime}/g_X > 48.2$ TeV, $81.6$ TeV and $137.2$ TeV respectively and $e^-e^+ \to f \bar{f}$ process can not constrain $L_{\mu}-L_{\tau}$ scenario where $e^\pm$ have no direct coupling with $Z^\prime$.

{\it \bf $B-3L_i$ scenarios:} We consider anomaly free $B-3L_i$ scenario where one of the three generations of the $\ell_L^i$ and $e_R^i$ is charged under $U(1)_{B-3L_i}$ gauge group while rest of the two generations are not. The particle content is given in Tab.~\ref{tab:charges3-1}. The RHNs being SM-singlet in this scenario follow the same footprints of the charged leptons. For example, if the first generation of $\ell_L^i$ and $e_R^i$ are charged under $U(1)_{B-3L_i}$ then the first generation of $N^i$ has $-3$ charge under this gauge group whereas second and third generations are not making the scenario anomaly free. Under this gauge group SM Higgs is uncharged but the SM quarks have $\frac{1}{3}$ charge. 
\begin{table}[h]
	\begin{center}
		\begin{tabular}{| c| c || c | c | c| c|}
			\hline
			\hspace{0.5cm}Fields \hspace{0.5cm}   & \hspace{0.5cm}$SU(3)_c\otimes SU(2)_L\otimes U(1)_Y$\hspace{0.5cm} & \hspace{0.5cm} $U(1)_{B-3L_{e}}$ \hspace{0.5cm}&$U(1)_{B-3 L_{\mu}}$ \hspace{0.5cm} &$U(1)_{B-3 L_{\tau}}$ \hspace{0.5cm} \\
			\hline \hline
			$q_L^i$ \ \             & $(3, 2, \frac{1}{6})$\ \      &$\frac{1}{3}$ &$\frac{1}{3}$&$\frac{1}{3}$ \\[0.1cm]
            $u_R^i$ \ \             & $(3, 1,  \frac{2}{3})$\ \      &$\frac{1}{3}$ &$\frac{1}{3}$&$\frac{1}{3}$ \\[0.1cm]
           $d_R^i$ \ \             & $(3, 1, -\frac{1}{3})$\ \      &$\frac{1}{3}$ &$\frac{1}{3}$&$\frac{1}{3}$ \\[0.1cm]
            \hline \hline
			$\ell_L^i$ \ \             & $(1, 2, -\frac{1}{2})$\ \         &\{-3,0,0\}&\{0,-3,0\}&\{0,0,-3\} \\[0.1cm]
			$ e_R^i$ \ \        & $(1, 1, -1)$\ \                  &\{-3,0,0\} &\{0,-3,0\}&\{0,0,-3\} \\[0.1cm]
            \hline \hline
			$H$       \ \  & $(1, 2, -\frac{1}{2})$\ \               & 0&0&0  \\[0.1cm]
			\hline \hline 
			$ N^i$ \ \           & $(1,1,0)$ \ \ 					    &\{-3,0,0\}&\{0,-3,0\}&\{0,0,-3\} \\[0.1cm]
			$\Phi$			\ \ &$(1,1,0)$ 				  \ \	    &6&6&6 \\
			\hline \hline
		\end{tabular}
	\end{center}
	\caption{$B-3L_i$ extensions of the SM where quarks and $i$th fermions charged under $B-3L_i$ scenario.}
	\label{tab:charges3-1}
\end{table} 
We introduce an SM-singlet BSM scalar $\Phi$ with charge $6$ under $U(1)_{B-3L_i}$ gauge group which generates Majorana mass of $i$th RHNs, whereas the remaining generations of RHNs could have a Majorana mass term directly following the Yukawa interaction 
\bea
{\cal L}^{\rm Yukawa} &=& - Y_u^{\alpha \beta} \overline{q_L^\alpha} H u_R^\beta
                                - Y_d^{\alpha \beta} \overline{q_L^\alpha} \tilde{H} d_R^\beta
				 - Y_e^{\alpha \beta} \overline{\ell_L^\alpha} \tilde{H} e_R^\beta
		- \sum_{\alpha \neq i} \sum_{\beta \neq i} Y_\nu^{\alpha \beta} \overline{\ell_L^\alpha} H N_R^\beta
       -Y_\nu^{ii} \overline{\ell_L^i} H N_R^i \nonumber \\
				&&-\frac{1}{2}Y_N^{ii} \Phi \overline{(N_R^i)^c} N_R^i+ {\rm h.c.}-
                \sum_{j \neq i}\frac{1}{2}\tilde{M}_N^{jj} \overline{(N_R^j)^c} N_R^j~~~~~~~
\label{Yuk-3l}
\eea
The Dirac and Majorana masses are generated following $U(1)_{B-3L_i}$ and the electroweak symmetry breaking $(\tilde{m}_D^{\alpha \beta} = \frac{Y_\nu^{\alpha \beta} v}{\sqrt{2}}, \tilde{m}_D^{ii}= \frac{Y_\nu^{ii} v}{\sqrt{2}}, \tilde{m}_N^{ii}= \frac{Y_N^{ii} v_\Phi}{\sqrt{2}})$ which evolve seesaw mechanism to originate tiny neutrino mass and flavor mixing as Eq.~(\ref{num-1}) and below the equation. The structure of mass matrix is restricted by $U(1)_{B-3L_i}$ symmetry as in the $U(1)_{L_i - L_j}$ case. The scalar potential of this scenario is given in Eq.~(\ref{pot1x}) and following Eq.~(\ref{scalar-1}) we find the $Z^\prime$ mass $M_{Z^\prime}= 6 g_X v_\Phi$ with $v_\Phi \gg v$. The interaction Lagrangian between $Z^\prime$ and SM fermions is
\bea
\mathcal{L}^{f} = -g_X (\overline{q_L^i}\gamma^\mu  Q_L^{q}  q_L^i+ \overline{q_R^i} \gamma^\mu Q_R^{q}  q_R^i+\overline{f_L^i}\gamma^\mu Q_L^{\ell^i}  f_L^i
+\overline{f_R^i}\gamma^\mu Q_R^{\ell^i}  f_R^i) Z_\mu^\prime
\label{Lag1-1}
\eea
where $Q_{L(R)}^{q(\ell^{i})}$ is left (right) handed charge of the quarks ($i$th lepton) under $U(1)_{B-3L_i}$. According to Eq.~(\ref{Lag1-1}) we calculate the partial decay width of the $Z^\prime$ into a pair of quarks, leptons, light neutrinos and RHNs as
\bea
 \Gamma(Z^\prime \to q \bar{q})&=&  \frac{M_{Z^\prime}^{} g_{X}^2}{36 \pi} \left( 1 + 2 \frac{m_q^2}{M_{Z^\prime}^2} \right) \sqrt{ 1 - \frac{4 m_q^2}{M_{Z^\prime}^2} }, \; \;
 \Gamma(Z^\prime \to \bar{\ell^i} \ell^i)= \frac{3 M_{Z^\prime}^{} g_{X}^2}{4 \pi} \left( 1 + 2 \frac{m_\ell^2}{M_{Z^\prime}^2} \right) \sqrt{ 1 - \frac{4 m_\ell^2}{M_{Z^\prime}^2} },\nonumber \\
\Gamma(Z^\prime \to \nu^i \nu^i)&=&  \frac{3 M_{Z^\prime}^{} g_{X}^2}{8 \pi}, \; \;
\Gamma(Z^\prime \to N_R^i N_R^i)= \frac{3 M_{Z^\prime}^{} g_{X}^2}{8 \pi}  \left( 1 - \frac{4 {M_N^i}^2}{M_{Z^\prime}^2} \right)^{\frac{3}{2}}
\eea
respectively where $m_{q(\ell)}$ is the quark(lepton) mass. The constraint on $B- 3L_{e}$ scenario from LEP-II is $M_{Z^\prime}/g_X > 2.33$ TeV comparing the limits on the effective scales using $e^-e^+ \to f \bar{f}$ process for $M_{Z^\prime} > > \sqrt{s}$ following Eqs.~(\ref{eq1})-(\ref{Lim}). Following the same manner prospective bounds at the $250$ GeV, $500$ GeV and $1$ TeV ILC can be estimated $M_{Z^\prime}/g_X >$ 16.07 TeV, 27.20 TeV and 45.73 TeV respectively. Using this method $B-3L_{\mu, \tau}$ scenarios can not be constrained because $e^\pm$ have no direct coupling with $Z^\prime$. \footnote{We assumed that kinetic mixing parameter between $Z$ and $Z^\prime$ bosons is negligibly small for simplicity. Due to a sizable kinetic mixings we will have additional processes which will be useful to estimate bounds on the model parameters specially in those cases where $Z^\prime$ has vanishing/ restricted interactions with the fermions. A detailed analysis in this aspect has been kept for a future study where the effect kinetic mixing will be enlightened.} 
\section{Theory of neutrino heating}
\label{neut-heat}
\begin{figure}[t!]
\includegraphics[width=140mm]{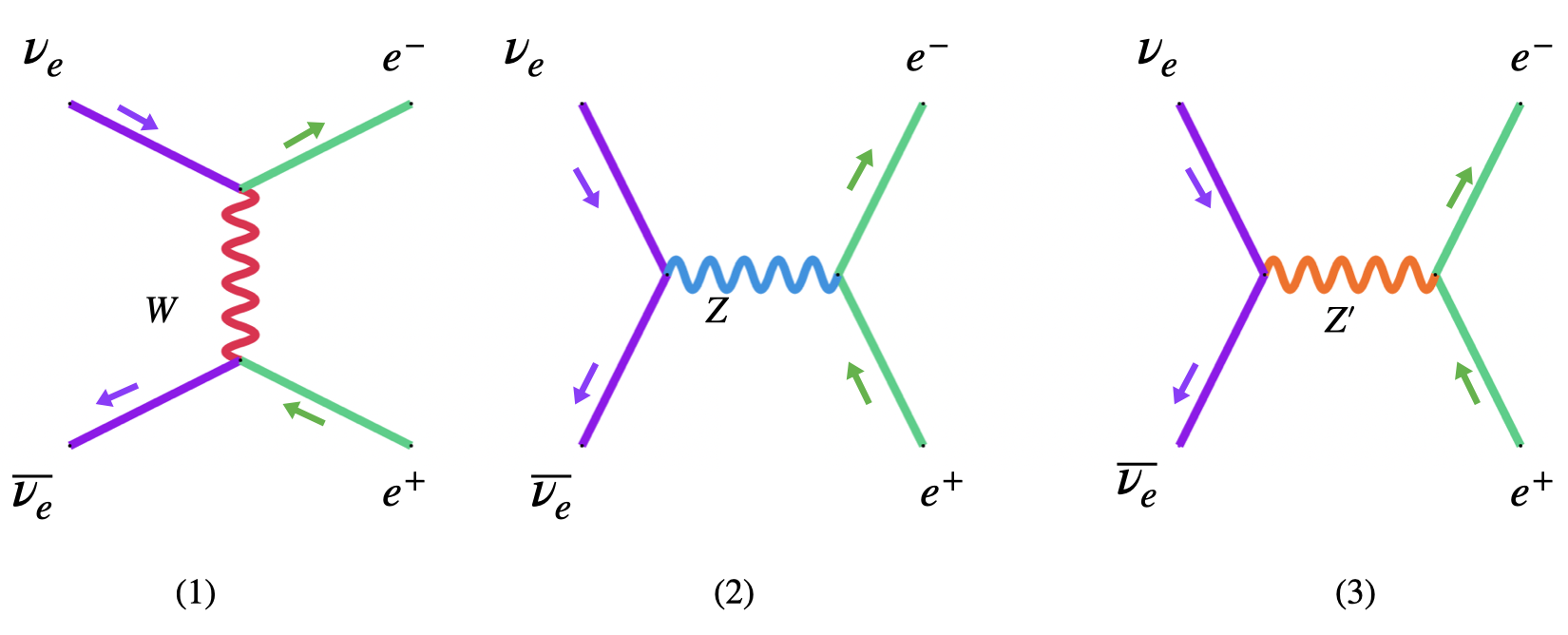}
\caption{$\nu \overline{\nu}$ scattering in SM and BSM scenario by $W^\pm$, $Z$ and $Z^\prime$ under $U(1)$ scenarios for $e^- e^-$ production.}
\label{ev}
\end{figure}
In the model in consideration, neutrino-antineutrino annihilation ($\nu\bar{\nu} \to e^+ e^-$) can continuously deposit energy in the outer envelope. The high energy electrons and positrons created in this annihilation process can subsequently produce photons, which can manifest as the observed GRBs in the fireball model \cite{Piran:1999kx,Salmonson:2002ci,Leng:2014dfa}. For our analysis, we make the simplifying assumption that the conversion of electron-positron pair energy into photons has 100$\%$ efficiency. That is, we assume all the energy deposited by the neutrinos into electron-positron pairs gets converted into photons that result in GRB. In our analysis, we did not consider the possibility of hydrodynamical shock waves in the medium created by the deposited energy of the annihilating neutrino pairs rather than gamma rays. Our model in consideration specifically deals with the scenario where the energy deposited in $e^+ e^-$ pair through neutrino annihilation is fully converted into gamma rays according to the fireball model. This maximal efficiency assumption simplifies the model by avoiding the need to account for energy losses during the intermediary particle physics processes and this enables us to place conservative limits on the new physics parameters

Following the GRB model discussed in \cite{Salmonson:2002ci}, we denote the ratio of the energy of the $e^+ e^-$ fireball plasma ($\varepsilon$) to the baryonic rest mass energy ($M$) with $\rho_{\varepsilon} \equiv \varepsilon/M$. More baryon loading, or small $\rho_{\varepsilon}$ (that is $\rho_{\varepsilon} \lesssim 10^5$ \cite{Salmonson:2002ci}), corresponds to a large opacity due to electrons associated with protons. In this scenario, baryons absorb all the fireball energy. Conversely, for less baryon loading ($\rho_{\varepsilon} \gtrsim 10^5$), the opacity due to $e^+ e^-$ pairs is significant, resulting in a burst of high energy photons \cite{Salmonson:2002ci}. Since the focus here is on neutrino pair annihilation in the region $r \gtrsim R_{\nu}$, the baryon loading is minimal, leading to substantial energy deposition due to neutrino pair annihilation.

In this section, we will calculate the energy deposited through the $\nu\bar{\nu} \to e^+ e^-$ process in the $U(1)$ extensions of SM in some metric background which will be used to constrain the new physics parameters using the observed GRB energies. In general $U(1)$ extensions neutrino-antineutrino interact with electron-positron through $W$, $Z$ and $Z^\prime$ bosons where first two comes from purely SM process as given in Fig.~\ref{ev}. In this case $W$ mediated $t-$ channel process contributes only in $\nu_e \bar{\nu_e}$ annihilation, however, it does not contribute in $\nu_{\mu,\tau} {\bar \nu_{\mu, \tau}}$ annihilation processes. We study $\nu\nu \to e^- e^+$ process in the light of GRB221009A to probe the effect of $Z^\prime$ in energy deposition rate defining 
\bea
\{\sigma(\nu\bar{\nu} \rightarrow e^+ e^-) \}= (\sigma |v_{\nu} -v_{\bar{\nu}}|) E_{\nu} E_{\bar{\nu}} = \Big[\sigma (\nu\bar{\nu} \rightarrow e^+ e^-)\Big] (E_{\nu} E_{\bar{\nu}}- \Vec{p}_{\nu}.\Vec{p}_{\bar{\nu}})^2
\label{Xsec}
\eea
where the term $\sigma (\nu\bar{\nu} \rightarrow e^+ e^-)$ within the box bracket is the symbolic neutrino annihilation cross section into electron-positron pair in the center of mass frame, $E_{\nu(\bar{\nu})}$ and $\Vec{p}_{\nu (\bar{\nu})}$ are the energy and three momenta of neutrino (anti-neutrino), respectively \cite{Goodman:1986we}. In the general $U(1)_X$ scenario we calculate the contribution of $\nu_e \bar{\nu}_e \rightarrow e^+ e^-$ process in Eq.~(\ref{Xsec}) as 
\bea
\{\sigma(\nu_e \bar{\nu}_{e} \rightarrow e^+ e^-)\} &=& \Big[ \frac{G_F^2}{3 \pi} (1 + 4 \sin^2\theta_w +8 \sin^4\theta_w) + \frac{g_X^4}{6 \pi M_{Z^\prime}^4} \left(\frac{x_H}{2}+ x_{\Phi}\right)^2 \left\{ \left(\frac{x_H}{2}+ x_{\Phi}\right)^2+(x_H+x_{\Phi})^2 \right\}+\frac{4 G_F g_X^2}{3 \sqrt{2} \pi M_{Z^\prime}^2} \nonumber \\
&& \left(\frac{x_H}{2}+x_{\Phi}\right) \left\{\sin^2\theta_w \left(\frac{3 x_H}{2}+2 x_{\Phi}\right)-\frac{1}{2} \left(\frac{x_H}{2}+ x_{\Phi}\right)\right\} + \frac{4 G_F g_X^2}{3\sqrt{2}\pi M_{Z^\prime}^2} \left(\frac{x_H}{2}+ x_{\Phi} \right)^2 \Big] 
(E_{\nu} E_{\bar{\nu}}- \Vec{p}_{\nu}.\Vec{p}_{\bar{\nu}})^2~~~~~~~
\label{eq:crosssec-ee}
\eea
where the first, second, third and fourth terms are coming from SM $(W, Z)$, $Z^\prime$ mediated, interference between $(Z, Z^\prime)$ and $(W, Z^\prime)$ processes, respectively. From $\nu_{\mu/\tau} \bar{\nu}_{\mu/\tau} \rightarrow e^+ e^-$ process we calculate its contribution in Eq.~(\ref{Xsec}) as 
\bea
\{\sigma(\nu_{\mu/\tau} \bar{\nu}_{\mu/\tau}\rightarrow e^+ e^-)\} &=&
\Big[ \frac{G_F^2}{3 \pi} (1 - 4 \sin^2\theta_w +8 \sin^4\theta_w) + \frac{g_X^4}{6 \pi M_{Z^\prime}^4}\left(\frac{x_H}{2}+x_{\Phi}\right)^2 \left\{\left(\frac{x_H}{2}+x_\Phi\right)^2+(x_H+x_\Phi)^2\right\}+ \nonumber \\
&& \frac{4 G_F g_X^2}{3 \sqrt{2} \pi M_{Z^\prime}^2} \left(\frac{x_H}{2}+x_{\Phi}\right) \left\{\sin^2\theta_w \left(\frac{3 x_H}{2}+2x_{\Phi}\right)-\frac{1}{2} \left(\frac{x_H}{2}+x_{\Phi}\right)\right\} \Big] (E_{\nu} E_{\bar{\nu}}- \Vec{p}_{\nu}.\Vec{p}_{\bar{\nu}})^2,
\label{eq:crosssec-mutau} 
\eea
where the first, second and third terms are coming from SM $(Z)$, $Z^\prime$ mediated and interference between $(Z, Z^\prime)$ processes, respectively for $U(1)_X$ scenario. We have disagreements with the $Z^\prime$ mediated cross sections given in \cite{Poddar:2022svz} for the $U(1)_X$ scenario where  $U(1)$ charges of the left and right handed fermions were not properly utilized at the time of calculating cross sections. In our case we have properly utilized the general $U(1)_X$ charges. Taking $\sin^2\theta_w=0.23121$ we estimate $(1 + 4 \sin^2\theta_w +8 \sin^4\theta_w)=2.3525$ and $(1 - 4 \sin^2\theta_w +8 \sin^4\theta_w)=0.592825$ where $\theta_w$ is the Weinberg angle. Here $G_F=1.663787\times 10^{-5}$ GeV$^{-2}$ \cite{ParticleDataGroup:2020ssz} is the Fermi constant. Similarly we calculate contribution from 
$\nu_e \bar{\nu}_e \to e^+ e^-$ process in $U(1)_{xq-\tau_R^3}$ model. Its contribution in Eq.~(\ref{Xsec}) is
\bea
\{\sigma(\nu_{e} \bar{\nu}_{e} \rightarrow e^+ e^-)\} &=&  \Big[
\frac{G_F^2}{3 \pi} (1+ 4 \sin^2\theta_w+8 \sin^4\theta_w)+ \frac{g_X^4 9 x^2}{6 \pi M_{Z^\prime}^4} \{9x^2+(1-6x)^2\}+   \\ & &
\frac{4 g_X^2 G_F}{3 \sqrt{2} \pi M_{Z^\prime}^2} (-3x) \left\{(-3x)\left(-\frac{1}{2}+\sin^2\theta_w \right)+
(1-6x) \sin^2\theta_w \right\}+\frac{4 g_X^2 G_F}{3 \sqrt{2} \pi M_{Z^\prime}^2} 9x^2   
\Big](E_{\nu} E_{\bar{\nu}}- \Vec{p}_{\nu}.\Vec{p}_{\bar{\nu}})^2 \nonumber
\eea
whereas contributions from $\nu_{\mu, \tau} \bar{\nu}_{\mu, \tau} \to e^+ e^-$ process in Eq.~(\ref{Xsec}) are
\bea
\{\sigma(\nu_{\mu/\tau} \bar{\nu}_{\mu/\tau} \rightarrow e^+ e^-)\} &=&  \Big[
\frac{G_F^2}{3 \pi} (1- 4 \sin^2\theta_w+8 \sin^4\theta_w)+ \frac{g_X^4 9 x^2}{6 \pi M_{Z^\prime}^4} \{9x^2+(1-6x)^2\}+ \nonumber \\
& &\frac{4 g_X^2 G_F}{3 \sqrt{2} \pi M_{Z^\prime}^2} (-3x) \left\{(-3x)\left(-\frac{1}{2}+\sin^2\theta_w \right)+ (1-6x) \sin^2\theta_w \right\}   
\Big](E_{\nu} E_{\bar{\nu}}- \Vec{p}_{\nu}.\Vec{p}_{\bar{\nu}})^2.
\label{eq:crosssec-ee-2}
\eea
Following the same manner, we calculate contribution from $\nu_e \bar{\nu}_e \to e^+ e^-$ process in Eq.~(\ref{Xsec}) in context of $U(1)_{q+x u}$ model as
\bea
\{\sigma(\nu_{e} \bar{\nu}_{e} \rightarrow e^+ e^-)\} &=&  \Big[\frac{G_F^2}{3 \pi} (1+4\sin^2\theta_w+8 \sin^4\theta_w)+ \frac{g_X^4}{6 \pi M_{Z^\prime}^4} \left(1+ \frac{(2+x)^2}{9} \right)+ \frac{4 g_X^2 G_F}{3\sqrt{2} \pi M_{Z^\prime}^2} \left\{ \left(-\frac{1}{2}+\sin^2\theta_w \right)+ \right.  \nonumber \\ 
&& \left. \frac{(2+x)}{3} \sin^2\theta_w \right\} + \frac{4 g_X^2 G_F}{3\sqrt{2} \pi M_{Z^\prime}^2}\Big] (E_{\nu} E_{\bar{\nu}}- \Vec{p}_{\nu}.\Vec{p}_{\bar{\nu}})^2
\label{eq:crosssec-ee-1}
\eea
and the contribution from $\nu_{\mu, \tau} \bar{\nu}_{\mu, \tau} \to e^+ e^-$ process in Eq.~(\ref{Xsec}) as
\bea
\{\sigma(\nu_{\mu/\tau} \bar{\nu}_{\mu/\tau} \rightarrow e^+ e^-)\} &=&  \Big[\frac{G_F^2}{3 \pi} (1-4\sin^2\theta_w+8 \sin^4\theta_w)+\frac{g_X^4}{6 \pi M_{Z^\prime}^4} \left\{1+ \frac{(2+x)^2}{9} \right\}
+ \frac{4 g_X^2 G_F}{3\sqrt{2} \pi M_{Z^\prime}^2} \left\{ \left(-\frac{1}{2}+\sin^2\theta_w \right)+ \right. \nonumber \\ 
&& \left. \frac{(2+x)}{3} \sin^2\theta_w \right\} \Big] (E_{\nu} E_{\bar{\nu}}- \Vec{p}_{\nu}.\Vec{p}_{\bar{\nu}})^2.
\label{eq:crosssec-ee-2}
\eea

From $L_i-L_j$ scenarios, the contribution of $\nu_e \bar{\nu}_e \to e^+ e^-$ process in Eq.~(\ref{Xsec}) is given by 
\bea
\{\sigma(\nu_e \bar{\nu}_e \to e^- e^+)\}=\Big[\frac{G_F^2}{3\pi} (1+ 4\sin^2\theta_w+8 \sin^4\theta_w)+\frac{g_X^4}{3\pi M_{Z^\prime}^4}+ \frac{4 G_F g_X^4}{3 \sqrt{2}\pi M_{Z^\prime}^2} \left(-\frac{1}{2}+2\sin^2\theta_w \right)+ \frac{4 G_F}{3\sqrt{2}\pi}\frac{g_X^2}{M_{Z^\prime}^2}\Big] (E_{\nu} E_{\bar{\nu}}-\vec{p}_\nu.\vec{p}_{\bar{v}})^2~~~~~
\label{LeLm1}
\eea
where the first, second, third and fourth terms are coming from SM $(W, Z)$, $Z^\prime$ mediated, interference between $(Z, Z^\prime)$ and $(W, Z^\prime)$ processes, respectively in $L_e-L_{\mu, \tau}$ models manifesting interaction between $Z^\prime$ and the first generation SM lepton. Contributions from $\nu_\mu \bar{\nu}_\mu \to e^+ e^-$ processes in Eq.~(\ref{Xsec}) can be written as
\begin{equation}
\{\sigma(\nu_\mu \bar{\nu}_\mu \to e^- e^+)\}= \Big[\frac{G_F^2}{3\pi} (1- 4\sin^2\theta_w+8 \sin^4\theta_w)+\frac{g_X^4}{3\pi M_{Z^\prime}^4}-\frac{4 G_F g_X^2}{3\sqrt{s} \pi M_{Z^\prime}^2} \left(-\frac{1}{2}+2\sin^2\theta_w \right) \Big] (E_{\nu} E_{\bar{\nu}}-\vec{p}_\nu.\vec{p}_{\bar{v}})^2
\label{LeLm2}
\end{equation}
where the first, second and third terms are coming from SM $(Z)$, $Z^\prime$ mediated and interference between $(Z, Z^\prime)$ processes, respectively. The contribution from $\nu_{e,\tau} \bar{\nu}_{e, \tau} \to e^+ e^-$ processes in Eq.~(\ref{Xsec}) will be same as those given in Eqs.~(\ref{LeLm1}) and (\ref{LeLm2}) from Tab.~\ref{tab:charges2}. From $B-3 L_{i}$ scenarios we find that $\nu_e \bar{\nu}_e \to e^- e^+$ process contribute in Eq.~(\ref{Xsec})
\bea
\{\sigma(\nu_e \bar{\nu}_e \to e^- e^+)\}=\Big[ \frac{G_F^2}{3 \pi} \Big(1+ 4 \sin^2\theta_W + 8 \sin^4\theta_W\Big)+\frac{27 g_X^4}{\pi M_{Z^\prime}^4}+ \frac{12 G_F g_X^2}{\sqrt{2} \pi M_{Z^\prime}^2} \Big(-\frac{1}{2}+2 \sin^2\theta_w\Big)+
\frac{12 G_F g_X^2}{\sqrt{2} \pi M_{Z^\prime}^2}\Big] (E_{\nu} E_{\bar{\nu}}-\vec{p}_\nu.\vec{p}_{\bar{v}})^2~~~~~
\label{B3Le}
\eea
from $B-3L_e$ scenario where the first, second, third and fourth terms are coming from SM$(W, Z)$, $Z^\prime$ mediated and interference between $(Z, Z^\prime)$ and $(W, Z^\prime)$ processes, respectively. Direct interaction between $Z^\prime$ and first generation lepton is not present in $L_{\mu}-L_{\tau}$ and $B-3L_{\mu,\tau}$ scenarios restricting their contributions in Eq.~(\ref{Xsec}).
\subsection{Neutrino trajectory in a general metric}
The line element in a general space-time metric with an off-diagonal $t\phi$ element is given by $ds^2 = g_{tt}\;dt^2 + g_{rr}\;dr^2  + g_{\theta\theta}\;d\theta^2 + g_{\phi\phi}\;d\phi^2 + 2g_{t\phi}\;dt\;d\phi$
neglecting tiny neutrino mass and restricting the dynamics to a plane with $\theta=\pi/2$. Hence the null geodesic equation for a massless particle is 
$g_{tt}\; \dot{t}^2 + g_{rr} \;\dot{r}^2 + g_{\phi\phi} \;\dot{\phi}^2 + 2 g_{t\phi}\;\dot{\phi}\dot{t} =0$. The Lagrangian of a massless particle is given by $\mathcal{L} = \frac{1}{2} g_{\mu\nu} \dot{x}^{\mu} \dot{x}^{\nu}$ 
and the generalized momenta can be derived as 
\begin{align}
    p_t = \frac{\partial \mathcal{L}}{\partial\dot{t}} = g_{tt}\;\dot{t} + g_{t\phi}\;\dot{\phi} = -E, \; \;
    p_r = \frac{\partial \mathcal{L}}{\partial\dot{r}} =g_{rr}\;\dot{r}, \;\;
    p_{\phi} = \frac{\partial \mathcal{L}}{\partial\dot{\phi}} =g_{\phi\phi}\;\dot{\phi} + g_{t\phi}\;\dot{t} = L,
    \label{eq:genmomenta3}
\end{align}
 where $E$ and $L$ are energy and angular momentum of the neutrino which can be simultaneously solved to obtain
\begin{align}
    \dot{t} = \frac{L\;g_{t\phi} + E\;g_{\phi\phi}}{g_{t\phi}^2 - g_{tt}\;g_{\phi\phi}}, \; \; \;
    \dot{\phi} = \frac{L\;g_{tt} + E\;g_{t\phi}}{g_{tt}\;g_{\phi\phi}-g_{t\phi}^2}.
    \label{eq:phidot}
\end{align}
Now the Hamiltonian of the neutrino is given by $2 \mathcal{H} = -E \;\dot{t} + L \;\dot{\phi} + g_{rr} \dot{r}^2 = \delta$ where $\delta = 0$ for null geodesics and solving the Hamilonian equation for $\dot{r}^2$, we obtain
\begin{equation}\label{eq:rdot2}
    \dot{r}^2 = \frac{L ^2 g_{tt} + E \left(2 L \;g_{t\phi} +E g_{\phi \phi }\right)}{g_{rr} \left(g_{t\phi}^2-g_{tt}\;g_{\phi \phi }\right)}
\end{equation}
and dividing Eq.~(\ref{eq:rdot2}) by $\dot{\phi}^2$, from Eq.~(\ref{eq:phidot}), we get 
\begin{equation}
\label{eq:drdphi1}
    \left(\frac{dr}{d\phi}\right)^2 = \frac{\left(g_{t\phi}^2-g_{tt} \;g_{\phi \phi }\right) \left(L^2 \; g_{tt}+ E \left(2 L\;  g_{t\phi}+ E\;g_{\phi \phi }\right)\right)}{g_{rr} \left(L \;g_{tt}+E\; g_{t\phi}\right)^2}.
\end{equation}
Here the local tetrad is defined by the equation $e^{\mu}_a \;  g_{\mu\nu} \; e^{\nu}_b= \eta_{ab}$, where $\eta_{ab}$ is the Minkowski metric and it can be expressed as  
\begin{align}
    e^{\mu}_i = \left(
    \begin{array}{cccc}
    \sqrt{\frac{g_{t\phi}^2}{g_{\theta \theta }}-g_{tt}} & 0 & 0 & 0 \\
    0 & \sqrt{g_{rr}} & 0 & 0 \\
    0 & 0 & \sqrt{g_{\theta \theta }} & 0 \\
    \frac{g_{t\phi}}{\sqrt{g_{\theta \theta }}} & 0 & 0 & \sqrt{g_{\theta \theta }} \\
    \end{array}
    \right)
\end{align}
where $\theta$ is the angle between the neutrino trajectory and tangent vector to that trajectory in terms of radial  $(V^r)$ and longitudinal $(V^\phi)$ components of velocity. We write
\begin{align}
    \tan \theta &= \frac{e^1_{r} V^r}{e^3_{\phi} V^{\phi} + e^3_{t}}= \frac{e^1_{r}}{e^3_{\phi}+ \frac{e^3_{t}}{V^{\phi}}} \left( \frac{dr}{d\phi}\right),
  \label{eq:tantheta2}
\end{align}
where $V^{\phi}= \dot{\phi}/\dot{t}$ and finally, using Eqs.~(\ref{eq:drdphi1}) and (\ref{eq:tantheta2}), we eliminate $dr/d\phi$ to obtain
\begin{equation}\label{eq:tanthetafinal}
    \tan \theta = \sqrt{\frac{g_{\theta \theta } \left(g_{\text{t$\phi $}}^2-g_{\text{tt}} g_{\phi \phi }\right) \left(b^2 \; g_{\text{tt}}+2 b\; g_{\text{t$\phi $}}+g_{\phi \phi }\right)}{\left(-b \;g_{\theta \theta } \;g_{\text{tt}}+b g_{\text{t$\phi $}}^2+g_{\text{t$\phi $}} \left(g_{\phi \phi }-g_{\theta \theta }\right)\right){}^2}} 
\end{equation}
defining the impact parameter $b=\frac{L}{E}$. Eq.~(\ref{eq:tanthetafinal}) will be used to calculate the neutrino trajectory in different space-time metric. We consider 4-velocity of a particle near the source as $u^{\mu} = (u^t ,0 ,0 ,u^{\phi})$. Hence we define the rotational velocity of the source under consideration as $\Omega = \frac{d\phi}{dt} = \frac{u^{\phi}}{u^{t}}$ and define $u_{\mu} = (u_t ,0 ,0 ,\Omega\; u_t) $. Using the normalization condition ($u_{\mu} u^{\mu} =  1$) we obtain $u^t = (g_{tt} + \Omega^2\;g_{\phi\phi} + 2 \Omega \; g_{t\phi})^{-1/2}$. Defining the frequency of a massless particle observed by a distant observer as $\omega = u^{\mu} \frac{dx_{\mu}}{d\lambda} = u^{\mu} g_{\mu\nu} \frac{dx^{\nu}}{d\lambda}$ where $\lambda$ is some affine parameter along the trajectory of the massless particle, we have 
\begin{align}
\omega &= \frac{-E + \Omega L}{\left(g_{tt} + \Omega^2\;g_{\phi\phi} + 2 \Omega \; g_{t\phi}\right)^{1/2}}
\end{align}
for a neutrino emitted with constant $r$ and $\theta$ using Eq.~(\ref{eq:genmomenta3}). Therefore the red-shift factor including effect of rotation is given by 
\begin{align}
\tilde{z}(r)=\left(g_{tt} + \Omega^2\;g_{\phi\phi} + 2 \Omega g_{t\phi}\right)^{1/2} 
\label{red1}
\end{align}
where $\Omega$ is responsible for the rotational red-shift. 
\subsection{Energy deposition rate from $\nu\bar{\nu} \rightarrow e^+ e^-$ processes}
To examine the effect of neutrino annihilation, we make the assumption that neutrinos are emitted from a defined surface called the neutrinosphere. The neutrinosphere is the radius at which neutrinos are able to stream freely \cite{Bethe:1990mw,Janka:2017vlw}, defined by the equation 
\bea
\int_{R_{\nu}}^{\infty} 1/\lambda_{\nu} d\ell = 2/3
\eea
where $\lambda_{\nu}$ is the neutrino mean free path, $R_{\nu}$ is the neutrinosphere radius. In general, the neutrinosphere radius $R_{\nu}$ depends on the neutrino energy, since $\lambda_{\nu} \propto 1/(\rho e_{\nu}^2)$, where $\rho$ is the surrounding matter density and $e_{\nu}$ is the neutrino energy. Assuming a thermal distribution for the neutrino energies, we take the neutrinosphere to have a fixed average value of $R_{\nu} \approx 20\mbox{ km}$, which represents the effects of the varying neutrino energies and matter densities by taking a mean value over thermal distribution. With this assumption, we evaluate energy deposition rate for $\nu\bar{\nu} \to e^- e^+$ process in two different ways in which the input parameter changes: (i) the luminosity of the neutrino observed at $r=\infty$ ($L_{\nu}^{\infty}$), i.e., luminosity at the observer frame and (ii) the local temperature of the neutrinosphere $(T_{\nu})$.

{\bf (i) Input parameter $L_{\nu}^{\infty}$:} Energy deposition rate per unit volume from $\nu\bar{\nu} \to e^- e^+$ processes is
\begin{equation}
    \dot{q}(r) = \int d^3\Vec{p}_{\nu}\;d^3\Vec{p}_{\bar{\nu}} \; f_{\nu}(\Vec{p}_{\nu},\Vec{r})\; f_{\bar{\nu}}(\Vec{p}_{\bar{\nu}},\Vec{r})\; \{\sigma(\nu\bar{\nu} \rightarrow e^+ e^-)\}\; \frac{E_{\nu} + E_{\bar{\nu}}}{E_{\nu}E_{\bar{\nu}}}
    \label{qdot}
\end{equation}
where $f_{\nu}(\vec{p}_{\nu},\Vec{r})$ is Fermi-Dirac distribution function which could be defined as $f_{\nu}= \frac{2}{h^3}(1/(1+ e^{\frac{E_\nu}{k_B T}}))$ \cite{Goodman:1986we}. Writing $\Vec{p}_{\nu (\bar{\nu})} = E_{\nu(\bar{\nu})} \hat{\Omega}_{\nu(\bar{\nu})}$ we get $d^3\Vec{p}_{\nu(\bar{\nu})} = E_{\nu(\bar{\nu})}^2 dE_{\nu(\bar{\nu})} d\hat{\Omega}_{\nu(\bar{\nu})}$ in the direction of solid angle. Taking the energy factor from Eq.~(\ref{Xsec}) and evaluating the energy integral \cite{Salmonson:1999es} we get 
\begin{equation}
  \int dE_{\nu}\;dE_{\bar{\nu}} \; f_{\nu}(E_{\nu})\; f_{\bar{\nu}}(E_{\bar{\nu}}) E_{\nu}^3 E_{\bar{\nu}}^3 (E_{\nu}+E_{\bar{\nu}})= \frac{21 (k_BT)^9 \zeta (5)}{128 \pi ^2}
\end{equation}
where $k_B$ is the Boltzmann constant, $T$ stands for neutrino and antineutrino temperatures and $\zeta(5) = 1.03693$ (Riemann zeta function). The angular integration in Eq.~(\ref{qdot}) can be written as $\Theta = \int d\Omega_{\nu}d\Omega_{\bar{\nu}}(1-\Omega_{\nu}\Omega_{\bar{\nu}})^2$ (where $\Omega$ is a unit vector) and taking $\Omega = (\mu, \sqrt{1-\mu^2} \cos\phi,\sqrt{1-\mu^2} \sin\phi )$ and defining $d\Omega = d\mu\;d\phi$ and $\mu = \sin \theta$, the angular integral can be evaluated as $\Theta = \frac{2\pi^2}{3}(1-x)^4(x^2+4x+5)$ where $x = \sin\theta$ which could be obtained from Eq.~(\ref{eq:tanthetafinal}). Now, taking the red-shift into account, we relate the temperature and luminosity of neutrinos at the neutrinosphere ($R_{\nu}$) to those for an observer away from the source following
\begin{align}
    T_{\nu}(r) = \frac{\tilde{z}(R_{\nu})}{\tilde{z}(r)} T_{\nu}(R_{\nu}) \;\;\;\text{and}\; \;\;
    L^{\infty}_{\nu}  = \tilde{z}(R_{\nu})^2 L_{\nu}(R_{\nu}).
        \label{eq:temp2}
\end{align}
where $\tilde{z}(R_{\nu})$ and $\tilde{z}(r)$ are the red-shift factors of the neutrinos at the neutrinosphere and the distance of the neutrinos from the source.  Assuming a black-body emission, we write the luminosity of neutrinos at neutrinosphere as 
\begin{equation}\label{eq:temp3}
    L_{\nu}(R_{\nu}) = 4\pi R_{\nu}^2 \frac{7}{16} a T^4_{\nu}(R_{\nu}) 
\end{equation}
where $a$ is the radiation constant, and using Eq.~(\ref{eq:temp2}) in $L_{\nu}(R_{\nu})$ we obtain 
\begin{equation}
    T_{\nu}^9(r) =  \frac{\tilde{z}(R_{\nu})^{9/2}}{\tilde{z}(r)^{9}} (L_{\nu}^{\infty})^{9/4} \left(\frac{7}{4}\pi a\right)^{-9/4} R_{\nu}^{-9/2} 
\end{equation}
which helps us to estimate energy deposition rate in SM. The total energy deposition rate over whole volume is given by $\dot{Q} = \int_{R_{\nu}}^{\infty} \sqrt{g} \;dr d\theta d\phi \; \dot{q}(r)$, where $\dot{q}(r)$ is defined in Eq.~(\ref{qdot}). 

Then, the energy deposition rate in the case of SM in a Newtonian background for the $\nu_i \bar{\nu_i} \rightarrow e^+ e^-$ process is given by 
\bea
    \dot{Q}^{N}_{L_{\nu_i}^{\infty}} &= \int_{1}^{\infty} y^2 dy \frac{28\pi}{64}(k_B)^9 \zeta(5) \left[ \frac{ G_F^2}{3 \pi} (1 \pm 4 \sin^2\theta_w +8 \sin^4\theta_w) \right] 
 \times (L_{\nu}^{\infty})^{9/4} \left(\frac{7}{4}\pi a\right)^{-9/4} R_{\nu}^{-3/2} \times (1-x_{N})^4(x_{N}^2+4x_{N}+5)~~~~~~~
    \label{enq1}
\eea
where we have defined $r = y  R_{\nu}$ and $x_{N} = \sin \theta_N = \sqrt{1-\frac{1}{y^2}}$ is the trajectory equation for a neutrino emitted tangentially ($\theta =0$) from the neutrinosphere in a Newtonian background. The quantity in the boxed bracket comes from the total cross section of $\nu_i \bar{\nu_i} \rightarrow e^+ e^-$ process in center of mass frame where $i$ denotes the generations of neutrinos and the contribution from `plus (minus)` sign originates from $\nu_{e(\mu, \tau)}$. Adding up the contributions from three generations of neutrinos and performing $y-$integration, we find energy deposition rate  
\begin{align}
\dot{Q}^{N}_{L_{\nu}^{\infty}} = \frac{28\pi}{192}(k_B)^9 \zeta(5) \left[ \frac{ G_F^2}{3 \pi} (3 - 4 \sin^2\theta_w +24 \sin^4\theta_w) \right] \times (L_{\nu}^{\infty})^{9/4} \left(\frac{7}{4}\pi a\right)^{-9/4} R_{\nu}^{-3/2}.
\end{align}
We consider $\Delta t$ as the timescale of neutrino energy deposition which is approximately $1s$ for a typical neutrino burst mechanism \cite{Goodman:1986we}, $L_{\nu}^{\infty} = 10^{53}$ erg/s and $R_{\nu} = 20$ km, respectively. Now we define the enhancement for the metric with respect to the Newtonian case as 
\begin{equation}
\label{eq:I_L_infinity}
    \mathcal{I}_{L^{\infty}_{\nu}} =  \frac{\dot{Q}^{\rm GR}_{L^{\infty}_{\nu_i}}}{\dot{Q}^{N}_{L^{\infty}_{\nu_i}}} = \frac{\int_{1}^{\infty} y^2 dy  \sqrt{g_{rr}(r)} \tilde{z}(R_{\nu})^{9/2} \tilde{z}(r)^{-9} (x_{\rm GR}-1)^4(x_{\rm GR}^2+4x_{\rm GR}+5)}{\int_{1}^{\infty} y^2 dy (x_{N}-1)^4(x_{N}^2+4x_{N}+5)}.
\end{equation}
 where cross-section and constants will cancel out from numerator and denominator resulting to enhancement of energy deposition rate purely due to metric (GR) under consideration relative to the Newtonian metric.

{\bf (ii) Input parameter $T_\nu$:} The reaction rate of neutrinos per unit volume is given by 
\begin{equation}
    \frac{d^2N}{dtdV} = \int n_{\nu}(E_{\nu}) n_{\bar{\nu}}(E_{\bar{\nu}}) E_{\nu}^3 E_{\bar{\nu}}^3 dE_{\nu} dE_{\bar{\nu}} \int d\Omega_{\nu} d\Omega_{\bar{\nu}} \sigma(\nu \bar{\nu} \rightarrow e^+ e^-) (1-\vec{\Omega}_{\nu} \cdot \vec{\Omega}_{\bar{\nu}})^2 
    \label{asano}
\end{equation}
where the number density can be defined as $n(E_{\nu(\bar{\nu})}) = \frac{2}{(2\pi)^3} \frac{1}{1+ \exp(E_{\nu(\bar{\nu})}/k_B T)}$ following the Fermi-Dirac statistics \cite{Asano_2000} and it is conserved along the neutrino trajectory. Here the total energy deposition rate of neutrinos can be obtained multiplying  Eq.~(\ref{asano}) by $(E_{0 \nu}+E_{0 \bar{\nu}})$, where $E_{0 \nu (\bar{\nu})}$ is the energy of (anti)neutrino in an observer frame at infinity. Now evaluating the angular integral and the energy integral over $E_{0 \nu (\bar{\nu})}$ for an observer at infinity, we find the density of the energy deposition rate as 
\begin{equation}
    \frac{d\dot{Q}_{\nu_i}}{dV} = \frac{\sigma(\nu_i\bar{\nu}_i \rightarrow e^+ e^-)}{g_{00}(r)^4} \frac{2\pi^2}{3} (x_{\rm GR}-1)^2(x_{\rm GR}^2 + 4x_{\rm GR} + 5) \frac{21(k_B T_{\rm eff})^9}{128\pi^2} \zeta(5) \tilde{z}(R_{\nu})^{9}    
    \label{qdot1}
\end{equation}
where $x_{\rm GR}$ is the trajectory function corresponding to some background metric and $T_{\rm eff}$ is the effective temperature of (anti)neutrino at neutrinosphere, which is defined as $T_{\rm eff}= \frac{T_{0}}{\tilde{z}(R_{\nu})}$ with $T_{0}$ being (anti)neutrino temperature observed at infinity. Integrating over a volume element $\sqrt{-g} \;d^3x$ we obtain energy deposition rate as
\begin{align}\label{eq:Asano-1}
    \dot{Q}_{\nu_i} = \sigma(\nu_{i}\bar{\nu}_{i} \rightarrow e^+ e^-) \zeta(5) \tilde{z}(R_{\nu})^{9} \frac{21(k_B T_{\rm eff})^9}{128\pi^2}  \int dr d\theta d\phi \sqrt{g} \frac{1}{\tilde{z}(r)^{8}} \frac{2\pi^2}{3} (x_{\rm GR}-1)^2(x_{\rm GR}^2 + 4x_{\rm GR} + 5).
\end{align}
where $g$ is determinant of the metric. Hence calculating the energy deposition rate from neutrinos in SM under Newtonian background we obtain
\begin{align}
\dot{Q}_{\nu_i} = \sigma(\nu_i\bar{\nu}_i \rightarrow e^+ e^-) \zeta(5) \frac{21(k_B T_{\rm eff})^9}{128\pi^2} R_{\nu}^3 \frac{8\pi^3}{9}
\label{qdot3}
\end{align}
for a single generation and the contribution from three generation case can simply be obtained by summing up contribution from each generation of neutrinos. Now using $T_{\nu}$ as an input parameter we define a quantity 
\begin{equation}
    \mathcal{I}_{T_{\nu}} = \frac{\dot{Q}^{\rm GR}_{T_{\nu_i}}}{\dot{Q}^{N}_{T_{\nu_i}}} = \frac{\int_{1}^{\infty} y^2 dy\;\sqrt{g}\;\tilde{z}(R_{\nu})^{9} \tilde{z}(r)^{-8} (x_{\rm GR}-1)^4(x_{\rm GR}^2+4x_{\rm GR}+5)}{\int_{1}^{\infty} y^2 dy (x_{N}-1)^4(x_{N}^2+4x_{N}+5)}
    \label{asano-4}
\end{equation}
where $x_{\rm GR}$ is trajectory function for any metric, $\dot{Q}^{N}_{T_{\nu_i}}$ and $\dot{Q}^{\rm GR}_{T_{\nu_i}}$ are energy deposition rates from $\nu_i$ in Newtonian and some metric backgrounds, respectively. 
\subsection{Estimation of energy deposition rates in SM and BSM}
To estimate the energy deposition rates in the SM and BSM cases, we first consider the effect of using different metric backgrounds. Although GRBs have been observed extensively, their sources remain poorly understood and Since SM interactions cannot account for the observed GRB energies using a Newtonian metric background, we consider two metric that are solutions of GR - the Sc metric, which is one of the simplest solution and the HT metric, which specifically models rotating NS. We also know that the GR predictions have been tested with high accuracy on different scales, ranging from the solar system and astrophysical to cosmological scales. Therefore, we consider : (i) Sc, (ii) HT scenarios to study deposition rates.
 
{ \bf (i) Sc:} The Sc metric is given by $g_{tt} = -\left(1 - 2\frac{M}{r}\right)$, $g_{rr} = \left(1 - 2\frac{M}{r}\right)^{-1}$, $g_{\theta\theta} = r^2$ and $g_{\phi\phi} = r^2 \; \sin^2\theta$.
Using this metric we find the energy deposition rate for the neutrinos
\begin{align}
    \dot{Q}_{\nu_{i}} = \int_{1}^{\infty} y^2 dy \sqrt{g_{rr}(r)} \frac{7\pi}{16}(k_B)^9 \zeta(5) \sigma_{\nu_i\bar{\nu}_i} \frac{\tilde{z}^{9/2}(R_{\nu})}{\tilde{z}^{9}(r)} (L_{\nu}^{\infty})^{9/4} \left(\frac{7}{4}\pi a\right)^{-9/4} R_{\nu}^{-3/2}(x_{\rm Sc}-1)^4(x_{\rm Sc}^2+4 x_{\rm Sc}+5)
    \label{eq1x1}
\end{align}
where $\tilde{z}$ is the red-shift with $\Omega=0$ and $\sigma_{\nu_i\bar{\nu}_i}=\sigma(\nu_e \bar{\nu}_e \to e^- e^+)$. Using Eq.~(\ref{eq:tantheta2}) in the above expression, we define the trajectory function $(x_{\rm Sc})$ of neutrinos in Sc background as
\begin{align}
 x_{\rm Sc} = \left( \sin\theta \right)^{\rm Sc} = \sqrt{\frac{(p-2)\;y^3 - p \;y+2}{(p-2)\;y^3}}
\end{align}
where $p= R_{\nu}/M$ and $y= r/R_{\nu}$ are dimensionless quantities with $R_{\nu}$ and $M$ being the radius of the neutrinosphere and mass of the source of GRB. If $\sigma_{\nu_i \bar{\nu_i}}$ involves SM (BSM) contribution, then Eq.~(\ref{eq1x1}) is denoted as $\dot{Q}_{\nu}^{\rm SM (BSM)}$ summing over the contributions of three generations of neutrinos. 

{ \bf (ii) HT:} Considering dipole approximation and ignoring the higher order terms, the HT metric is given by $g_{tt}= -\left(1 - 2\frac{M}{r} - \frac{J^2}{r^4}\right)$, $g_{rr}= \left(1 - 2\frac{M}{r} - \frac{J^2}{r^4}\right)^{-1} \left(1-5\frac{J^2}{r^4}\right)$, $g_{\theta\theta}= r^2,\; g_{\phi\phi}= r^2 \sin^2\theta$, $g_{t\phi}= -2 \frac{J}{r} \sin^2\theta$ under dipole approximation. Using these metric components and $\tilde{z}$, we find energy deposition rate for neutrinos as
\begin{align}
    \dot{Q}_{\nu_{i}} = \int_{1}^{\infty} y^2 dy \sqrt{g_{rr}(r)} \frac{7\pi}{16}(k_B)^9 \zeta(5) \sigma_{\nu_i\bar{\nu}_i} \frac{\tilde{z}^{9/2}(R_{\nu})}{\tilde{z}^{9}(r)} (L_{\nu}^{\infty})^{9/4} \left(\frac{7}{4}\pi a\right)^{-9/4} R_{\nu}^{-3/2}(x_{\rm HT}-1)^4(x_{\rm HT}^2+4x_{\rm HT}+5)
    \label{eq1x2}
\end{align}
where $x_{\rm HT}$ is the trajectory function of neutrinos for HT metric and using Eq.~(\ref{eq:tantheta2}) it can be defined as
\begin{align}
x_{\rm HT} = \sqrt{1-\frac{\left(p^2-2\;j\right)^2 \left(j^2 \;(p\; y-3)+p^3 y^3 \;(2-p \;y)\right)}{\left(j^2 (p-3)-(p-2)\;p^3\right) \left(p^2 \;y^3-2 j\right)^2}}
\end{align}
with $p=R_{\nu}/M$ and $j=J/M^2$ are dimensionless quantities while $R_{\nu}$, $M$ and  $J$ are the radius of the neutrinosphere, gravitational mass of the source and angular momentum of the source, respectively.  If $\sigma_{\nu_i \bar{\nu_i}}$ involves SM (BSM) contribution, then Eq.~(\ref{eq1x2}) is denoted as $\dot{Q}_{\nu}^{\rm SM (BSM)}$ summing over the contributions of three generations of neutrinos. 
\begin{figure}[h]
    \centering
    \includegraphics[scale=0.45]{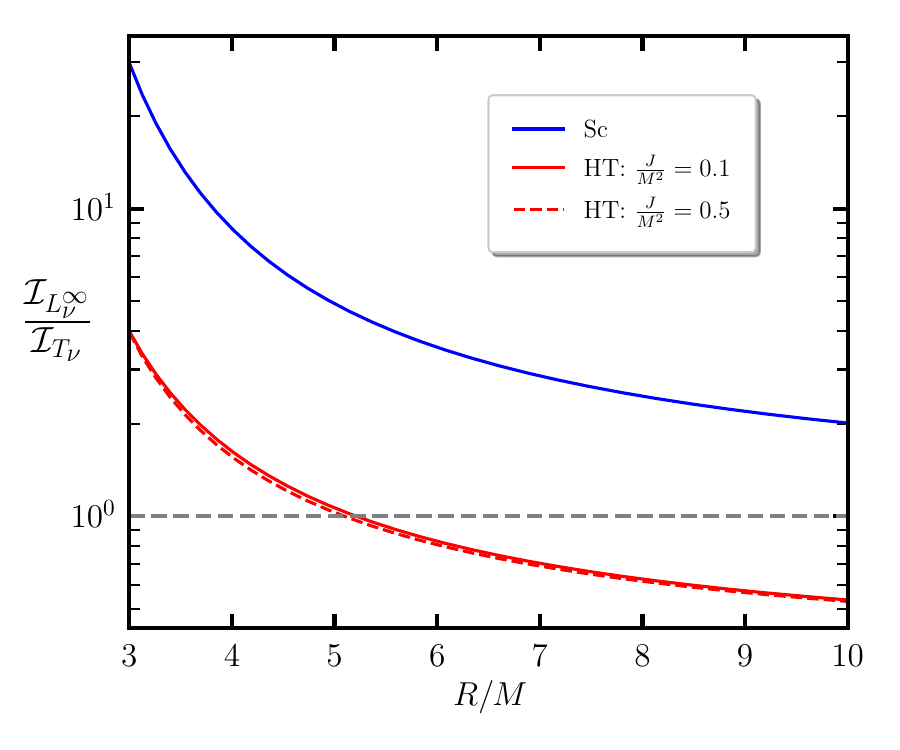}
    \includegraphics[scale=0.45]{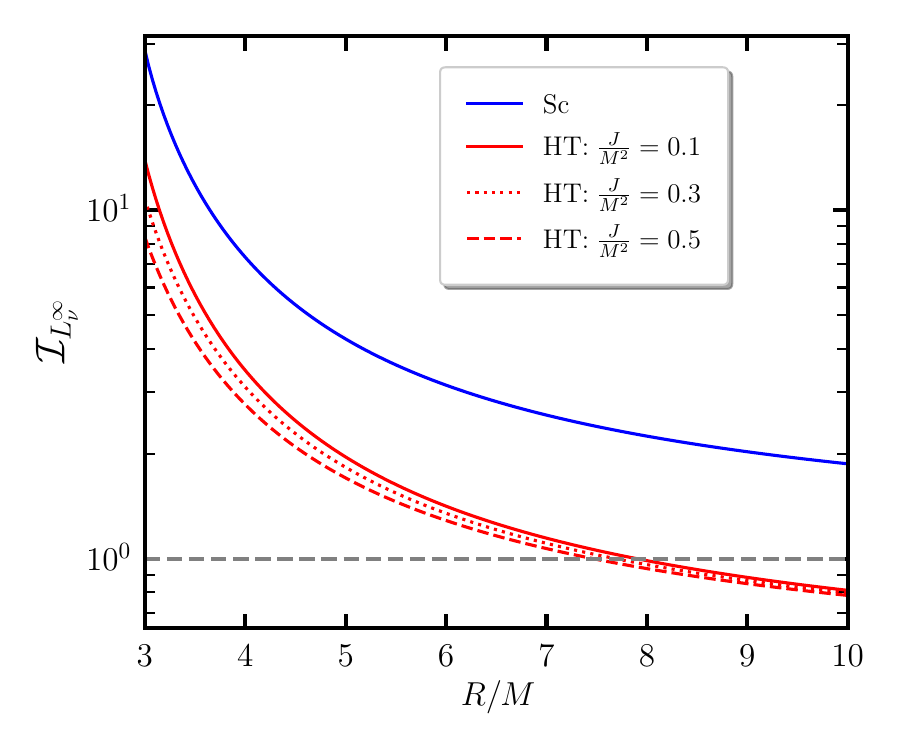}
    \caption{ $\frac{\mathcal{I}_{L_{\nu}^{\infty}}}{\mathcal{I}_{T_{\nu}}}$ (left panel) and $\mathcal{I}_{L^{\infty}_{\nu}}$ (right pane) for Sc and HT (with different $\frac{J}{M^2}$) metrics as a function of $\frac{R}{M}$.}
    \label{fig:sal_vs_asano}
\end{figure}

Now, taking the ratio of Eq.~(\ref{eq:I_L_infinity}) to Eq.~(\ref{asano-4}) we find $\frac{\mathcal{I}_{L_{\nu}^{\infty}}}{\mathcal{I}_{T_{\nu}}}$ to compare the two approaches based on $L_{\nu}^{\infty}$ and $T_{\nu}$ in the left panel of Fig.~\ref{fig:sal_vs_asano}. For Sc metric $\frac{\mathcal{I}_{L_{\nu}}^{\infty}}{\mathcal{I}_{T_{\nu}}}> 1$ with $1 \leq \frac{R}{M} \leq 10$, however, for HT metric $\frac{\mathcal{I}_{L_{\nu}}^{\infty}}{\mathcal{I}_{T_{\nu}}} < 1$ for $\frac{R}{M} > 5.21$ where dependence on the dimensionless quantity $\frac{J}{M^2}$ is weak for HT. Method based on $L_{\nu}^{\infty}$ does not depend on local physics of GRB. Therefore we use this method for further analyses. Now in the right panel of Fig.~\ref{fig:sal_vs_asano} we investigate the effect of Sc and HT metrics in Eq.~(\ref{eq:I_L_infinity}) where $\mathcal{I}_{L^{\infty}_{\nu}}$ decreases with $\frac{R}{M}$ for both the metrics. For HT metric $\mathcal{I}_{L^{\infty}_{\nu}} > 1$,  for $\frac{R}{M} < 8$ with strong dependence on $\frac{J}{M^2}$ while for $\frac{R}{M}> 8$, the quantity $\mathcal{I}_{L^{\infty}_{\nu}} < 1$ irrespective of $\frac{J}{M^2}$. The probability of neutrinos to interact reduces due to the effect of angular momentum and rotational red-shift due to the non zero off-diagonal element in the metric making $\mathcal{I}_{L^{\infty}_{\nu}}$ in HT case lower than the Sc case.  

To estimate bounds on $M_{Z^{\prime}}/g_X$, we use the energy deposition rate calculated from various metrics following 
\begin{equation}
    \dot{Q}_{\rm{BSM}}^{\rm GR} \leq \dot{Q}_{\rm{obs}} - \dot{Q}_{\rm{SM}}^{\rm GR}
\label{bounds-1-1}    
\end{equation}
where the left side denotes BSM contribution from $Z^\prime$ and its interference with $W$ and $Z$ bosons, while on the right side we have the observed energy deposition rate minus the SM contribution assuming a GR metric as gravitational background. To estimate $\dot{Q}_{\rm obs}$, we use the brightest GRB event GRB221009A \cite{Burns:2023oxn} whose isotropic energy was $E_{\gamma,\rm iso} \simeq 1.2\times 10^{55}$ erg \cite{2022GCN.32668....1F,2022GCN.32636....1V,Insight-HXMT:2023aof,Frederiks:2023bxg,2022GCN.32636....1V,Williams:2023sfk,Lesage:2023vvj}, where the true energy can be calculated as $E_{\rm true} = (1-\cos\theta_j) E_{\gamma,\rm iso}$ \cite{Fong:2015oha} with $\theta_j=1.5^{\degree}$ for GRB221009A \cite{Burns:2023oxn}. Using $\dot{Q}_{\rm obs}= E_{\rm true}/t$, we get the observed energy deposition rate as $4\times 10^{51}$erg/s with $t=1$s being the time frame within which neutrinos deposit energy \cite{Goodman:1986we} . In cases where $\dot{Q}_{\rm SM}^{\rm GR}$ is within 10\% of $\dot{Q}_{\rm obs}$, we estimate bounds requiring $\dot{Q}_{\rm BSM}^{\rm GR} \leq \sigma$, where $\sigma$ is observational uncertainty. We take $\sigma \simeq 0.1(0.01)  \;\dot{Q}_{\rm obs}$ for 10\%(1\%) current (prospective) uncertainty, respectively.

\begin{figure}[h!]
\includegraphics[scale=0.35]{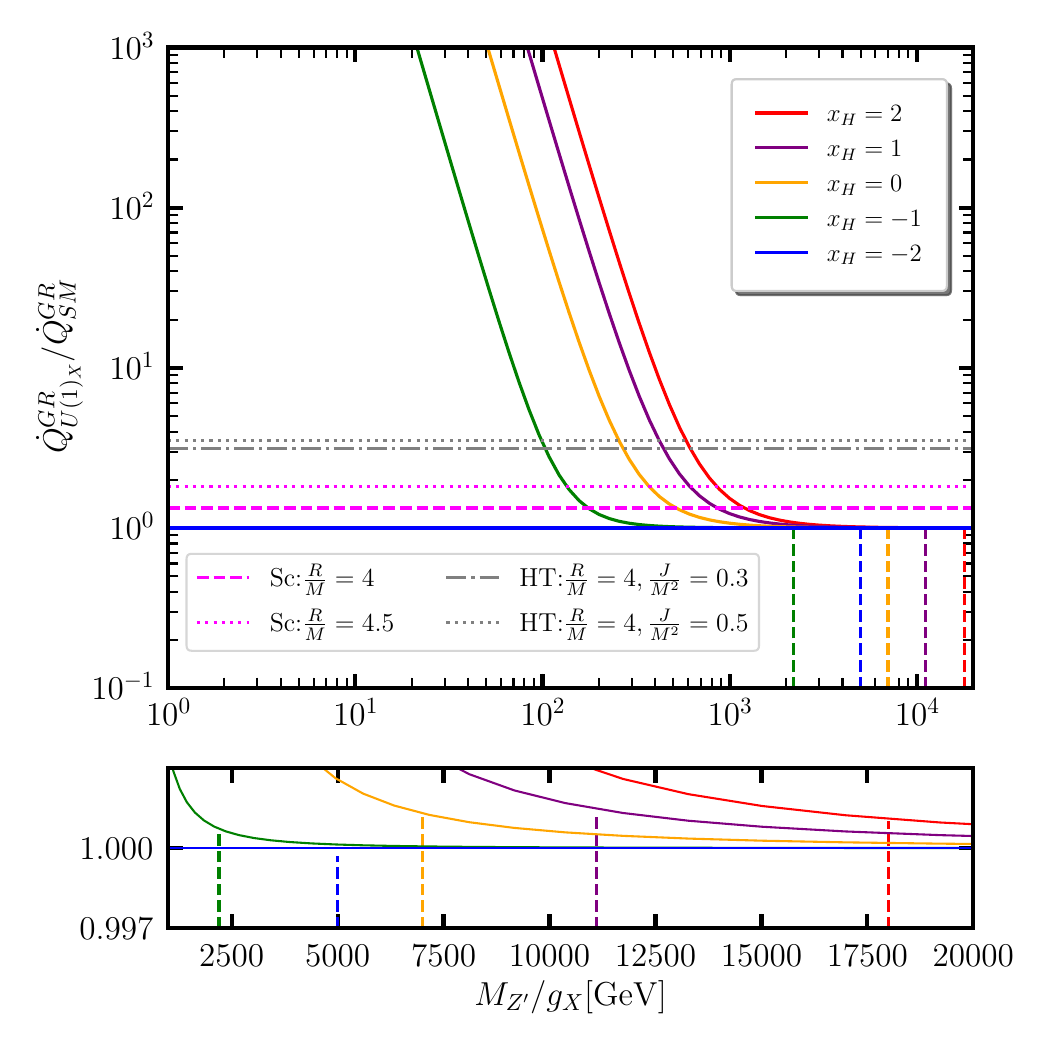}%
\includegraphics[scale=0.35]{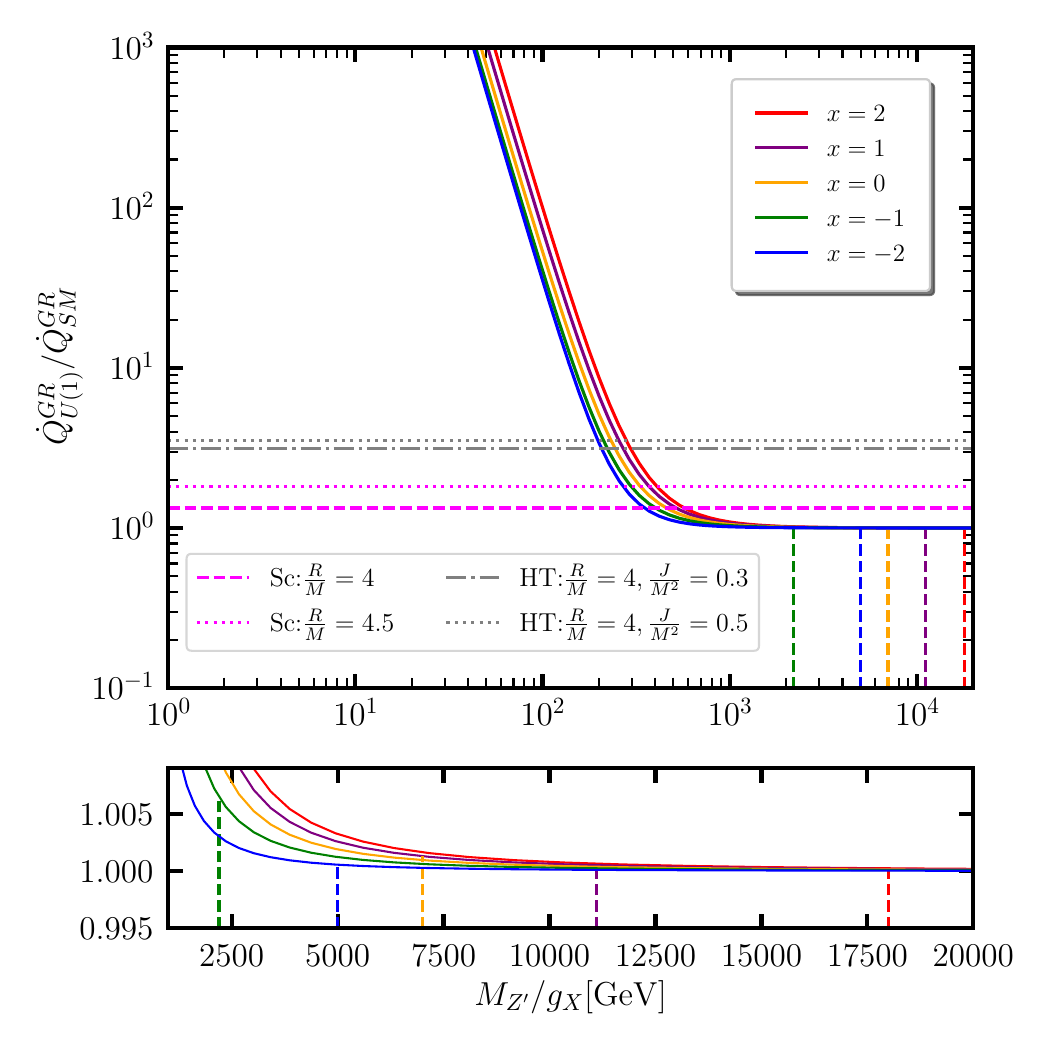}%
\includegraphics[scale=0.35]{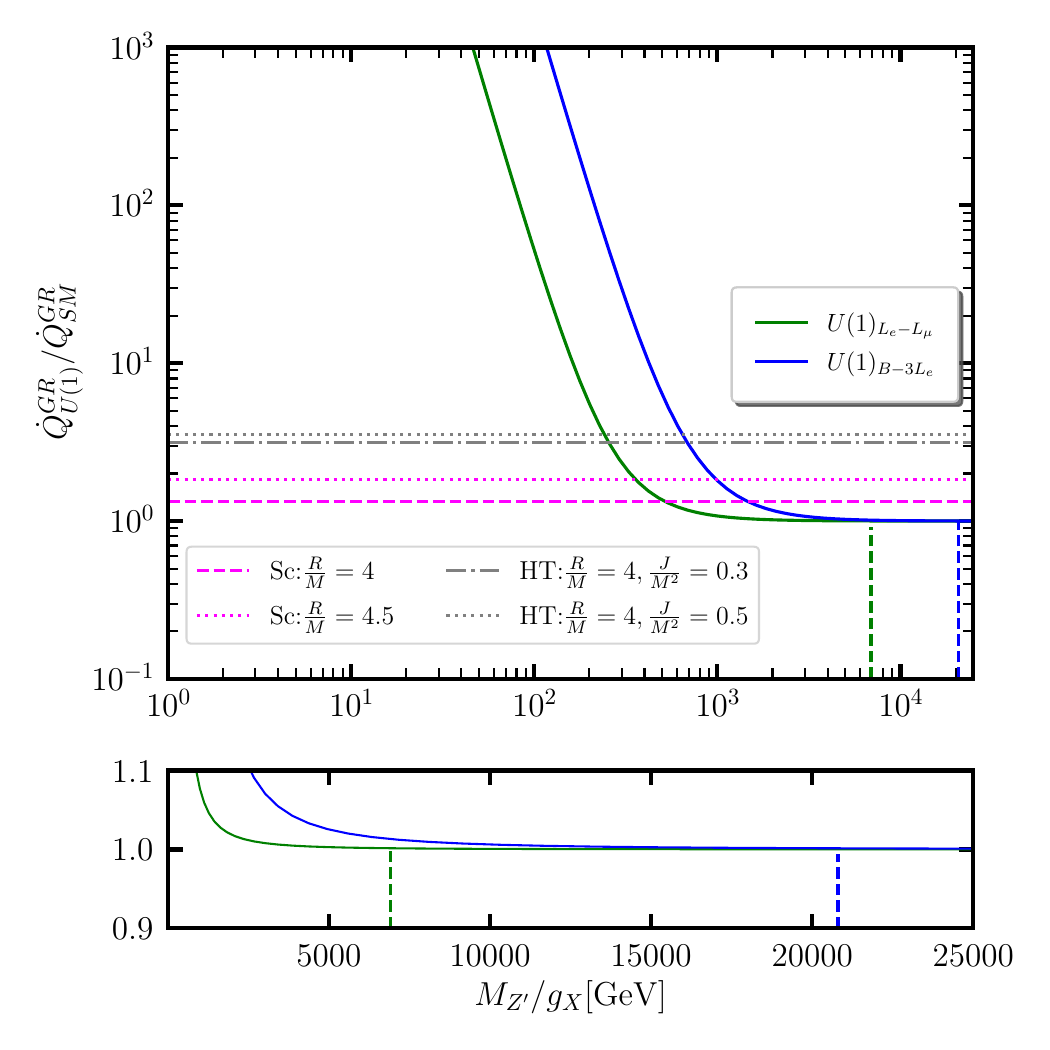}
\caption{Enhancement in energy deposition rates for different $U(1)$ scenarios compared to SM for Sc and HT cases with respect to $M_{Z^\prime}/g_X$. Perpendicular lines show LEP-II bounds for $M_{Z^\prime} > 209$ GeV. }
\label{fig:limx1}
\end{figure}

Using Eq.~(\ref{bounds-1-1}) we can calculate $\dot{Q}_{\rm BSM/ SM}^{\rm GR}$ following Eq.~(\ref{eq1x1}) for Sc and HT cases. We show the estimated bounds on the ratio of energy deposition rates $(\dot{Q}_{\rm BSM}^{\rm GR}/\dot{Q}_{\rm SM}^{\rm GR})$ from Eq.~(\ref{bounds-1-1}) with respect to $M_{Z^\prime}/g_X$ represented by different horizontal colored lines in Fig.~\ref{fig:limx1} for different metrics and different $U(1)$ scenarios. We compute the same directly estimating the energy deposition rates for BSM and SM scenarios under a particular metric background being represented by different curved lines in Fig.~\ref{fig:limx1} for different $U(1)$ scenarios. If the horizontal lines move downwards, the effect is dominated by the corresponding metric and if moves upward then enhancement can be obtained from BSM scenarios.

The cross-over between the horizontal and curved lines shows bounds on $M_{Z^\prime}/g_X$ for different $U(1)$ scenarios and metrics. Since metric dependent part cancels out in $\dot{Q}_{\rm BSM}^{\rm GR}/\dot{Q}_{\rm SM}^{\rm GR}$, these curved lines represent the ratio of total BSM cross-section to SM cross section asymptotically reaching $1$ with increasing $M_{Z^\prime}/g_X$. The estimated bounds with $10(1)\%$ uncertainty are shown in Tab.~\ref{tab:lim}. Bounds on $M_{Z^\prime}/g_X$ become stronger with a lower $R/M$ ratio compared to a higher $R/M$ ratio. From Fig.~\ref{fig:limx1} we find that HT case provide stronger limits on $M_{Z^\prime}/g_X$ compared to Sc case. Therefore in Tab.~\ref{tab:lim} we show HT case only. 

In $U(1)_X$ scenario for $x_H=-2$ there is no coupling between $\ell_L$ and $Z^\prime$. Therefore we do not find any bound in this case, whereas $x_H=0$ is the B$-$L case. However, in the case of $U(1)_{q+xu}$ scenario we consider $x_H=-1$, $0$, $1$ and $2$ respectively where $x_H=1$ reproduces B$-$L case. In flavored scenarios we consider $L_e- L_{\mu, \tau}$ cases where $Z^\prime$ couples with the first generation lepton doublet from the SM and similar scenario can be observed in $B-3L_e$. Therefore in these flavored scenarios only we obtain bounds on $M_{Z^\prime}/ g_X$ from GRB which are shown in Tab.~\ref{tab:lim}. 

In Fig.~\ref{fig:limx1} we show perpendicular lines which are the LEP-II bounds on $M_{Z^\prime}/g_X$ for different $U(1)$ scenarios considering $M_{Z^\prime} >> \sqrt{s}$. LEP-II does not affect the bounds obtained from GRB directly. However, limits estimated on $M_{Z^\prime}/g_X$ from LEP-II will be strong for $M_{Z^\prime} > \sqrt{s}$. Therefore the perpendicular lines resemble indirect bounds varying $\dot{Q}_{\rm BSM}^{\rm GR}/\dot{Q}_{\rm SM}^{\rm GR}$ for a fixed $M_{Z^\prime}/g_X$. In this analysis, we consider large VEV $(v_\Phi=M_{Z^\prime}/2 g_X > v)$ scenario for different $U(1)$ models. Therefore we will further utilize only such values of $M_{Z^\prime}/g_X$ from Tab.~\ref{tab:lim} satisfying this condition.  
\begin{table}
	\begin{center}
\begin{tabular}{ |c|c|c| } 
\hline
  & $\dot{Q}_{\rm{obs}}=4\times 10^{51}$ erg &$\dot{Q}_{\rm{obs}}=4\times 10^{51}$ erg\\ 
  &$M_{Z^\prime}/g_X$(GeV)&$M_{Z^\prime}/g_X$(GeV)\\
  \hline
 $U(1)_X$ &$\frac{R}{M}^{\rm HT}=3$,$\frac{J}{M^2}=0.37$&$\frac{R}{M}^{\rm HT}=5$,$\frac{J}{M^2}=0.37$\\
  \hline
  $x_H=2$&2006.79(6038.55)&489.047 \\ 
  $x_H=1$&1435.64(4312.68)&351.975 \\
  $x_H=0$&858.423(2563.21)&214.773 \\ 
  $x_H=-1$&257.813(635.772)&84.6959 \\
\hline
 $U(1)_{q+xu}$ &$\frac{R}{M}^{\rm HT}=3$,$\frac{J}{M^2}=0.37$&$\frac{R}{M}^{\rm HT}=5$,$\frac{J}{M^2}=0.37$ \\
  \hline
  $x=2$&957.09(2875.12)&234.65 \\ 
  $x=1$&858.423(2563.21)&214.773 \\
  $x=0$&750.559(2209.4)&195.638\\ 
  $x=-1$&633.64(1792.27)&179.399\\
  \hline
 $U(1)_{B-3L_e}$ & $\frac{R}{M}^{\rm HT}=3$,$\frac{J}{M^2}=0.37$&$\frac{R}{M}^{\rm HT}=5$, $\frac{J}{M^2}=0.37$ \\
 \hline
 &2566.55(7964.7)&529.94 \\
  \hline
$U(1)_{L_e-L_{(\mu, \tau)}}$&$\frac{R}{M}^{\rm HT}=3$, $\frac{J}{M^2}=0.37$&$\frac{R}{M}^{\rm HT}=5$,$\frac{J}{M^2}=0.37$\\
\hline
  &886.23(2711.32)&200.92 \\
  \hline
\end{tabular}
	\end{center}
	\caption{Bounds on $M_{Z^\prime}/g_X$ for HT case with $10(1)\%$ precision under various $U(1)$ scenarios and different charges.}
\label{tab:lim}
\end{table}
\section{Neutrino-DM scattering from cosmic blazar and active galaxy}
\label{BLAGN}
To investigate $\nu$-DM scattering in general $U(1)$ scenarios, we extend models with SM-singlet potential DM candidates considering three types of DM candidates: (i) scalar, (ii) Dirac and (iii) Majorana as explained below:

\noindent
{\bf (i) Scalar DM:} We introduce a complex scalar (CS) DM candidate $\Phi_1=\{1,1,0, Q_\chi\}$ in general $U(1)_X$ model with an odd $Z_2$ parity where remaining particles are even under $Z_2$ transformation 
and $\Phi_1$ interacts with other fields of the model exchanging $Z^\prime$. However, $\Phi_1$ can interact with scalar sector of the model through the potential following
\begin{equation}
V \supset \lambda_{\rm mix_1} (H^\dagger H)(\Phi_1^\ast \Phi_1)+\lambda_{\rm mix_2} (\Phi^\dagger \Phi)(\Phi_1^\ast \Phi_1). 
\end{equation}
We assume $\lambda_{\rm mix_{1,2}}$ to be very small in the line of scalar mixing considered in Eq.~(\ref{pot1x}) taken to be small. Like the $U(1)_X$ scenario, $\Phi_1$ can be introduced in $U(1)_{qx+u}$, $U(1)_{B-3L_i}$ and $L_i-L_j$ scenarios. The interaction between $\Phi_1$ and $Z^\prime$ and neutrinos are given by
\begin{equation}
\mathcal{L}_{\rm DM}^{\rm scalar} = Q_\chi g_X Z^\prime_\mu \Big\{\Phi_1^\ast (\partial_\mu \Phi_1) - (\partial_\mu \Phi_1^\ast)  \Phi_1\Big\}+ m_{\rm{DM}}^2 \Phi_1^\ast \Phi_1+ g_X Q_\ell \overline{\nu_L} \gamma^\mu \nu_L Z^\prime_{\mu}. 
\label{LDM-1}
\end{equation}
where $Q_\chi$ is the general $U(1)$ charge of $\Phi_1$, the second term is the mass term where $m_{\rm DM}$ is the mass of $\Phi_1$ and $Q_\ell$ being the general $U(1)$ charge of the neutrinos following Eq.~(\ref{Lag1}). 
\begin{figure}[h]
\includegraphics[width=80mm]{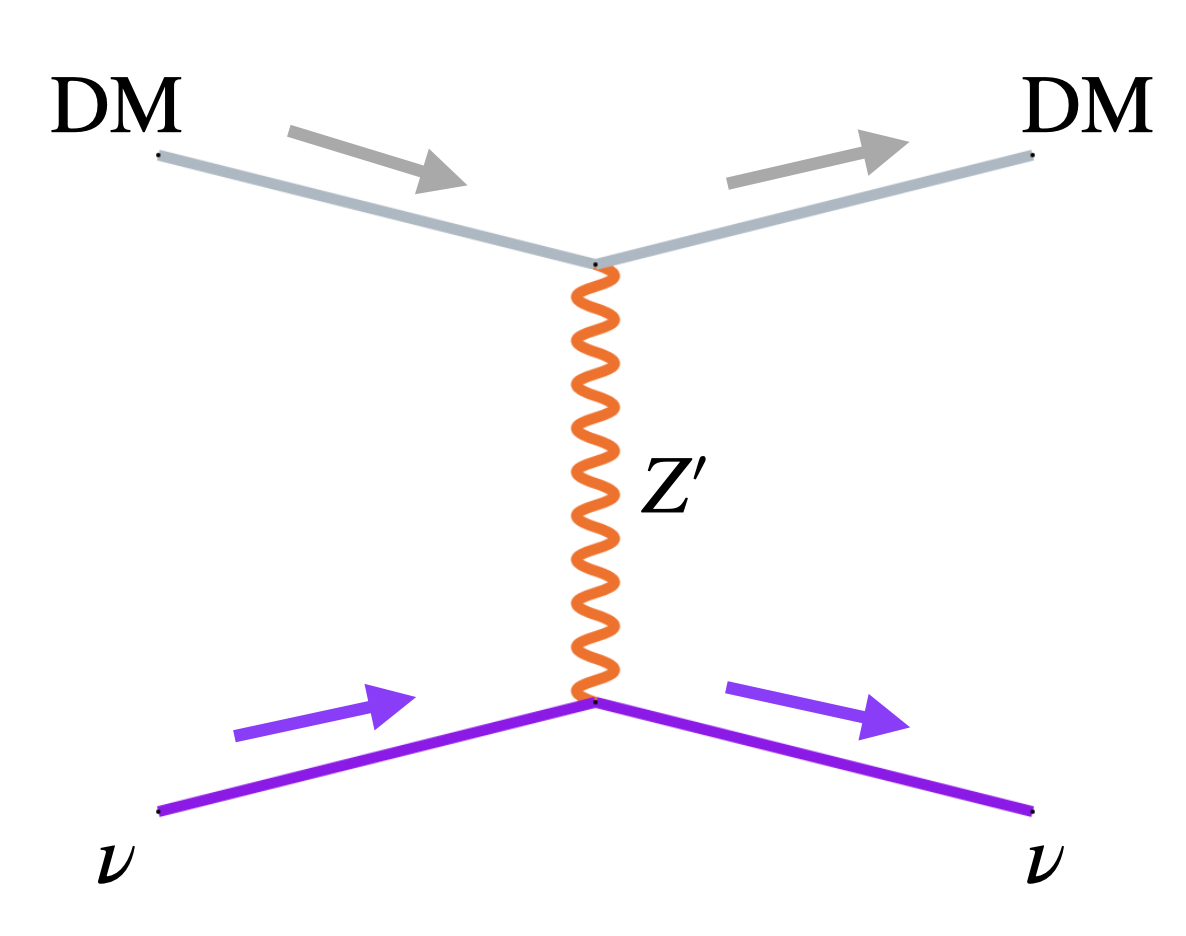}
\caption{Neutrino-DM scattering in $t-$channel mediated by $Z^\prime$ boson in laboratory frame and DM represents either of complex scalar, Majorana and Dirac type particle. }
\label{vDM1}
\end{figure}
We estimate the differential scattering cross section for the $\nu-$DM process in $t-$channel mediated by $Z^\prime$ in the laboratory frame following Fig.~\ref{vDM1} as
\begin{equation}
\frac{d \sigma}{d E_{\nu}^\prime} = \frac{g_{X}^4 Q_{\ell}^2 Q_{\chi}^2}{8 \pi} \frac{m_{\rm DM}}{E_{\nu}^2}    
\frac{(2 E_{\nu} E_{\nu}^\prime- m_{\rm DM}(E_{\nu}-E_{\nu}^\prime)}{ \{M_{Z^\prime}^2 + 2\; m_{\rm DM} (E_{\nu}-E_{\nu}^\prime)\}^2} 
\label{DM-s}
\end{equation}
where incoming and outgoing energies of the neutrino are $E_\nu$ and $E^\prime_\nu$ respectively. For $U(1)_X$ scenario, considering $x_\Phi=1$ we find that $Q_{\ell}=(-\frac{x_H}{2}-1)$. For $x_H=-2$, there is no interaction between $\nu_L$ and $Z^\prime$ which is $U(1)_R$ case. In this case we consider gauge kinetic mixing to be extremely small due to simplicity. In $U(1)_{q+xu}$ scenario we find the charge of the neutrino as $-1$ and it can be found in Tab.~\ref{tab:charges}. For the flavored $Z^\prime$ cases neutrino charges could be either $+1$ or $-1$ in $L_i-L_j$ as shown in Tab.~\ref{tab:charges2}. In $B-3L_i$ scenario, the charge of the $i$th generation neutrino is $-3$ and the remaining two generations of the neutrinos are uncharged as given in Tab.~\ref{tab:charges3-1}. 

\noindent
{ \bf (ii) Majorana DM:} Majorana fermions could be considered as potential DM candidates. Among three RHNs in $U(1)$ scenarios, one generation could have odd $Z_2$ parity to ensure stability whereas remaining fields in the model could be even under $Z_2$. We consider $N^3_R$ as a potential DM candidate while neutrino mass and flavor mixing will be governed by $N^{1,2}_R$. Being charged under $U(1)_X$, DM interacts with the SM sector through $Z^\prime$ where contribution from BSM scalar is small due to the assumption of low scalar mixing. We write the interaction Lagrangian of the DM candidate and neutrinos from Eqs.~(\ref{LYk}) and (\ref{Lag1}) as
\bea
-\mathcal{L}_{\rm DM_1}^{\rm Majorana} = \{ \frac{1}{2}Y_{N}^3 \Phi \overline{({N_{R}^3})^c} N^3_{R}+ h.c.\} + \frac{1}{2}g_X Q_{\chi} N^3 \gamma^\mu \gamma_5  N^3 Z_\mu^{\prime}+ g_X Q_\ell \overline{\nu_L} \gamma^\mu \nu_L Z^\prime_{\mu}
\label{LDM-M1}
\eea
where the first term generates DM mass after general $U(1)$ breaking, the second term represents $Z^\prime-$DM interaction and the third term represents $\nu-Z^\prime$ interaction. Here $Q_{\ell}$ is the general $U(1)$ charge of $\nu_L$. Hence we estimate $\nu-$DM differential scattering cross section in laboratory frame in $t-$ channel mediated by $Z^\prime$ from Fig.~\ref{vDM1} as 
 \begin{equation}
        \frac{d \sigma}{d E_{\nu}^\prime} = \frac{g_{X}^4 Q_{\ell}^2 Q_{\chi}^2\;}{8\pi}  \frac{m_{\rm DM}}{ E_{\nu}^2}  
        \frac{(E_{\nu}^2 + {E^\prime_{\nu}}^2  +  m_{\rm DM}(E_{\nu}-E_{\nu}^\prime))}{ \{M_{Z^\prime}^2 + 2\; m_{\rm DM} (E_{\nu}-E^\prime_{\nu})\}^2}
        \label{DM-M}
    \end{equation}
where incoming and outgoing neutrino energies are $E_\nu$ and $E_\nu^\prime$, respectively. In case of $U(1)_X$ scenario, $Q_\chi=-x_\Phi=-1$ and $Q_\ell=(-\frac{x_H}{2}-1)$ which could vanish for $x_H=-2$, could be $-\frac{1}{2}$ and $-1$ for $x_H=-1$ and $0$, respectively. Neutrino charge in $U(1)_{q+xu}$ scenario is $Q_\ell=- 1$ whereas the DM charge can be found as $Q_{\chi}= \frac{-4+x}{3}=-1$ being the RHN charges under $U(1)_{q+xu}$.  Variation of charges in $U(1)_X$ and $U(1)_{q+xu}$ cases could be found in Tab.~\ref{tab:charges}.

There is another interesting aspect in general $U(1)_X$ scenario where two Higgs doublets could be introduced with different $U(1)_X$ charges protecting the second Higgs doublet $(H_2=\{1,2, -\frac{1}{2}, -\frac{x_H}{2}+3\})$ from any direct interaction with the SM fermions. After solving the gauge and mixed gauge-gravity anomalies we find that the $U(1)_X$ charge assignments of the three generations of RHNs follow a different pattern in which the first two generations have $(-4,-4)$ charge $(N_R^{1,2}=\{1,1,0,-4\})$ and the third generation has $5$ charge ${N_R^3=\{1,1,0, Q_\chi= 5\}}$ under $U(1)_X$. Therefore $N_R^3$ can be considered as a potential DM candidate without any additional $Z_2$ symmetry. The $U(1)_X$ charge assignment of the SM fermions are exactly same as $U(1)_X$ case shown in Tab.~\ref{tab:charges} with $x_\Phi=1$. In this model we introduce three SM singlet scalar fields $\Phi_{(A, B, C)}=\{1,1,0, (8,-10,-3)\}$. The corresponding Lagrangian can be written as
\begin{equation}
-\mathcal{L}_{\rm DM_2}^{\rm Majorana} = \{ \frac{1}{2}Y_{N}^3 \Phi_B \overline{({N_{R}^3})^c} N^3_{R}+ h.c.\} + \frac{5}{2} g_X  N^3 \gamma^\mu \gamma_5  N^3 Z_\mu^{\prime}+ g_X Q_\ell \overline{\nu_L} \gamma^\mu \nu_L Z^\prime_{\mu}
\label{LDM-M2}
\end{equation}
where the first term generates DM mass after general $U(1)$ breaking, the second term represents $Z^\prime-$DM interaction and the third term shows $\nu-Z^\prime$ interaction with $Q_{\ell}$ being the general $U(1)_X$ charge of $\nu_L$. In this scenario $N^{1,2}_R$ will participate in neutrino mass generation mechanism after the $U(1)_X$ symmetry breaking \footnote{The Yukawa interaction can be written as $-\mathcal{L}_{\nu}= \frac{1}{2}\sum_{i=1}^{2}Y_{N}^{i} \Phi_A \overline{({N_{R}^{i}})^c} N^{i}_{R}+ \sum_{m=1}^{3} \sum_{n=1}^{2}Y_{D}^{mn} \overline{\ell_{L_m}} H_2 N_{R_n}+ h.c.$ \cite{Okada:2018tgy}. Here we discuss only the part relevant to the $\nu-$DM interaction. The other constraints of this model will remain exactly the same as general $U(1)_X$ scenario due to the common charge assignments of SM fermions.}. We estimate the $\nu-$DM differential scattering in $t-$ channel mediated by $Z^\prime$ in laboratory frame to be 25 times more what we have from Eq.~(\ref{DM-M}) with $Q_\chi=-x_\Phi=-1$ because $Q_\chi=5$ in this scenario. Due to larger cross section we use this scenarios referring the Majorana DM (M) to estimate bounds from $\nu-$DM scattering.

In $L_i-L_j$ case $i$th and $j$th generation neutrinos have $\pm1$ charge $U(1)$ charges where the remaining flavor is not charged under $U(1)_{L_i-L_j}$. The field content could be found in Tab.~\ref{tab:charges2} without a DM candidate. We consider $N^1_R$ as a potential Majorana DM (M) candidate with odd $Z_2$ parity whereas remaining fields are even $Z_2$ allowing $N^{2, 3}_R$ to participate in the neutrino mass generation mechanism. Hence in this case $U(1)$ charge of the DM candidate will be $Q_{\chi}=1$. From Eq.~(\ref{DM-M}) we find  $\frac{d\sigma}{d E_\nu^\prime}$ depends on $Q_{\chi}^2 Q_{\ell}^2$ which is $1$ in this case and therefore signs of charges will not matter. We consider $B-3L_i$ scenario where $i$th generation neutrino has $-3$ change under general$U(1)$ gauge group while the remaining two generations are not charged. Detailed particle contents without DM candidate are given in Tab.~\ref{tab:charges3-1}. We introduce an odd $Z_2$ parity only for the $i$th generation RHN of the model and the remaining two generations participate in neutrino mass generation and flavor mixing mechanisms. Needless to mention that DM candidates in flavored cases will follow the interactions from Eq.~(\ref{LDM-M1}) and differential scattering cross section from Eq.~(\ref{DM-M}), respectively.  

\noindent
{\bf (iii) Dirac DM:} We extend the general $U(1)$ field contents with weekly interacting $\chi_{L,R}= \{1,1,0, Q_\chi \}$ which will potentially be Dirac type DM (D1) candidate participating in $\nu-$DM scattering. We assign the general $U(1)$ charge for the DM to ensure its stability following the interaction Lagrangian as 
\begin{align}
\mathcal{L}_{\rm DM}^{\rm Dirac}= i \overline{\chi_L} \gamma^{\mu} (\partial_{\mu}+ i g_X Q_{\chi} Z^\prime_{\mu}) \chi_L +  i \overline{\chi_R} \gamma^{\mu} (\partial_{\mu}+ i g_X Q_{\chi} Z^\prime_{\mu}) \chi_R + (m_{\rm DM} \overline{\chi_L} \chi_R +h. c.) +g_X Q_\ell \overline{\nu_L} \gamma^\mu \nu_L Z^\prime_{\mu}
\end{align}
with $\chi= \chi_L+\chi_R$ we obtain 
\begin{align}
\mathcal{L}_{\rm DM}^{\rm Dirac}= i \overline{\chi} \gamma^{\mu} (\partial_{\mu}+ i g_X Q_{\chi} Z^\prime_{\mu}) \chi + m_{\rm DM} \overline{\chi} \chi.
\end{align}
To ensure the stability of the DM candidate, we prevent some charges for the DM candidate prohibiting some couplings. To ensure stability of DM we find that $Q_\chi \neq \{\pm 3 x_\Phi, \pm x_\Phi\}$ $=\{\pm 3,\pm1\}$, taking $x_\Phi=1$ for $U(1)_X$, $Q_\chi\neq  \pm 3\Big( \frac{-4+x}{3}\Big),~\pm\Big(\frac{-4+x}{3}\Big),~\pm\Big(\frac{2+x}{3}\Big)$ for $U(1)_{x+qu}$, $Q_\chi \neq \pm3,~\pm2,~\pm1,~0$ for $L_i-L_j$ and $Q_\chi \neq \pm9,~\pm6,~\pm3,~0$ for $B-3L_i$ scenarios respectively. Except these charges other possibilities could be allowed.  
In this scenario we consider a UV complete theory which might allow the neutrino to mix with the DM candidate through non-renormalizable, higher dimensional operators for odd $Q_\chi$. Therefore we can safely choose $Q_\chi$ as either even numbers or fractional numbers. The $\nu-$DM scattering process depends on $x_H$ coming from $Q_{\ell}$, the $U(1)_X$ charge of the neutrino. For simplicity we consider the gauge kinetic mixing to be very small. We estimate the $\nu-$DM differential scattering cross section in $t-$channel mediated by $Z^\prime$ in laboratory frame following Fig.~\ref{vDM1} as
\bea
 \frac{d \sigma}{d E_{\nu}^\prime} = \frac{g_{X}^4 Q_{\ell}^2 Q_{\chi}^2\;}{8\pi}  \frac{m_{\rm DM}}{ E_{\nu}^2}  
        \frac{(E_{\nu}^2 + {E^\prime_{\nu}}^2  -  m_{\rm DM}(E_{\nu}-E_{\nu}^\prime))}{ \{M_{Z^\prime}^2 + 2\; m_{\rm DM} (E_{\nu}-E^\prime_{\nu})\}^2}
        \label{DM-D}
\eea
where $E_{\nu}$ and $E_{\nu}^\prime$ are the energies of the incoming and outgoing neutrinos and $m_{\rm DM}$ is DM mass. In case of $U(1)_X$ scenario considering $x_{\Phi}=1$, $Q_{\ell}=-\frac{x_H}{2}-1$ which manifests chiral scenario while in case of $U(1)_{q+x u}$ scenario $Q_{\ell}=-1$, respectively. In flavored scenarios, the $U(1)$ charges of $\nu_L$ are given in Tabs.~\ref{tab:charges2} and \ref{tab:charges3-1} for $L_i-L_j$ and $B-3L_i$ scenarios, respectively. 

Neutrino-DM scattering can be constrained from different cosmic observations such as cosmic blazar TXS0506+056 and active galaxy NGC1068 data from IceCube  at the IceCube observatory in the south pole. 
Due to their high luminosity and variability, blazars have been studied extensively in astrophysics, particularly in the context of understanding the properties of the relativistic jets that are thought to be responsible for their emission. Recently, there has been growing interest in using blazars as astrophysical probes of BSM physics from the aspects of astroparticle physics. In our paper, we specifically focus on events from cosmic blazar TXS0506+056 and active galaxy NGC1068
to study $\nu-$DM interactions and constrain $g_X-M_{Z^\prime}$ plane for different $U(1)$ scenarios.
\subsection{Cosmic blazar TXS0506 + 056}
The cosmic blazar TXS0506+056 \cite{ICECUBE:2018cha}, located at a distance of approximately 4 billion light years from Earth, was the first blazar observed to emit a high-energy neutrino. A 290 TeV neutrino event, known as IC- 170922A, was observed by the IceCube experiment in 2017 and was verified to be coming from the cosmic blazar TXS0506+056. This has since spurred significant interest in using blazars as a tool for studying $\nu-$DM interactions. The neutrino flux emitted by the blazar can be evaluated using the lepto-hadronic model. In the lepto-hadronic model, the jets of blazars are thought to be composed of a plasma of relativistic electrons and protons in which the electrons are accelerated to very high energies which then emit synchrotron radiation. The expected neutrino flux is then given by \cite{Gasparyan:2021oad,Cline:2022qld}
\begin{equation}
    \log_{10} \frac{\Phi_{\nu}}{\text{cm}^2} = -\mathcal{F}_0 x- \frac{\mathcal{F}_1 x}{1 + \mathcal{F}_2 \vert x\vert^{\mathcal{F}_3}}
    \label{flux-1}
\end{equation}
where $\mathcal{F}_0= 13.22$, $\mathcal{F}_1 = 1.498$, $\mathcal{F}_2 = -0.00167$, $\mathcal{F}_3 = 4.119$, and $x = \log_{10} (E_{\nu}/\text{TeV})$ with $E_{\nu} \in [10^{-1.2},10^{4.2}]$ TeV being the energy of the neutrinos. 
\begin{figure}[h!]
    \centering
    \includegraphics[scale=0.545]{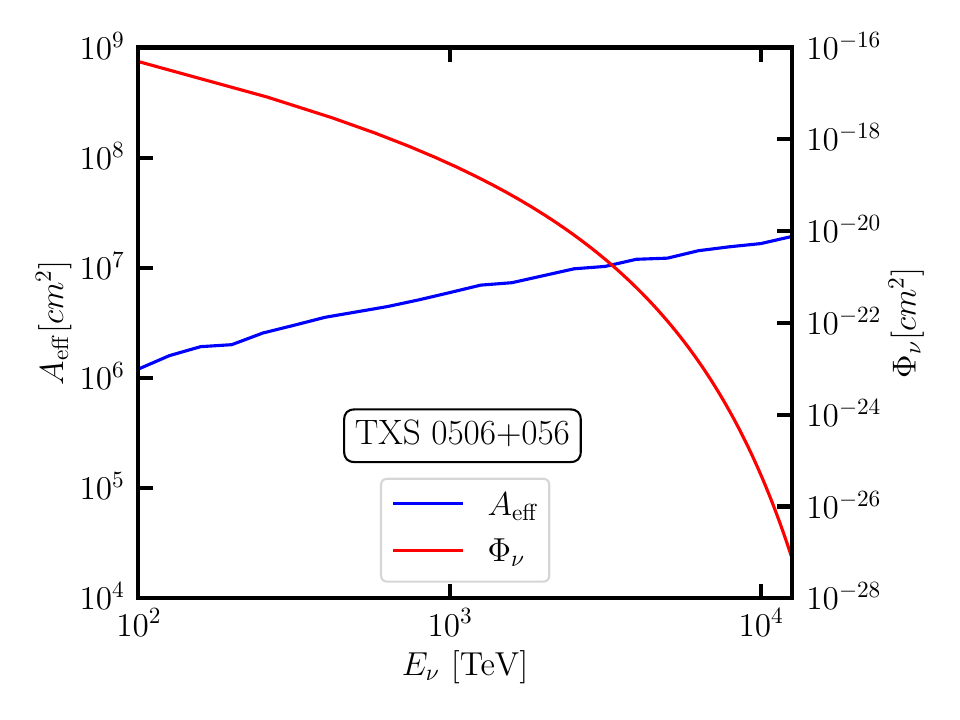}
    \includegraphics[scale=0.545]{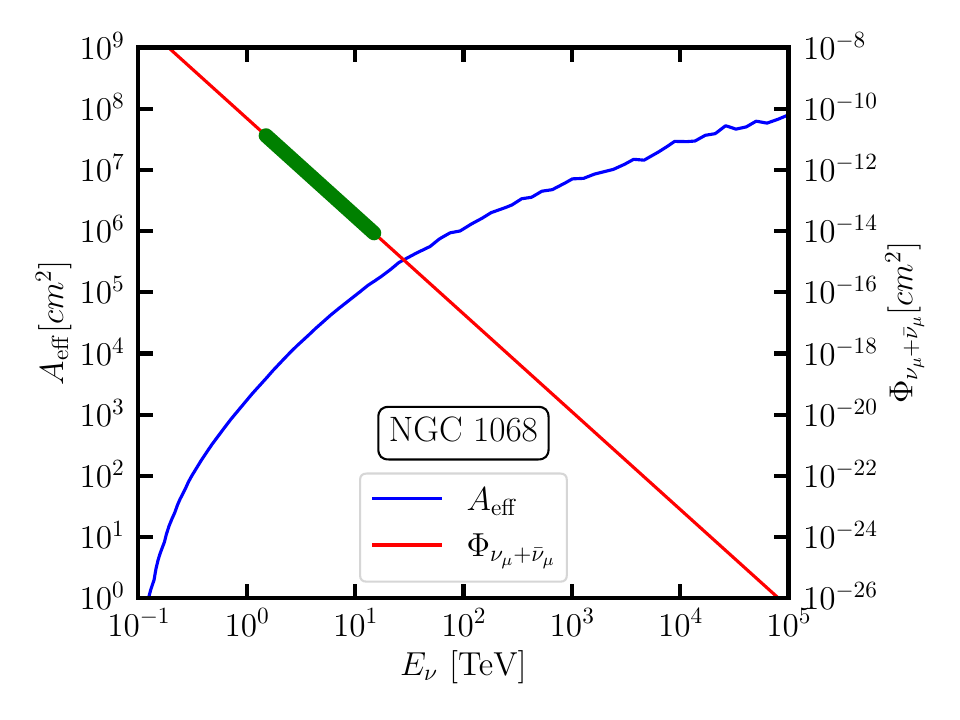}
    \caption{The flux predicted by the lepto-hadronic model and the effective area of detection obtained from IceCube data. The left panel shows an effective area for the blazar event TX0506 + 056 while the right panel shows the same quantities for NGC 1068. The region marked in green for NGC 1068 where IceCube reliably measures the flux.  The energy range corresponding to the green region is $E_{\nu} ~\in [1.5,15]$ TeV which has been considered in this paper.}
    \label{fig:blazar_intro}
\end{figure}
Then, the number of events observed at IceCube can be calculated using
\begin{equation}\label{eq:eventseqn}
    N_{\rm events} = t_{\rm obs} \int dE \; A_{\rm eff}(E) \; \Phi_{\nu}(E_{\nu})
\end{equation} 
where $A_{\text{eff}}$ is the effective area of detection of high energy neutrinos and it describes the probability for a neutrino to convert into a muon inside the detector of IceCube \cite{icecubedata}. In this analysis, we take $t_{\rm obs} = 898$ days for the IC86a campaign which lasted from Julian day 57161 - 58057. The total flux and effective area of detection of high energy neutrinos from the blazars are plotted in the left panel of Fig.~\ref{fig:blazar_intro} using Eq.~(\ref{flux-1}) and the IceCube data repository.

We consider a model in which potential DM candidate surrounds a central BH, so that neutrinos from the blazar will interact with DM. We need to compute column density which measures the DM density along the line of sight of the observer. This makes it possible to constrain $\nu$-DM scattering from blazar. We define the accumulated DM column density following \cite{Gondolo:1999ef} as
\begin{equation}
    \tau = \frac{\Sigma(r)}{m_{\rm DM}} = \int dr \frac{\rho_{\rm DM}}{m_{\rm DM}}
    \label{column-den}
\end{equation}
where $\rho_{\rm DM}$ is the spike density profile of DM and is given by $\rho_{\rm DM} (r) = \frac{ \rho^{\rm core} \tilde{\rho}(r)}{\rho^{\rm core} + \tilde{\rho}(r)}$. Here $\rho^{\rm core}$ is the maximum core density set by DM annihilation in the inner spike region and is given by the relation $\rho^{\rm core} \simeq \frac{m_{\rm DM}}{\langle \sigma v\rangle_{\rm eff} \; t_{\rm BH}}$ where $t_{\rm BH} \simeq 10^9$ years being age of the BH following `Bachal-Wolf` solution from \cite{1976ApJ...209..214B} and $\langle \sigma v\rangle_{\rm eff}$ is the effective thermal averaged DM annihilation cross section. 
\begin{figure}[h!]
    \centering
   \includegraphics[scale=0.4]{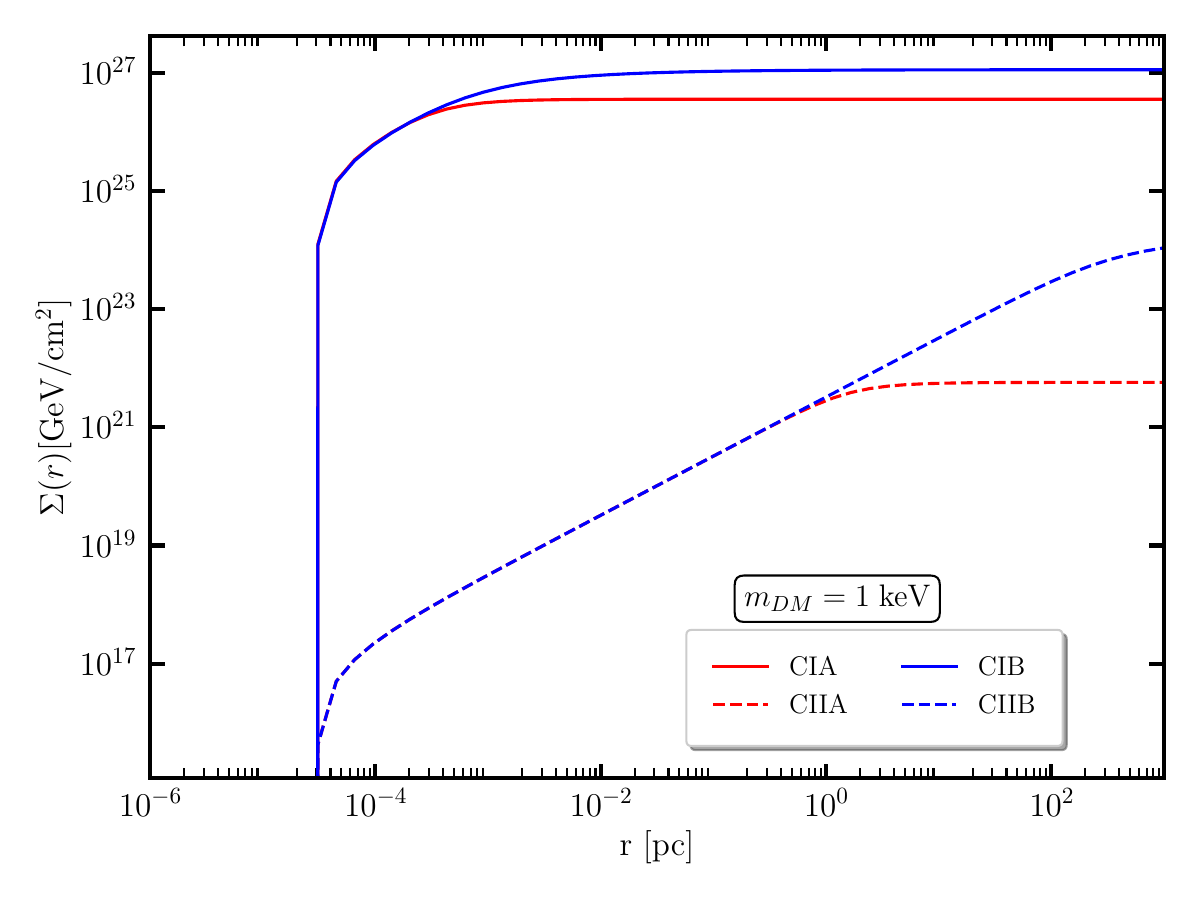}
    \includegraphics[scale=0.4]{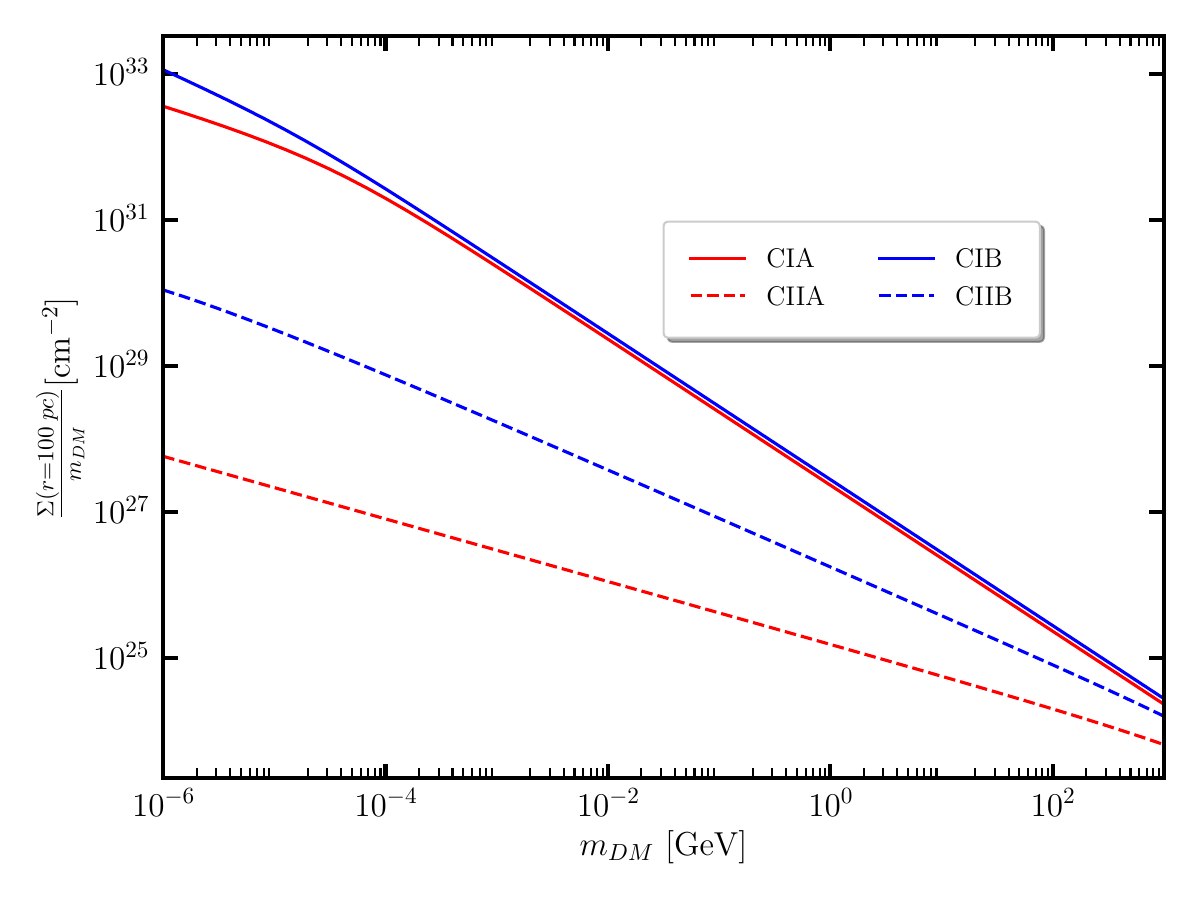}    
     \caption{Accumulated column density with respect to distance $r$ from the BH, for a fixed DM mass $m_{\rm DM} = 10^{-3}$ MeV and for different benchmark scenarios, is shown in the left panel, while accumulated column density per unit DM mass at $r=100$ pc with respect to DM mass is shown in the right panel.}
    \label{fig:Sigmavsr-1}
\end{figure}
The density profile $\tilde{\rho}(r)$ is defined as 
\bea
\tilde{\rho}(r) = \mathcal{N} \Big(1 - 4 \frac{R_{\rm Sc}}{r}\Big)^3 r^{-\alpha_{\rm sp}}
\eea
where $R_{\rm Sc} = 2 G M_{\rm BH}$ is the Schwarzschild radius of BH and $\alpha_{\rm sp}$ is the slope of DM spike profile and $\mathcal{N}$ being a renormalization constant which can be determined as  
\bea
\mathcal{N} = \frac{M_{\rm BH}}{4\pi \int_{4 R_{\rm Sc}}^{10^5 R_{\rm Sc}} r^{2-\alpha_{\rm sp}} \Big(1- 4\frac{R_{\rm Sc}}{r}\Big)^3 dr }
\eea
by requiring that the mass of the spike to be of the same order as $M_{\rm BH}$ \cite{Ullio:2001fb}. DM particles within $4 R_{\rm Sc}$ are captured by the BH\cite{Gorchtein:2010xa}. It has been estimated in \cite{Padovani:2019xcv} that BH mass of the blazar TXS0506+056 is $3.09 \times 10^8 M_{\odot}$. We consider that canonical thermally averaged generic WIMP type DM annihilation cross section has an upper limit of $\langle \sigma v\rangle \simeq 3\times 10^{-26}$ cm$^{3}$ s$^{-1}$ \cite{Fermi-LAT:2011vow}. Hence we take effective thermally averaged DM annihilation cross sections as $\langle \sigma v\rangle_{\rm eff}= 10^{-34}$ cm$^{3}$ s$^{-1}$ and $3\times 10^{-26}$ cm$^{3}$ s$^{-1}$ as benchmarks. Following \cite{1997ApJ...490..493N,Moore:1999nt,Klypin:2000hk,Power:2002sw,Gnedin:2003rj} we summarize, applying the assumptions of collisionless and phase space density conserving particle DM scenario, dissipationless galaxy simulations predict a power law cusp in DM density $\rho_{\rm DM} \propto r^{-\gamma}$ with $ 1 < \gamma < 1.5$ or even for $\gamma < 1$ \cite{Stoehr:2003hf} where $\gamma$ is the slope of initial profile. However, in the vicinity of a galactic center the mass in the inner core is dominated by the supermassive BH which may undergo an adiabatic growth in the central region of the BH due to an effect of a small speed accereting luminous and nonluminous objects \cite{Schodel:2002py,Genzel:2000mj,Ghez:1998ph,Lynden-Bell1971,Zhao:2001py}, the DM cusp may enhance spiking up the form $\rho_{\rm{DM}} \propto r^{-\alpha_{\rm sp}}$ where $2.3 < \alpha_{\rm sp}= (9- 2\gamma)/(4-\gamma) < 2.5 $ \cite{Gondolo:1999ef}. It has been shown in \cite{Ullio:2001fb} that enhancement becomes weaker due to the instantaneous appearance of the BH being induced by mergers of progenitor halos resulting a slope in spike and identified as $\alpha_{\rm sp}=1.33$. On the other hand simulation studies showed that mergers of BHs in the progenitor halos may reduce the density of DM to a reduced power law $\rho_{\rm DM} \propto r^{-1/2}$ due to kinetic heating of the particles during merger \cite{Milosavljevic:2001vi,Volonteri:2002vz,Merritt:2002vj}. It further grows away from the central region of the DM distribution \cite{1971reas.book.....Z}.  According to the previous studies \cite{Navarro:1995iw,Navarro:2003ew,Reed:2003hp,Fukushige:2003xc,Diemand:2004wh} we identify $\gamma$ as the power spectrum index parametrizing the inner cusp of initial DM halo density.
Another aspect pointed out in \cite{Gnedin:2003rj} is that the galactic center may have a compact cluster of stars in addition to the supermassive BH. These stars may scatter DM particles causing an evolution to the DM distribution function leading to a quasi-equilibrium profile through two-body relaxation for the stars and DM. Demanding a steady state scenario, DM distribution function can be obtained as a power-law of energy and DM density can be obtained as a power-law of radius providing unique solutions being independent of initial conditions \cite{1976ApJ...209..214B,1977ApJ...216..883B,Spudich1976}. Considering DM mass to be negligibly smaller than stellar mass, the quasi-equilibrium solution further reduce to $\alpha_{\rm sp}=3/2$ for two-component system of DM and starts described by a collisional Fokker-Planck equation  \cite{1983ApJ...264...24M} with no energy flux. We consider two benchmark cases $\alpha_{\rm sp} =3/2$ and $7/3$, and the complete benchmark scenarios to be: (CIA) $\langle \sigma v\rangle_{\rm eff}= 10^{-34}\rm{cm}^{3} \rm{s}^{-1}$ $\alpha_{\rm sp}=7/3$; (CIIA) $\langle \sigma v\rangle_{\rm eff}= 3\times10^{-26}\rm{cm}^{3} \rm{s}^{-1}$, $\alpha_{\rm sp}=7/3$; (CIB) $\langle \sigma v\rangle_{\rm eff}= 10^{-34}\rm{cm}^{3} \rm{s}^{-1}$, $\alpha_{\rm sp}=3/2$; (CIIA) $\langle \sigma v\rangle_{\rm eff}= 3\times10^{-26}\rm{cm}^{3} \rm{s}^{-1}$, $\alpha_{\rm sp}=3/2$
Hence following Eq.~(\ref{column-den}) and applying $\rho_{\rm DM}$ for blazars we show the accumulated DM column density with respect to $r$ in the left panel of Fig.~\ref{fig:Sigmavsr-1}. In this case, the behaviour of the accumulated DM column density changes with $\alpha_{\rm sp}$ depending on $r$. For fixed $\langle \sigma v\rangle_{\rm eff}$, $\Sigma(r)$ changes for $\alpha_{\rm sp}$. The lower slope gives a denser profile after $r > 1$ pc. For fixed $\alpha_{\rm sp}$ we find that lower $\langle \sigma v\rangle_{\rm eff}$ gives a denser profile due to its presence in the denominator of $\rho^{\rm core}$. The density profile per unit DM mass is shown in the right panel of Fig.~\ref{fig:Sigmavsr-1} depending on DM mass. For fixed $\alpha_{\rm sp}$, the quantity $\Sigma(r)/m_{\rm DM}$ becomes different for $m_{\rm DM} < 10^{-3}$ GeV which will be useful to solve cascade equation for blazars to estimate constrains on $g_X-M_{Z^\prime}$ plane for different $U(1)$ scenarios. 
\subsection{Active galaxy NGC1068}
IceCube reported an excess of $79^{+22}_{-20}$ neutrinos identified with the active galaxy NGC1068 \cite{ICECUBE:2022der} at a significance of $4.2 \sigma$. The active galaxy NGC1068 is located at a distance of approximately 14 million light years from earth. Being based on the optical emission lines it is classified that, NGC1068 is a Seyfert type-2 galaxy characterized by its broad emission lines resulting from interaction between the radiation and surrounding gas. The active galaxy NGC1068 is identified to be the first steady source of neutrino emission. To study the effects of the $\nu$-DM scattering on the initial neutrino flux we solve the cascade equation defined as
\begin{equation}
    \frac{d\Phi}{d\tau} = -\sigma\;\Phi + \int_{E^\prime}^{\infty} dE \; \frac{d\sigma}{dE^\prime} \Phi(E) 
\end{equation}
where $E$ and $E^\prime$ are the initial and final energies of neutrino where $\tau = \Sigma(r)/m_{\rm DM}$ is the accumulated column density of DM along the line of sight. Defining a dimensionless quantity $x = \tau  \frac{m_{DM}}{\Sigma(r)}$ the equation can be rewritten as
\begin{equation}\label{eq:cascade}
    \frac{d\Phi}{dx} = -\sigma \frac{\Sigma(r)}{m_{\rm DM}}\;\Phi + \frac{\Sigma(r)}{m_{\rm DM}} \int_{E^\prime}^{\infty} dE \; \frac{d\sigma}{dE^\prime} \Phi(E) 
\end{equation}
where $x \in [0,1]$. Here $\sigma$ and $\Phi$ are the total cross section and flux, respectively. The cascade equation given in Eq.~(\ref{eq:cascade}) can be solved using the vectorization method as described in \cite{Vincent:2017svp}. In our analysis we fix $m_{\rm{DM}} = 3 M_{Z^\prime}$. Hence we estimate constraints on $g_X-M_{Z^\prime}$ plane for different $U(1)$ scenarios for NGC1068. The functional form of the neutrino flux can be written as 
\bea
\Phi_{\nu_{\mu} + \bar{\nu}_{\mu}} (E_{\nu})= \Phi_{\rm ref} \Big(\frac{E_\nu}{E_{\rm ref}}\Big)^{-a}.
\label{flux-2}
\eea
From \cite{icecubedata-agn} we find that Eq.~(\ref{flux-2}) reduces to $4.9032 \times 10^{-11}  E_{\nu}^{-3.196}$ where $E_{\nu} \in [0.1 ,10^4]$ TeV considering $E_{\rm ref}=1$ TeV and the central value of the observed events being 79 at IceCube. Using $\Phi_{\nu_{\mu} + \bar{\nu}_{\mu}} (E_{\nu})$ and corresponding $A_{\rm eff}$ in Eq.~(\ref{eq:eventseqn}) with $t_{\rm obs} =3186$ days \cite{ICECUBE:2022der}, we calculate the number of expected events in the form of muon neutrinos. The number of events reduces to 31 within the reliably measured IceCube range $E_\nu \in [1.5, 15]$ TeV. The flux $\Phi_{\nu_{\mu} + \bar{\nu}_{\mu}}$ and $A_{\rm eff}$ for the event NGC1068 can be directly obtained from the IceCube data \cite{icecubedata-agn} and the correspondence has been shown in the right panel of Fig.~\ref{fig:blazar_intro}. Here the green region corresponds to reliable measurement of the flux taking $\nu_\mu$ detected by IceCube detectors. Due to the effect of neutrino oscillation over an astrophysical distance, the flux for all three flavors will be enhanced by a factor of three \cite{Bustamante:2015waa,Arguelles:2015dca}. The $\nu-$DM scattering has a possibility to dissipate the energy of the neutrino from the source towards an observer on Earth. It has been pointed out in \cite{Cline:2022qld} that $\nu-$DM scattering could shift the flux peak of a spectrum to lower energies resulting in a larger amount of  expected number of neutrinos at the detectable energy range of IceCube. As a result, bounds on $\nu-$DM scattering cross section may be stronger than a flux having a peak at higher energy compared to the observed ones. Therefore we can use the flux given in Eq.~(\ref{flux-2}) as the initial flux to estimate conservative bounds on $\nu-$DM scattering cross section. 
\begin{figure}[h!]
    \centering
   \includegraphics[scale=0.4]{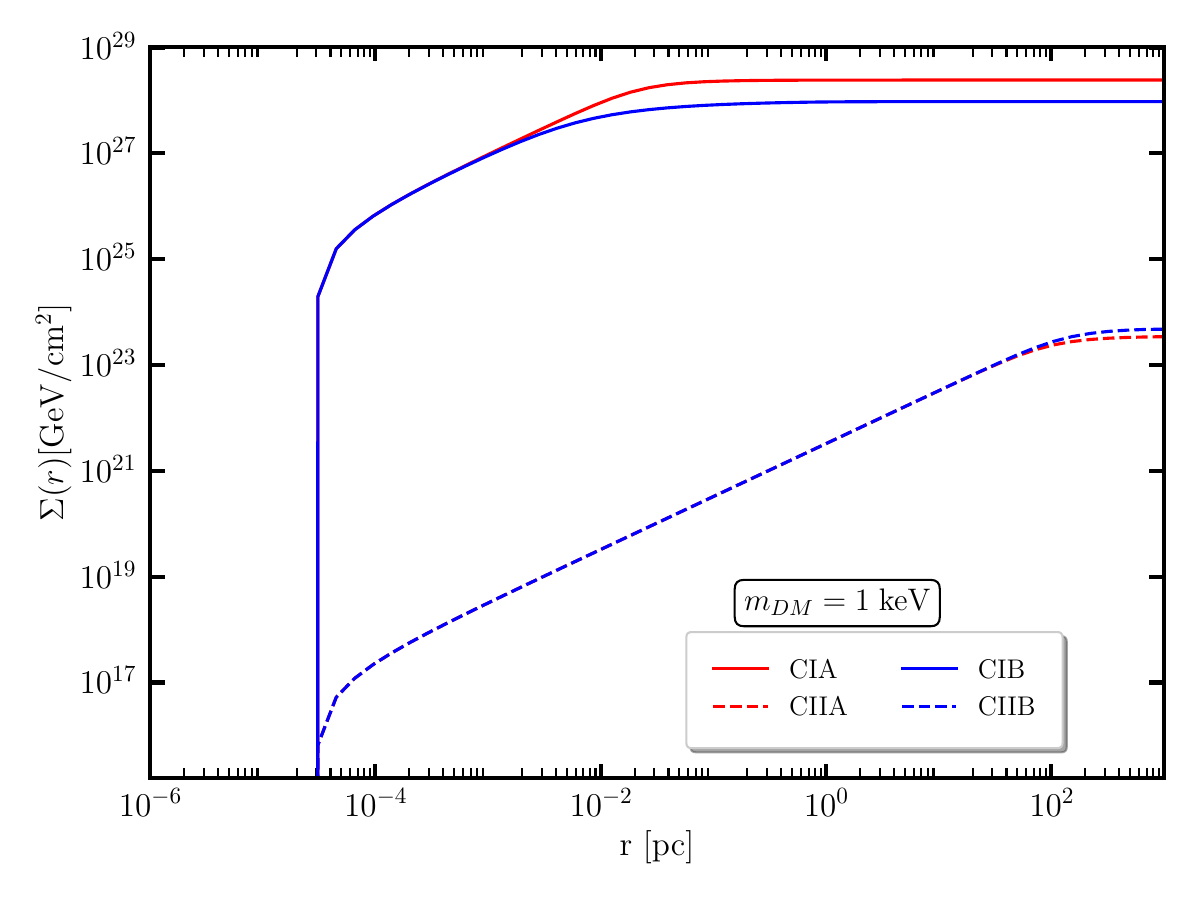}
    \includegraphics[scale=0.4]{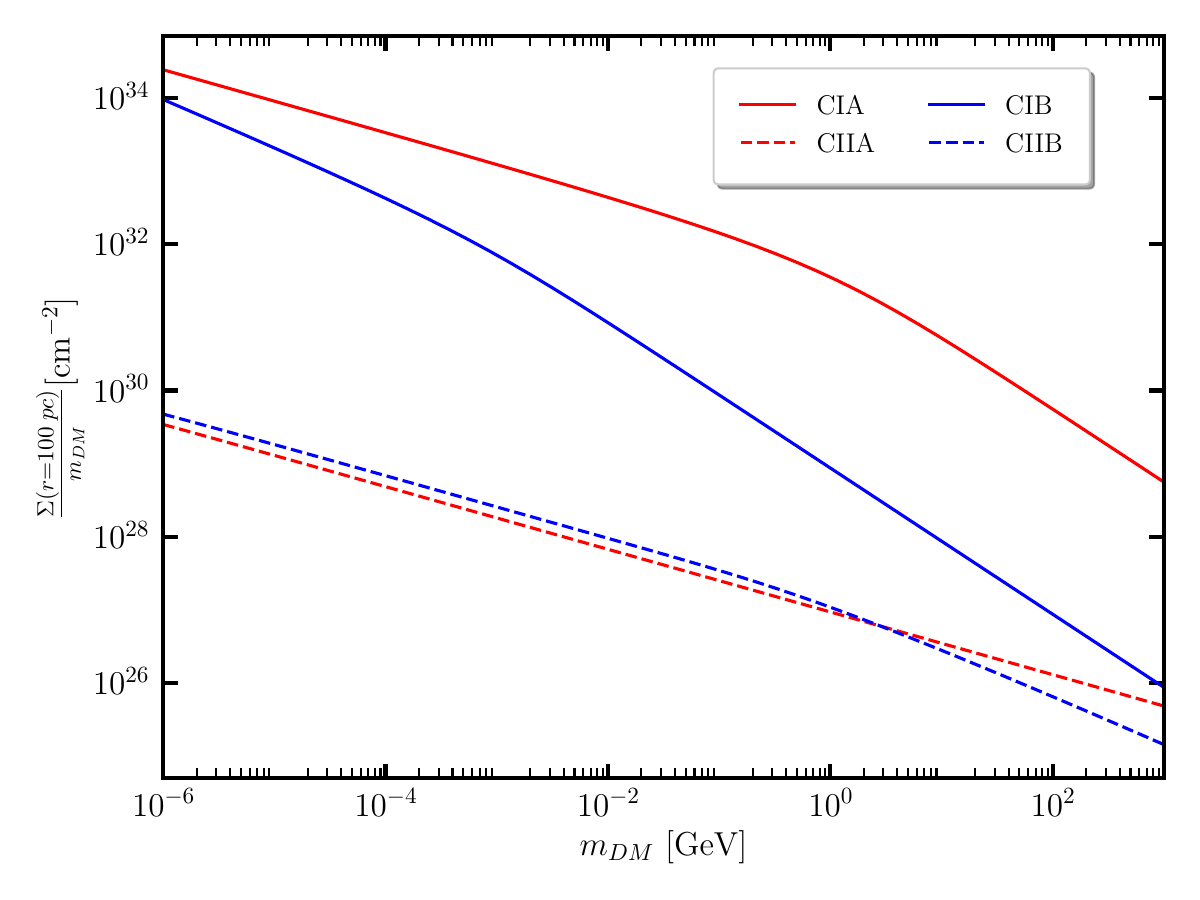}    
     \caption{Accumulated column density with respect to distance $r$ from the BH, for a fixed DM mass $m_{\rm DM} = 10^{-3}$ MeV and for different benchmark scenarios, is shown in the left panel, while accumulated column density per unit DM mass at $r=100$ pc with respect to DM mass is shown in the right panel.}
    \label{fig:Sigmavsr-2}
\end{figure}
To study $\nu-$DM scattering we introduce DM density profile. From \cite{Gondolo:1999ef} we come to know if the accretion of the BH is adiabatic and we neglect the relativistic effects so DM density profile can be expressed in the form $\rho_{\rm DM}\simeq \rho_{\rm sc} (\frac{r}{r_{\rm sc}})^{-\gamma}$, that is as a cusp in the region close to the BH where $\rho_{\rm sc}$ and $r_{\rm sc}$ being the scale density and scale radius, respectively. Then the DM density profile evolves into 
\bea
\tilde{\rho}_{\rm DM} (r) \simeq \rho_{R} \Big(1-4 \frac{R_{\rm Sc}}{r}\Big)^3 \Big(\frac{R_s}{r}\Big)^{\alpha_{\rm sp}}, 
\eea
where $R_s$ is the typical size of a spike profile \cite{Gondolo:1999ef,Quinlan:1994ed}. Due to the presence of stars in the inner region of a galaxy, DM may experience scattering which may vary $\alpha_{\rm sp}$ allowing us to consider two slopes $3/2$ and $7/3$ which is possible within a radius of influence $(r_I)$ inside a supermassive BH \cite{Merritt:2003qk,Gnedin:2003rj,Merritt:2006mt,Shapiro:2022prq}. The radius of influence defines a region where the gravitational effect of the supermassive BH affects the movement of neighboring stars and radius of influence is generally less than the typical size of the spike profile. Now we define the DM spike density profile for $\alpha_{\rm sp}=3/2$ as 
\begin{align}
    \rho_{3/2}= 
\begin{cases}
    \rho_N \left(1-4\frac{R_{\rm Sc}}{r} \right)^3 \left(\frac{r_I}{r}\right)^{3/2}, &  r_i \leq r \leq r_I \\ 
    \rho^\prime_N \left(\frac{R_{s}}{r} \right)^{7/3},   &  r \geq r_I ~(\rm outer~profile)
\end{cases}
\end{align}
and that with $\alpha_{\rm sp}=7/3$ as 
\begin{align}   
\rho_{7/3}= \; \rho_N \left( 1 - 4\frac{R_{\rm Sc}}{r}\right)^3 \left(\frac{R_{s}}{r} \right)^{7/3}, \; r\geq r_i  
\end{align}
respectively. Here the inner radius of the spike is given by $r_i = 4 R_{\rm Sc}$ and $r_I=\frac{G M_{\rm BH}}{\sigma_v^2} \simeq 0.65$ kpc for a supermassive BH taking $\sigma_v$ as stellar velocity dispersion. The mass of the supermassive BH has been considered as $M_{\rm BH}= 10^{7} M_{\odot}$ being consistent with the estimation of the DM halo mass $\mathcal{O}(10^{11} M_{\odot})$ NGC1068 \cite{Baes:2003rt,Ferrarese:2002ct}. For $\alpha_{\rm sp} = 3/2$ we define the normalization density parameter as $\rho_N =  \mathcal{N} r_{I}^{-3/2}$ and that for $\alpha_{\rm sp}=7/3$ we define $\rho_N = \mathcal{N} R_{s}^{-7/3}$ respectively with 
\bea
\mathcal{N}=  \frac{M_{\rm BH}}{4\pi(f(\alpha_{\rm sp},r_I) - f(\alpha_{\rm sp},r_i))}
\eea
and 
\bea
f(\alpha_{\rm sp},r) = r^{-\alpha_{\rm sp}} \left(\frac{r^3}{3-\alpha_{\rm sp}} + \frac{12 R_{\rm Sc} r^2}{\alpha_{\rm sp} -2} - \frac{48 R_{\rm Sc}^2 r}{\alpha_{\rm sp}-1} + \frac{64 R_{\rm Sc}^3}{\alpha_{\rm sp}} \right)
\eea
while $M_{\rm BH} \simeq 4\pi \int_{r_i}^{r_I}  \rho_{\rm DM} (r)~r^2~dr$ \cite{Ullio:2001fb,Gorchtein:2010xa}. Here $\rho_{\rm DM}$ is given by 
\begin{align}
    \rho_{\rm DM}= 
\begin{cases}
    0, &  r\leq r_i \\
    \frac{\rho_{\alpha_{\rm sp}}(r) \rho^{\rm core}}{\rho_{\alpha_{\rm sp}}(r) +\rho^{\rm core}},               & r_i \leq r \leq R_{s}\\
    \frac{\rho_{\rm NFW}(r) \rho^{\rm core}}{\rho_{\rm NFW}(r) +\rho^{core}},               & r \geq R_{s}
    \label{Dm-den}
\end{cases}
\end{align}
where $\rho^{\rm core} \simeq \frac{m_{\rm DM}}{\langle \sigma v \rangle_{\rm eff} t_{\rm BH}}$. For the NFW halo profile we define $\rho_{\rm NFW}(r) = \rho_s \left(\frac{r_{\rm sc}}{r}\right) \left( 1 + \frac{r}{r_{\rm sc}}\right)^{-2}$ where $\rho_{\rm sc} = 0.35$ $\rm GeV/{cm}^3$ and the scale parameter for density profile is $r_{\rm sc} = 13$ kpc \cite{Cline:2023tkp,Gentile:2007sb} being consistent with the data of the supermassive BH in Milky Way \cite{EventHorizonTelescope:2022wkp} which has a mass alike one in NGC1068. In this context we mention that normalization density parameter $\rho_N^\prime \simeq \rho_N (r_I/R_s)^{7/3}=\mathcal{N} r_I^{5/6} R_s^{-7/3}$ requiring that at $r=r_I$ then $\rho_N^\prime \to \rho_{3/2}^{\prime}$. Matching the conditions of spike and outer part of the halo as $\rho_N^\prime, \rho_N \simeq \rho_{\rm sc} (R_s/r_{\rm sc} )^{-\gamma}$ for $\alpha=3/2$ and $7/3$ where $R_{\rm Sc} << r_I < R_s << r_{\rm sc}$,  we obtain
\begin{align}
    R_{s}= 
\begin{cases}
    \left( \frac{\mathcal{N}}{\rho_{\rm sc} r_{\rm sc}} \right)^{3/4} r_I^{5/8}, &  \alpha = 3/2 \\
    \left( \frac{\mathcal{N}}{\rho_{\rm sc} r_{\rm sc}} \right)^{3/4},      & \alpha = 7/3
    \label{Rs}
\end{cases}
\end{align}
For NGC1068 we consider $R_{\rm s} \simeq 0.7$ kpc which is greater than $r_I$ under the parameters used throughout this paper. Finally we define the accumulated column density of DM as $  \Sigma(r) = \int_{R_{\rm EM}}^r  dx \; \rho_{\rm DM}(x)$ where $R_{\rm EM}$ ($\approx 30 R_{\rm Sc} =2.8 \times 10^{-5}$ pc) is the position at which neutrinos are expected to be produced with respect to the BH \cite{Murase:2019vdl,Murase:2022dog}. The DM density $\rho_{\rm DM}$ is given by the expression in Eq.~(\ref{Dm-den}) and integrating over the radius for different benchmark cases (CIA, CIIA, CIB, CIIB) we show the column density in the left panel of Fig.~\ref{fig:Sigmavsr-2} starting from $r > R_{\rm EM}$. The accumulated column density for different DM models saturate to a constant value around $r\simeq 10$ pc and $100$ pc depending on $\alpha_{\rm sp}$ and $\langle \sigma v\rangle_{\rm eff}$. Hence we come to a conclusion that at the Earth where $r\simeq 14.4$ Mpc the accumulated column density is considered to be a constant. We show the accumulated column density per unit DM mass for CIA, CIIA, CIB and CIIB as this ratio is a relevant parameter in solving the cascade equation.
\section{Results and discussions}
\label{dis}
We study scattering and beam-dump experiments for different $U(1)$ extensions of the SM to estimate bounds on $g_X-M_{Z^\prime}$ plane and compare these results with those from GRB, cosmic blazars and active galaxy to highlight complementarity:

\subsection{Limits on TeV scale $Z^\prime$ from colliders}
\noindent
{\bf (i) LEP-II and ILC:} The scattering cross sections are calculated including $Z^\prime$ exchange diagram for $e^-e^+ \to f \overline{f}$ process at the $Z$ pole and compared with observed value from LEP-II \cite{ALEPH:2013dgf} to estimate upper limit on $g_X-M_{Z^\prime}$ plane. First we consider the hadronic final state using $e^+ e^- \to q \bar q$ process where the cross section is $\sigma = 41.544 \pm 0.037$ nb. Comparing the $Z^\prime$ mediated process for hadronic process we estimate upper limits at 90\% C.L and then we consider $e^-e^+ \to \ell^- \ell^+$ process with $R_\ell= \Gamma_{\rm had}/ \Gamma_\ell=20.768\pm0.025$ to estimate bounds from the dilepton final state in LEP-II. Upper bounds for the dijet and dilepton final states are shown in Figs.~\ref{fig:lim0}-\ref{fig:lim3} for $U(1)_X$ case in the upper panels where dilepton final state seems to provide strongest bound at the $Z$ pole. In this line we show the strongest bound from the dilepton final state for $U(1)_{q+xu}$ scenarios shaded by gray in the lower panels of Figs.~\ref{fig:lim0}-\ref{fig:lim3}. In the same line we estimate strongest limits for $L_{e}-L_{\mu, \tau}$ and $B-3L_e$ scenarios using dilepton and dijet final states shown in Figs.~\ref{fig:lim4} and \ref{fig:lim6} respectively. 

Using the fact that $M_{Z^\prime} >> \sqrt{s}$ we find limits on $M_{Z^\prime}/g_X$ for different $U(1)$ scenarios from LEP-II and prospective ILC at $\sqrt{s}=250$ GeV, 500 GeV and 1 TeV, respectively using the bounds given in Tab.~\ref{tab3}. The corresponding limits are shown in Figs.~\ref{fig:lim0}-\ref{fig:lim3} for different general $U(1)$ scenarios, in Fig.~\ref{fig:lim4} for $L_e-L_{\mu, \tau}$ and Fig.~\ref{fig:lim6} for $B-3L_e$ scenario respectively. In $L_e-L_{\mu,\tau}$ scenarios LEP-II provides strongest direct bounds. Prospective ILC bounds could be stronger for heavy $Z^\prime$ for different $U(1)$ scenarios depending on the charges. 

\noindent
{\bf (ii) LHC:} We estimate dilepton production cross section $(\sigma^{\rm{model}})$ at $\sqrt{s}=13$ TeV proton collider for a trial $U(1)$ coupling $g^{\rm{model}}$ under different $U(1)$ scenarios except $L_i-L_j$. Now we compare the cross section obtained in LHC $(\sigma^{\rm{CMS/ ATLAS}})$ from  CMS\cite{CMS:2021ctt} (ATLAS\cite{ATLAS:2019erb}) with $\frac{\Gamma}{M_{Z^\prime}}=3\%$ where $\Gamma$ is total decay width of $Z^\prime$ at $140(139)$ fb$^{-1}$ luminosity. Hence limits on different $U(1)$ scenarios can be estimated using 
\bea
g_X=g^{\rm{model}} \sqrt{\frac{\sigma^{\rm{CMS/ATLAS}}}{\sigma^{\rm{model}}}}.
\eea
The model cross section has been calculated using the narrow width approximation as 
\begin{equation}
\sigma^{\rm model}= \frac{8 \pi^2}{3}~\sum_{q,\bar{q}} \int dx \int dy~q(x, Q), \bar{q}(y, Q) \times \frac{\Gamma(Z^\prime \to q \bar{q})}{M_{Z^\prime}} \delta(\hat{s}-M_{Z^\prime}) \times \rm{BR}(Z^\prime \to 2\ell)
\end{equation}
where $q(x, Q)$ $(\bar{q}(y, Q))$ are the parton distribution functions of the quark (antiquark) and $\hat{s}= xys$ is invariant mass squared of the colliding quark at $\sqrt{s}$, the center of mass energy. We set the factorization scale at $Q=M_{Z^\prime}$ in the parton distribution function CTEQ6L\cite{Pumplin:2002vw}. These bounds are shown by CMS2$\ell$ and ATLAS2$\ell$ in Figs.~\ref{fig:lim0}-\ref{fig:lim3} for general $U(1)$ and in Figs.~\ref{fig:lim6} and \ref{fig:lim7} for the $B-3L_{i}$ cases where strongest direct upper bounds from CMS and ATLAS are almost comparable. We scale these bounds to a projected 3 ab$^{-1}$ luminosity using 
\bea
g_X^{\rm projected}\simeq g_X^{\rm{current}}\sqrt{\frac{139(140)~\rm fb^{-1}}{\mathcal{L}_{\rm projected}}}
\eea
shown by CMS$2\ell$-3 and ATLAS$2\ell$-3 respectively using dilepton final states \cite{CMS:2021ctt,ATLAS:2019erb,CMS:2016xbv,ATLAS:2017eiz} at the LHC.  
\subsection{Limit from dark photon search at LHC experiments}
\begin{figure}
\hspace*{-1.5cm}
\includegraphics[scale=0.51]{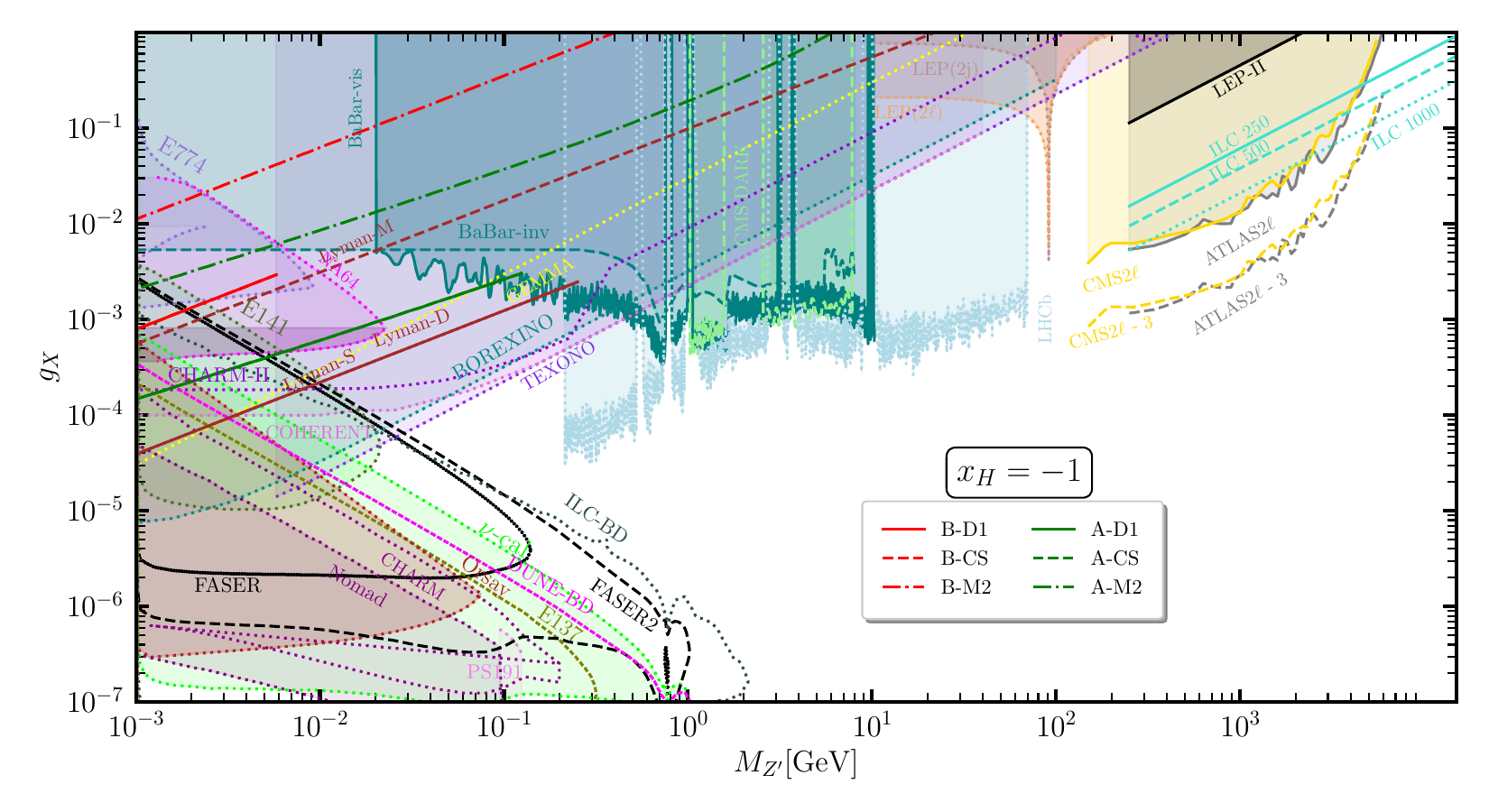}
\hspace*{-1.5cm}
\includegraphics[scale=0.58]{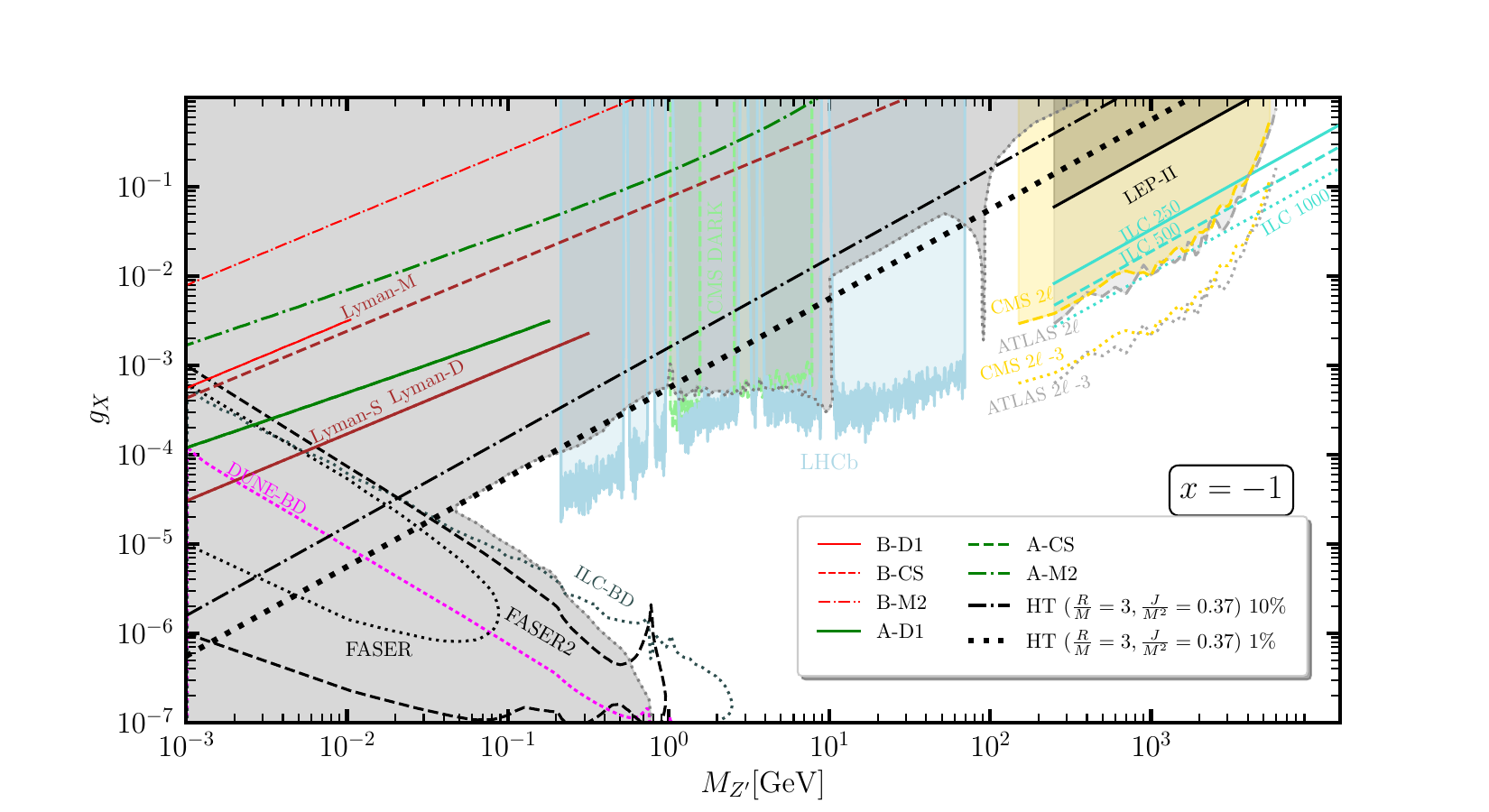}
\caption{Limits on $g_X-M_{Z^\prime}$ plane for $U(1)_X$ and $U(1)_{q+xu}$ scenarios for $x_H=-1$ (upper) and $x=-1$ (lower).}
\label{fig:lim0}
\end{figure}

\begin{figure}
\hspace*{-1.5cm}
\includegraphics[scale=0.56]{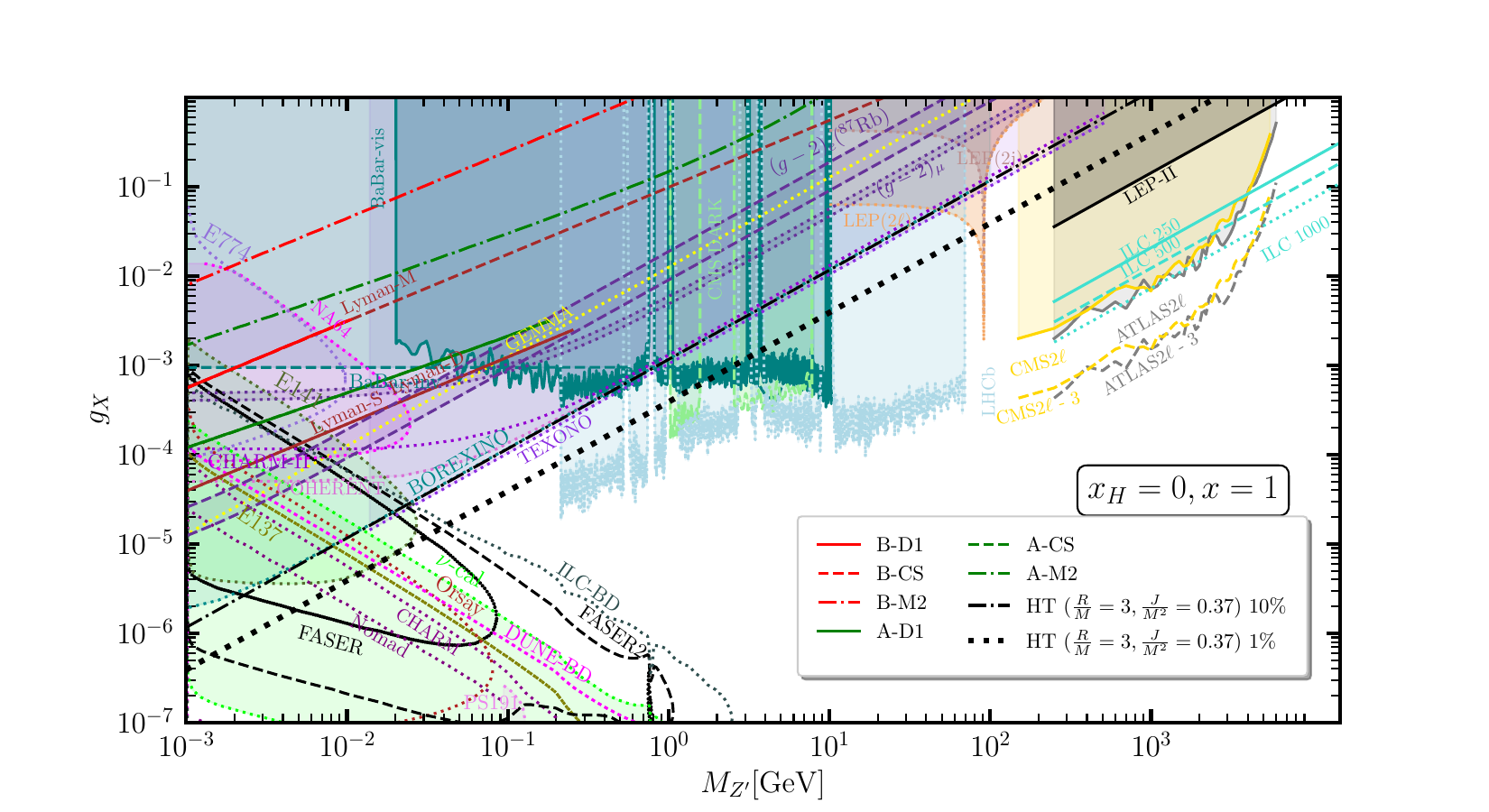}
\caption{Limits on $g_X-M_{Z^\prime}$ plane for $U(1)_X$ and $U(1)_{q+xu}$ extensions for $x_H=0$ (upper) and $x=1$ (lower).}
\label{fig:lim1}
\end{figure}

\begin{figure}
\hspace*{-1.5cm}
\includegraphics[scale=0.56]{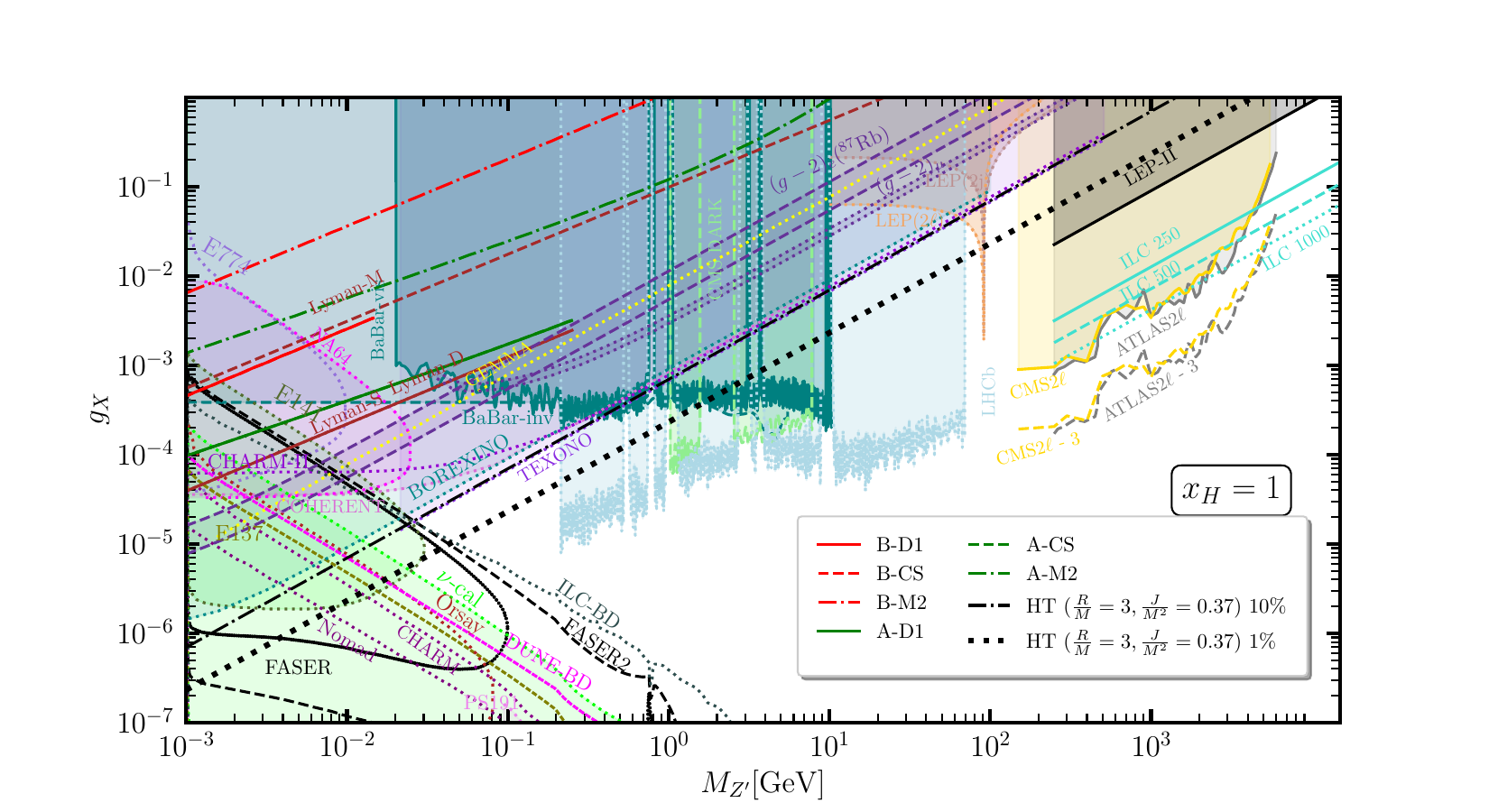}
\hspace*{-1.5cm}
\includegraphics[scale=0.56]{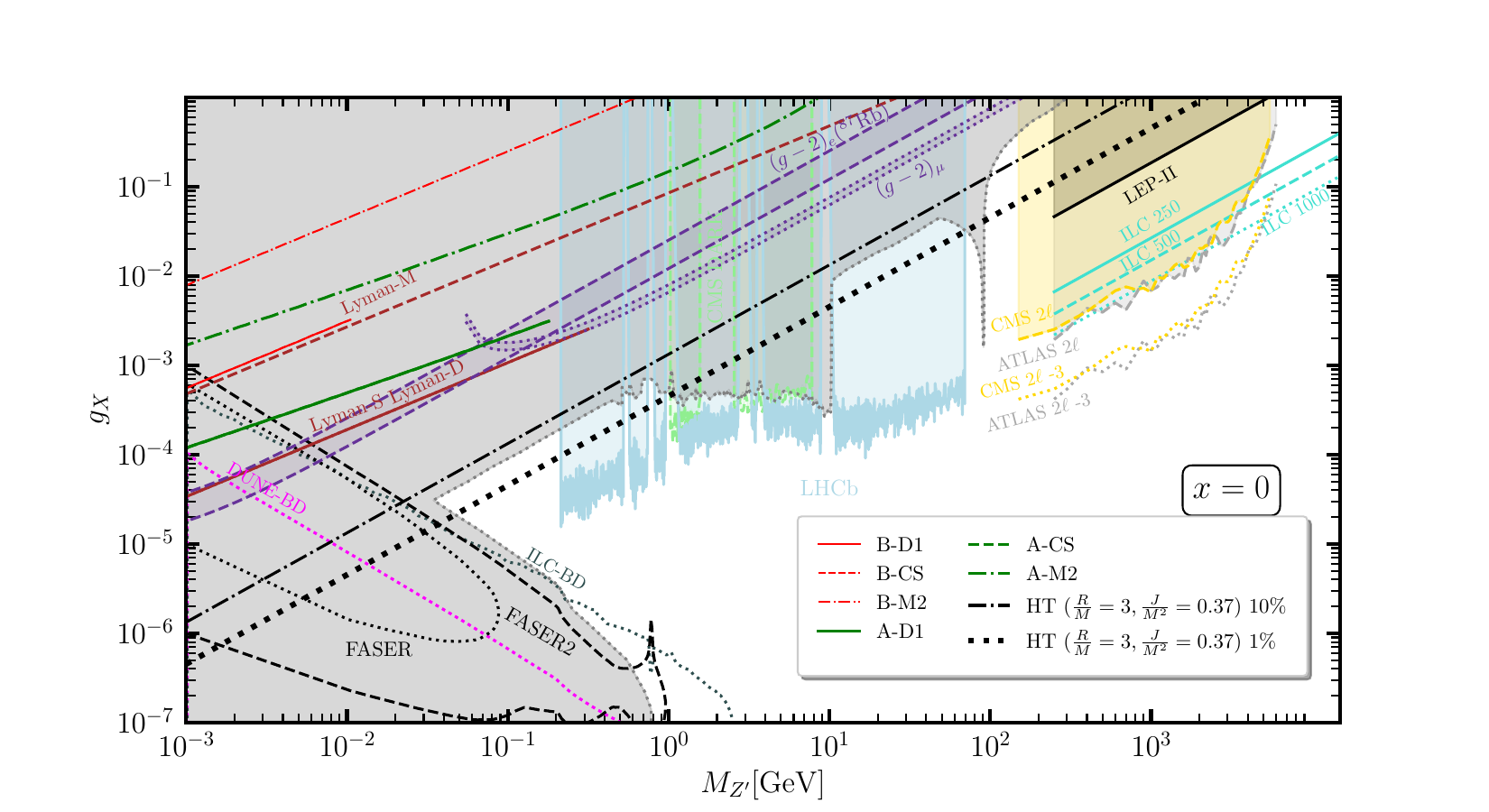}
\caption{Limits on $g_X-M_{Z^\prime}$ plane for the $U(1)_X$ and $U(1)_{q+xu}$ extensions with $x_H=1$ (upper) and $x=0$ (lower).}
\label{fig:lim2}
\end{figure}

\begin{figure}
\hspace*{-1.5cm}
\includegraphics[scale=0.56]{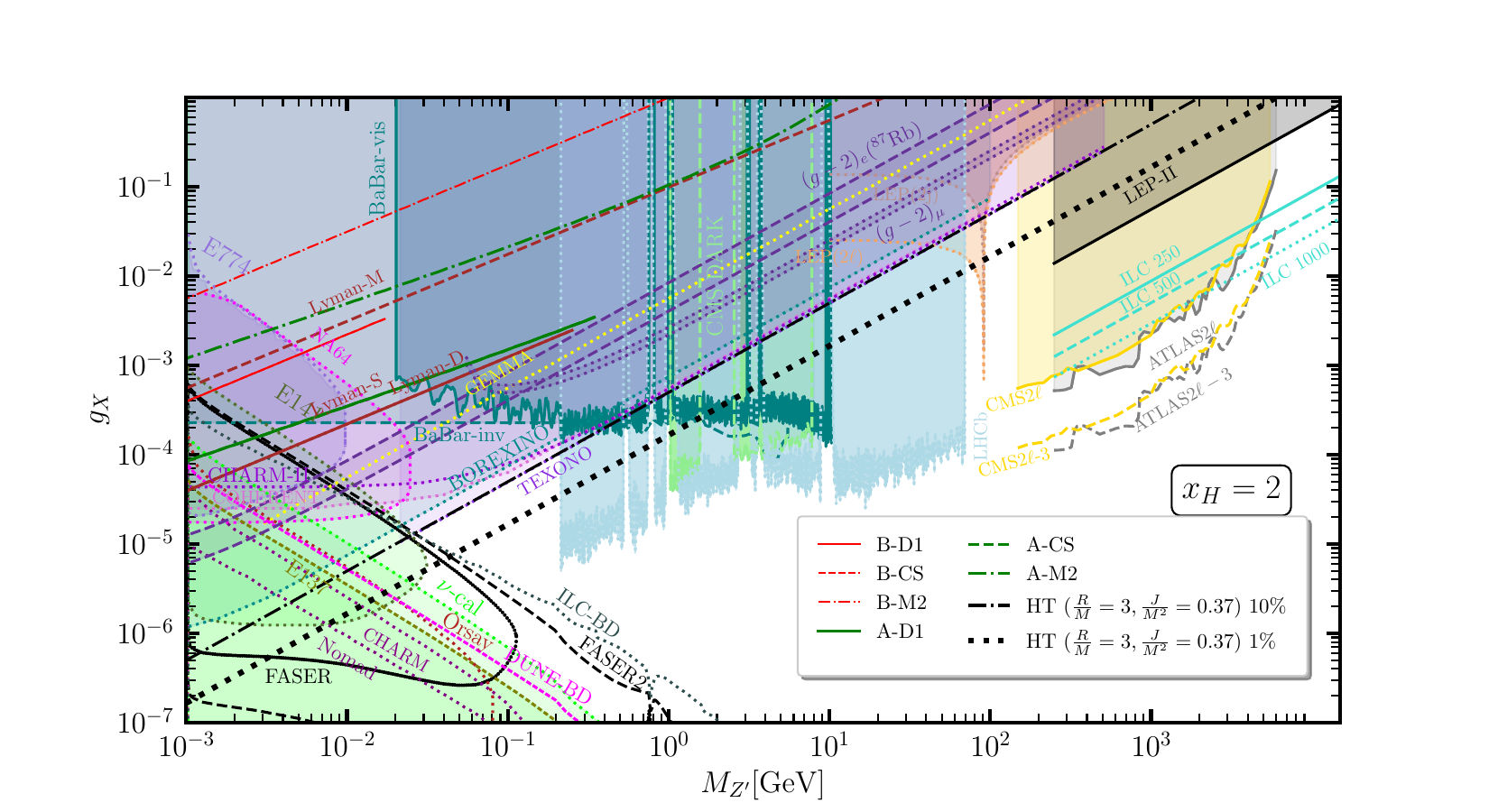}
\hspace*{-1.5cm}
\includegraphics[scale=0.56]{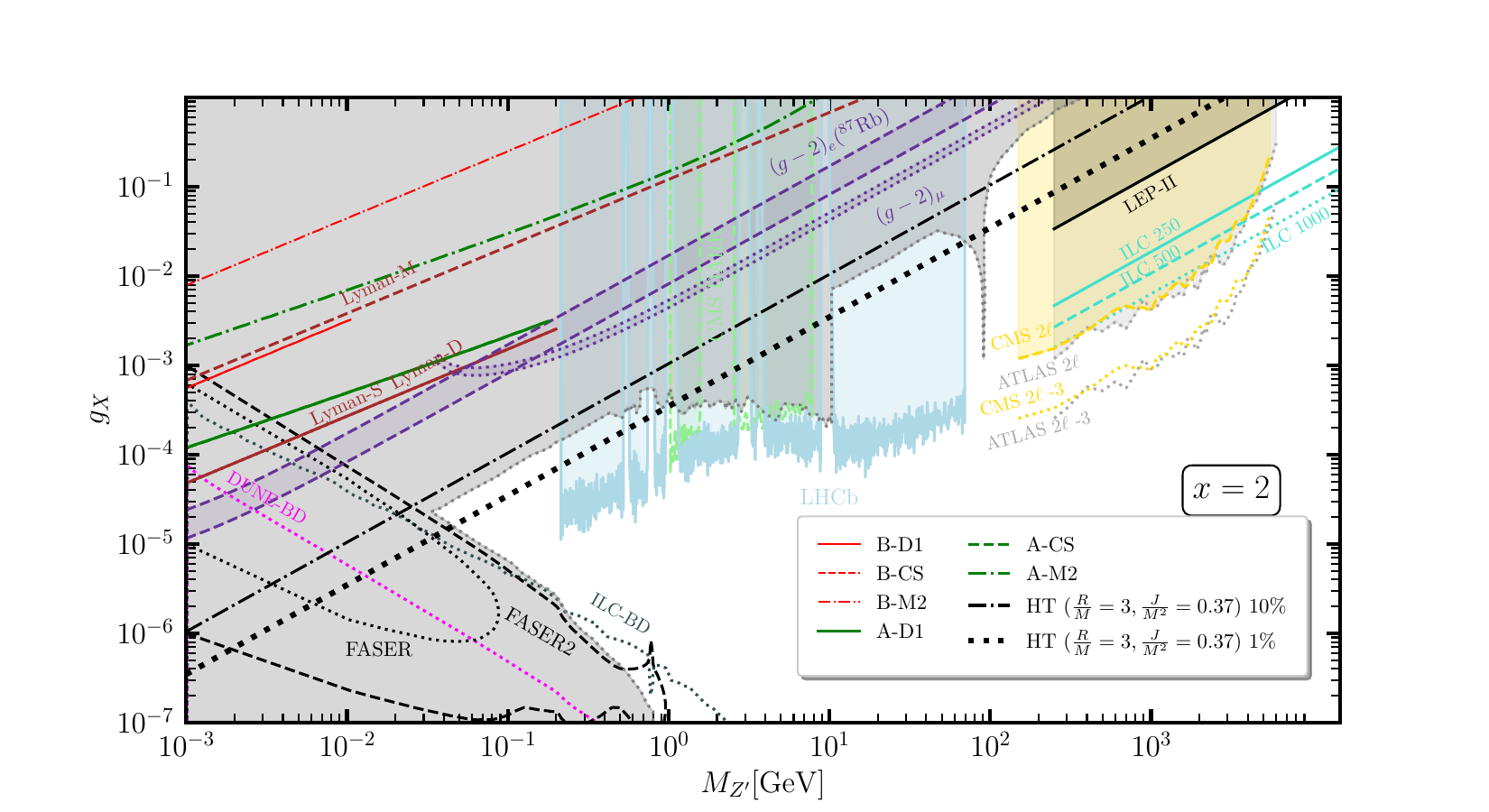}
\caption{Limits on $g_X-M_{Z^\prime}$ plane for the $U(1)_X$ and $U(1)_{q+xu}$ extensions with $x_H=2$ (upper) and $x=2$ (lower).}
\label{fig:lim3}
\end{figure}

\begin{figure}
\hspace*{-1.5cm}
\includegraphics[scale=0.56]{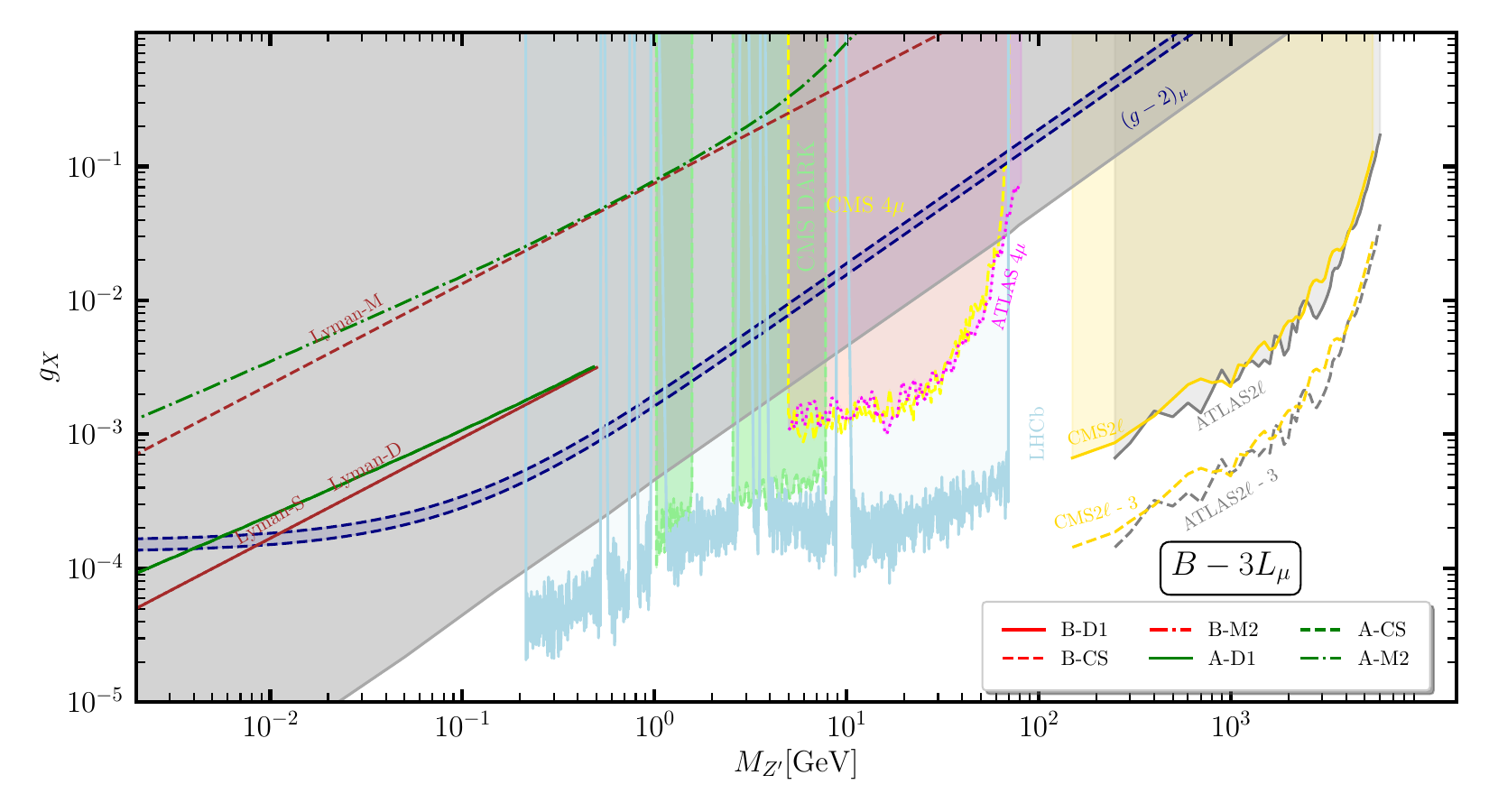}
\hspace*{-1.5cm}
\includegraphics[scale=0.56]{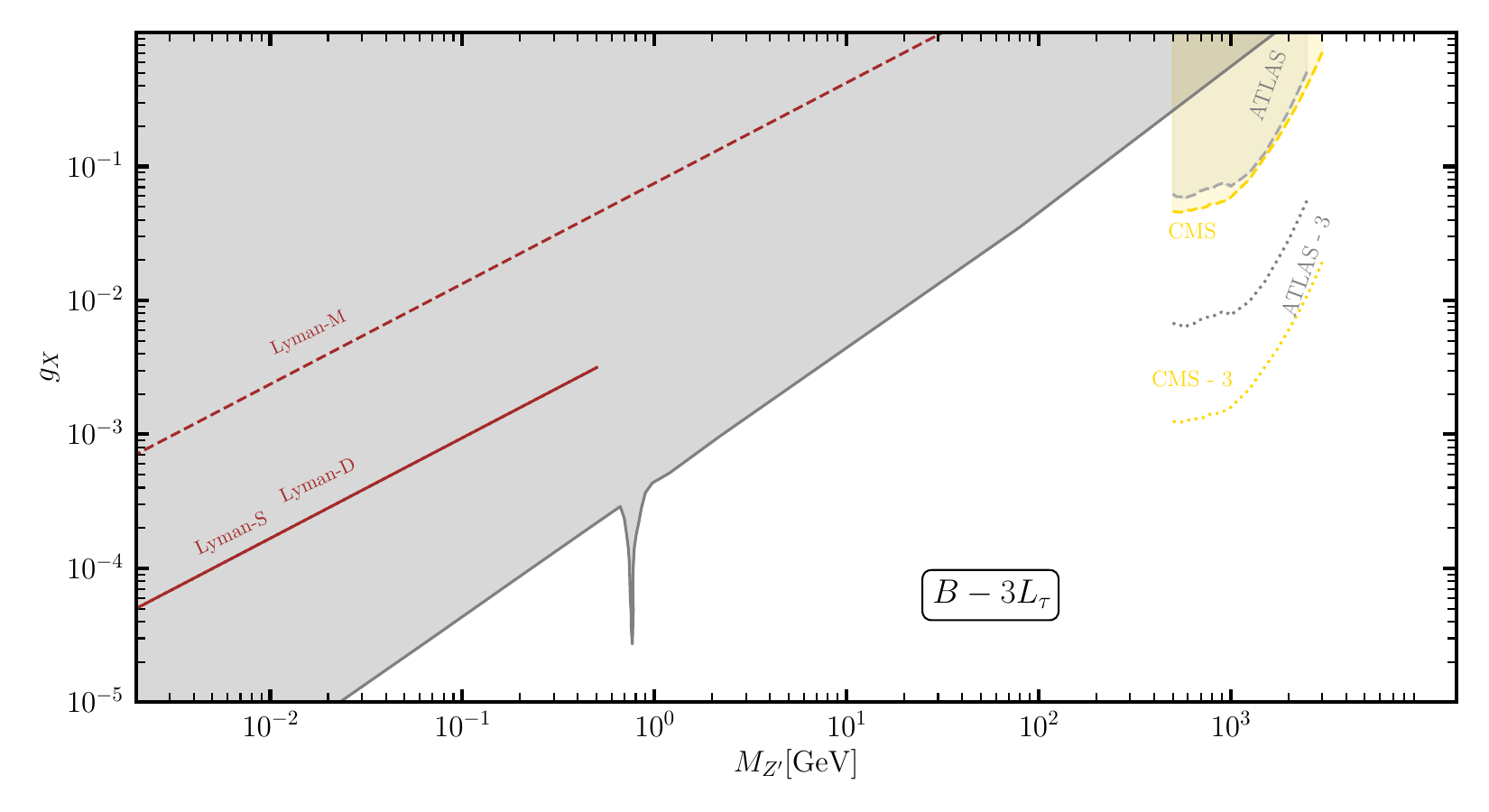}
\caption{Limits on $g_X-M_{Z^\prime}$ plane for $B-3L_\mu$ (upper) and $B-3L_\tau$ (lower) scenarios.}
\label{fig:lim7}
\end{figure}

\noindent
{\bf (i) LHC:} We estimate bounds on $g_X-M_{Z^\prime}$ plane from the results of CMS~\cite{CMS:2023slr} and LHCb~\cite{LHCb:2019vmc} experiments that search for dark photon $(A^\prime)$ decaying into $\mu^+ \mu^-$ which provide bounds on  dark photon mass $(m_{A^\prime})$ and kinetic mixing $(\epsilon)$. We estimate upper limit on $g_X$ as a function of $M_{Z^\prime}$ in use of the following rescaling 
 \begin{equation}
 g_X^{\rm max}(M_{Z^\prime} = m_{A^\prime}) = \epsilon^{\rm max}(m_{A^\prime}) e \sqrt{\frac{\sigma(pp \to A^\prime) BR(A^\prime \to \mu^+ \mu^-)}{\sigma(pp \to Z^\prime)BR(Z^\prime \to \mu^+ \mu^-)} },
 \end{equation}
 where $\sigma(pp \to A^\prime(Z^\prime) )$ is $A^\prime(Z^\prime)$ production cross section and $\epsilon^{\rm max} (m_{A^\prime})$ is the experimental upper limit depending on $m_{A^\prime}$. We estimate $A^\prime(Z^\prime)$ production cross sections using CalcHEP3.5 implementing relevant interactions to calculate limits on $g_X$ which are given in Fig:~\ref{fig:lim0}-\ref{fig:lim3} for $U(1)_X$ and $U(1)_{q+xu}$ and in Fig.~\ref{fig:lim7} for $B-3L_\mu$ scenario.   
 
\begin{figure}
\hspace*{-1.5cm}
\includegraphics[scale=0.56]{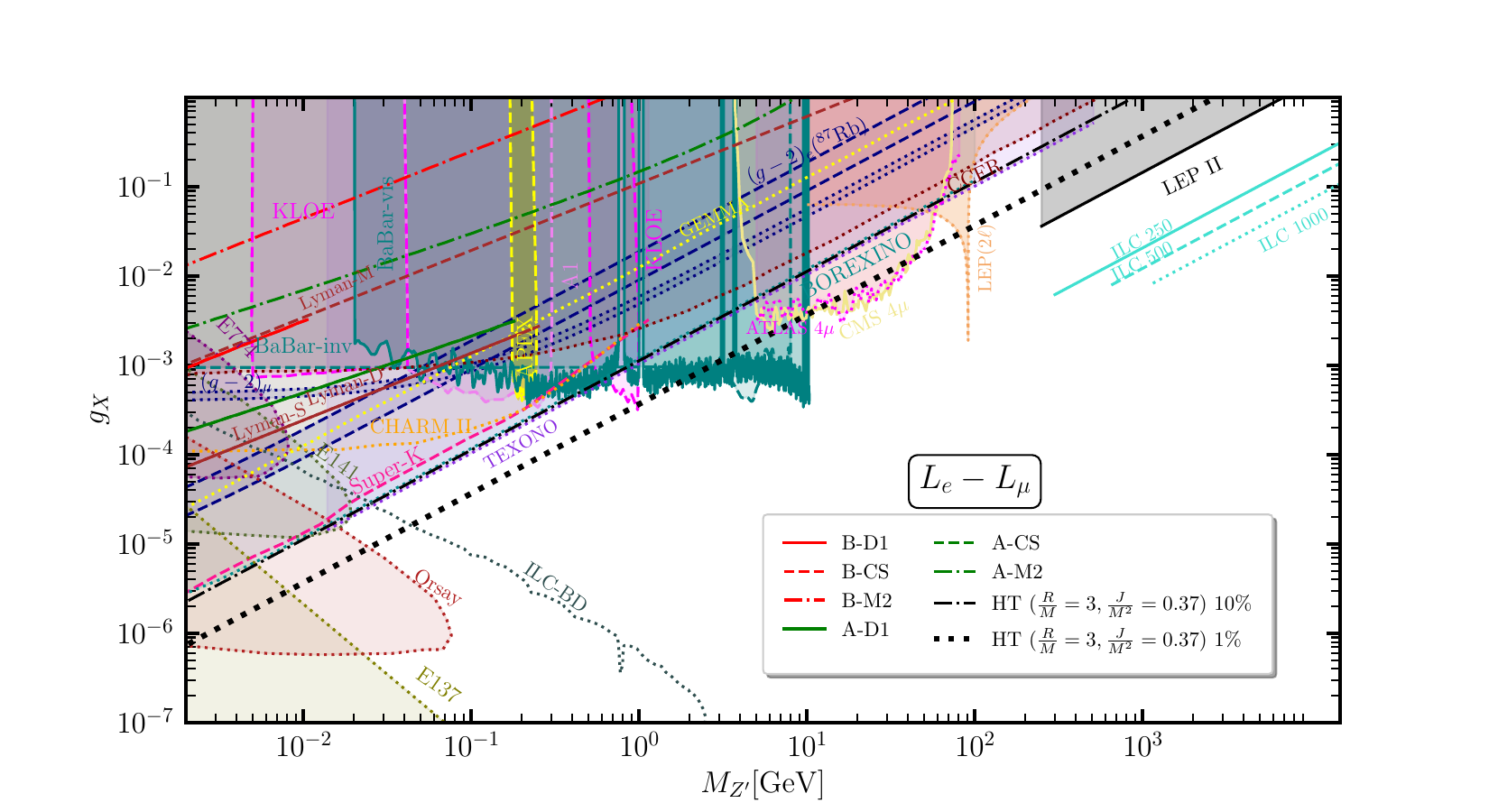}
\hspace*{-1.5cm}
\includegraphics[scale=0.56]{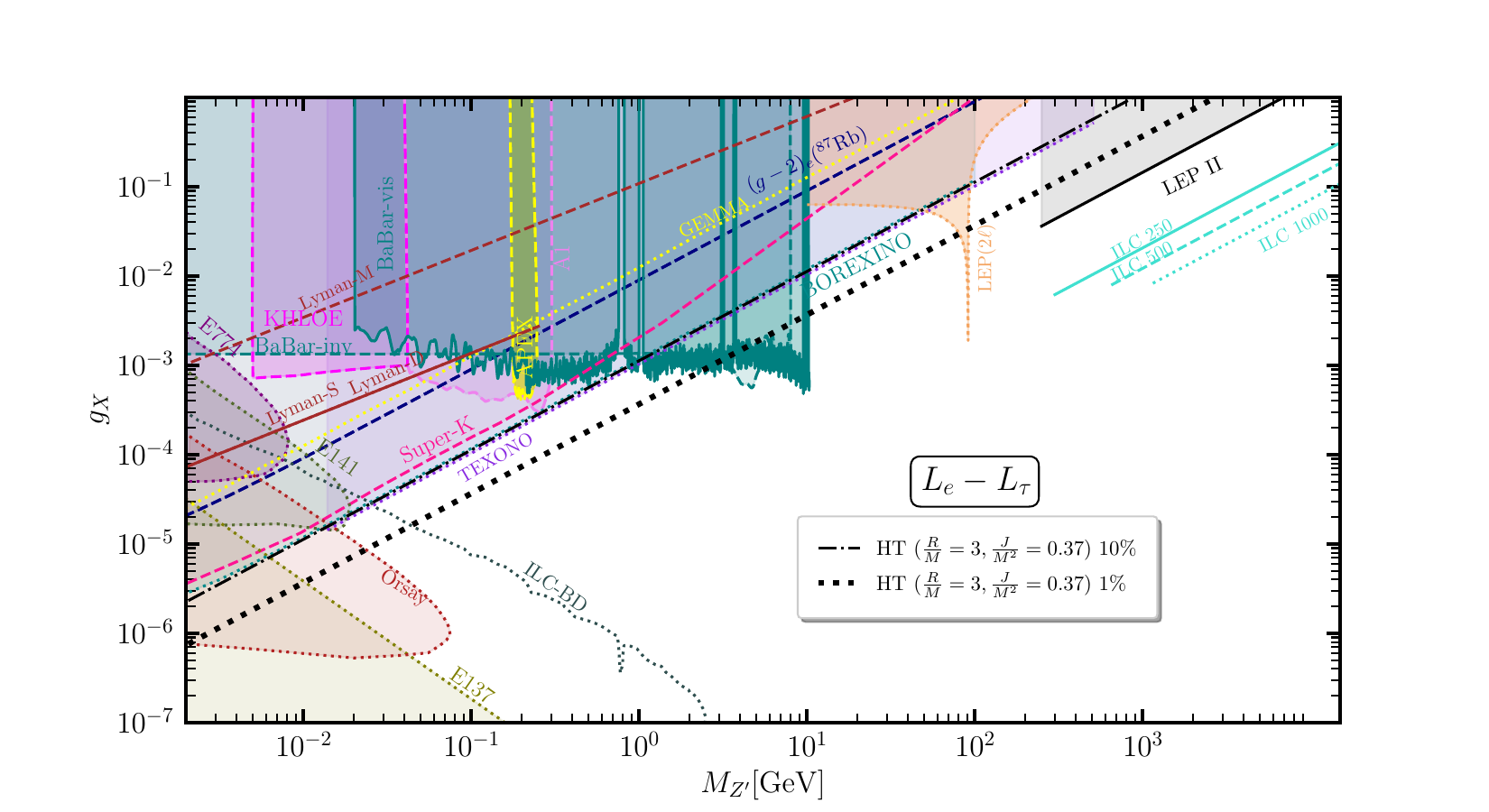}
\caption{Limits on $g_X-M_{Z^\prime}$ plane for the $L_e-L_\mu$ (upper) and $L_e-L_\tau$ (lower) scenarios.}
\label{fig:lim4}
\end{figure}

\begin{figure}
\hspace*{-1.5cm}
\includegraphics[scale=0.56]{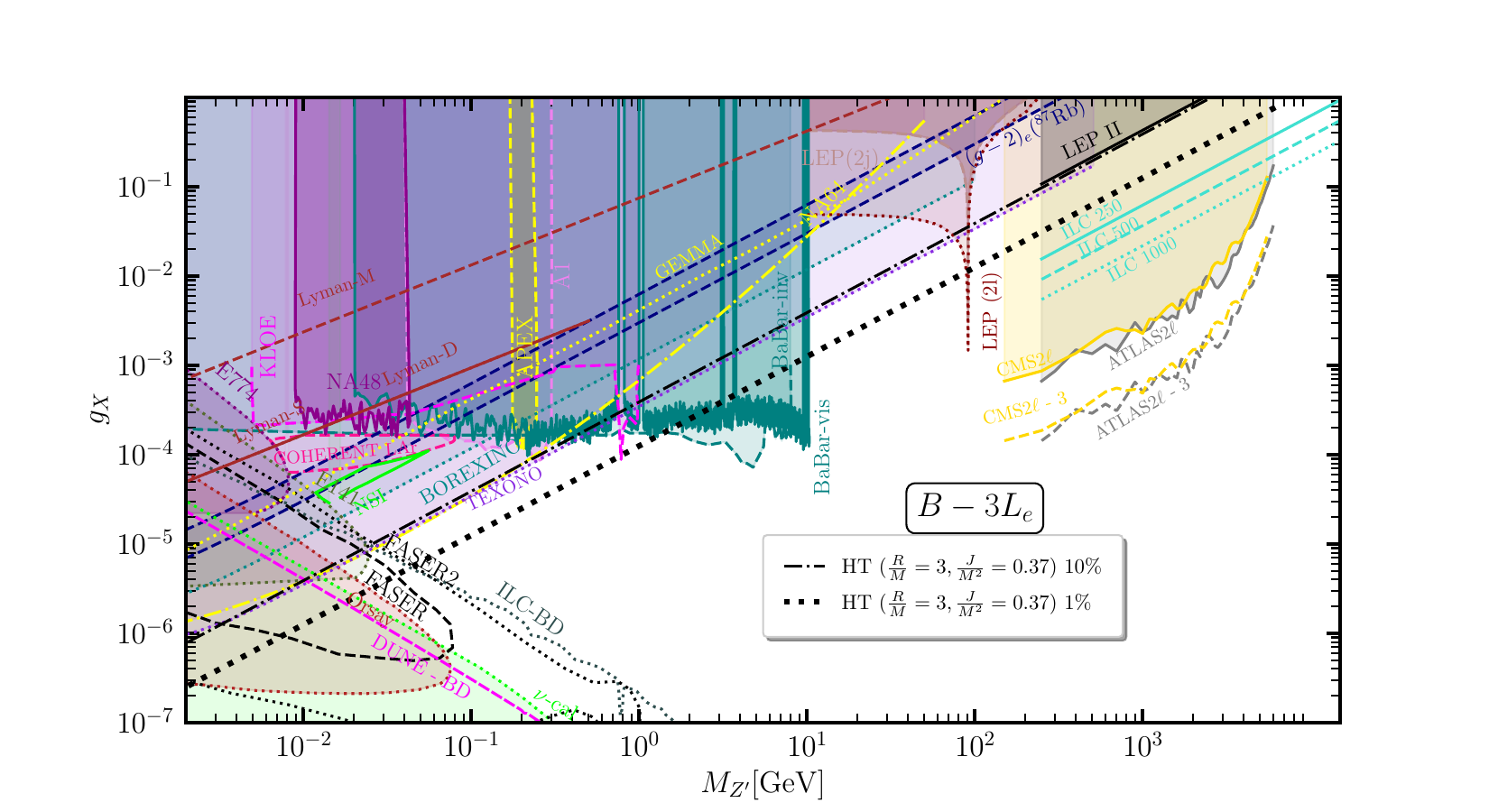}
\caption{Limits on $g_X-M_{Z^\prime}$ plane for $B-3L_e$ scenario.}
\label{fig:lim6}
\end{figure}

\noindent
{\bf (ii) BaBar:} At the BaBar experiment, $e^+ e^-$ collision produces $A^\prime$ via $e^+ e^- \to A^\prime \gamma$ process followed by $A^\prime \to  \{e^+e^-, \mu^+ \mu^-, \text{light mesons} \}$(visible) and $A^\prime \to $neutrinos (invisible) decay modes. Rescaling the upper limits from the B$-$L case \cite{BaBar:2014zli,BaBar:2017tiz} as a function of $M_{Z^\prime}$ we obtain the limits on $g_X$ in our models using visible mode as 
 \begin{equation}
  g_X^{\rm max}(M_{Z^\prime} ) = g_{B-L}^{\rm max}(M_{Z^\prime})  \sqrt{\frac{\sigma(e^+e^- \to \gamma Z^\prime_{B-L}) BR(Z^\prime_{B-L} \to \text{visible states})}{\sigma(e^+ e^- \to \gamma Z^\prime)BR(Z^\prime \to \text{visible states})} },
 \end{equation}
 where $Z^\prime_{B-L}$ is B$-$L gauge boson and that for invisible mode is 
 \begin{equation}
  g_X^{\rm max}(M_{Z^\prime} ) = g_{B-L}^{\rm max}(M_{Z^\prime})  \sqrt{\frac{\sigma(e^+e^- \to \gamma Z^\prime_{B-L}) BR(Z^\prime_{B-L} \to \bar \nu \nu)}{\sigma(e^+ e^- \to \gamma Z^\prime)BR(Z^\prime \to \bar \nu \nu) } },
 \end{equation}
where all neutrino modes are summed up. Bounds are shown in Figs.~\ref{fig:lim0}-\ref{fig:lim3} for $U(1)_X$ scenarios. In case of $U(1)_{q+xu}$ scenarios these bounds belong to the gray shaded region. Bounds for $L_{e}-L_{\mu, \tau}$ scenarios are shown in Fig.~\ref{fig:lim4} and those for B$-3L_e$ are shown in Fig.~\ref{fig:lim6}.
\begin{figure}
\hspace*{-1.5cm}
\includegraphics[scale=0.48]{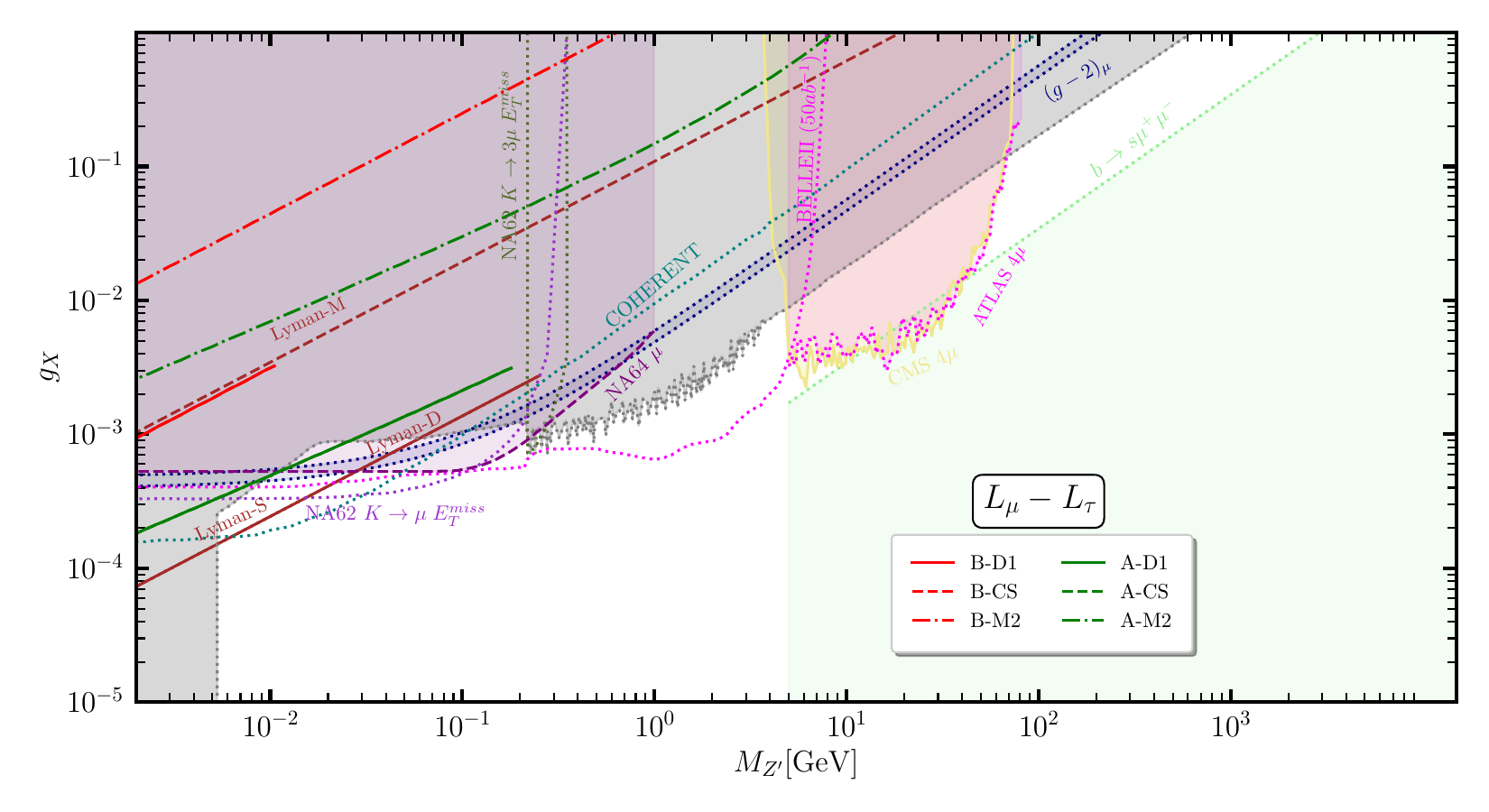}
\caption{Limits on $g_X-M_{Z^\prime}$ plane for the $L_\mu-L_\tau$ scenario.}
\label{fig:lim5}
\end{figure}
\subsection{Limits from electron-(anti)neutrino scattering in neutrino experiments}
The electron-(anti)neutrino scattering cross sections under the existence of $Z^\prime$ interactions are estimated to obtain bounds on $g_X-M_{Z^\prime}$ plane . The differential cross section can be written as 
\begin{equation}
\label{eq:nu-e-scattering}
\frac{d \sigma (\nu  e)}{d T}  =
\left. \frac{d \sigma (\nu  e)}{d T} \right|_{\rm SM} + 
\left. \frac{d \sigma (\nu  e)}{d T} \right|_{Z^\prime} + 
\left. \frac{d \sigma (\nu  e)}{d T} \right|_{\rm Int} 
\end{equation}
where $T$ is the $e^-$ recoil energy and the first, second and third terms from Eq.~(\ref{eq:nu-e-scattering}) denote the contributions from the SM, $Z^\prime$ and interference between the SM and $Z^\prime$ processes whereas the purely SM contribution is
\begin{align}
\label{eq:nu-e-scattering-SM}
\left. \frac{d \sigma (\nu  e)}{d T} \right|_{\rm SM} & = \frac{2 G_F^2 m_e}{\pi E_\nu^2} \left( a^2_1 E^2_\nu + a_2^2 (E_\nu - T)^2 - a_1 a_2 m_e T \right), 
\end{align}
where $E_\nu$ is the initial neutrino energy. Here the coefficients $a_1$ and $a_2$ are written as
\bea
a_1 &=& \left\{\sin^2 \theta_W + \frac12, \ \sin^2 \theta_W, \ \sin^2 \theta_W - \frac12, \ \sin^2 \theta_W \right\} \ {\rm for} \ \{\nu_e e, \bar \nu_e e, \nu_\beta e, \bar \nu_\beta e\}, \nonumber \\
a_2 &=& \left\{\sin^2 \theta_W, \ \sin^2 \theta_W +  \frac12, \ \sin^2 \theta_W, \ \sin^2 \theta_W -\frac12 \right\} \ {\rm for} \ \{\nu_e e, \bar \nu_e e, \nu_\beta e, \bar \nu_\beta e\},
\eea
where $\beta = \{\mu, \tau \}$. The contribution from $Z^\prime$ exchanging process is 
\begin{align}
\left. \frac{d \sigma (\overset{(-)}{\nu}_\alpha e )}{d T} \right|_{Z^\prime} & = \frac{g_X^4 (x_\ell)^2 m_e}{4 \pi E_\nu^2 (2m_e T + M^2_{Z^\prime})} [(2 E^2_\nu - 2 E_\nu T + T^2)(b^2_1 +b^2_2) \pm 2 b_1 b_2(2 E_\nu - T)T - m_e T(b^2_1 - b_2^2)],
\end{align}
where $b_1 = \frac{x_\ell + x_e}{2}$ and $b_2 = \frac{x_\ell - x_e}{2}$ with $x_{\ell, e}$ are the corresponding general $U(1)$ charges of the left and right handed leptons where negative sign stands from $\overline{\nu}$ processes. The contribution from interference between the SM and $Z^\prime$ are also obtained, depending on the processes, such that
\bea
\left. \frac{d \sigma (\nu_e e)}{d T} \right|_{\rm int}  &= & \frac{G_F g_X^2 x_\ell m_e}{\sqrt{2} \pi E^2_\nu (2m_e T +M^2_{Z^\prime}) } [2 E_\nu^2(b_1 + b_2) + (2 E^2_\nu - 2 E_\nu T + T^2)(b_1 c_1 + b_2 c_2)] \nonumber \\
&& + T (2E_\nu - T)(b_1 c_2 + b_2 c_1) - m_e T(b_1 -b_2 + b_1 c_1 - b_2 c_2)], \nonumber \\
\left. \frac{d \sigma (\bar{\nu}_e e)}{d T} \right|_{\rm int} & = & \frac{G_F g_X^2 x_\ell m_e}{\sqrt{2} \pi E^2_\nu (2m_e T +M^2_{Z^\prime}) } [2 (E_\nu - T)^2(b_1 + b_2) + (2 E^2_\nu - 2 E_\nu T + T^2)(b_1 c_1 + b_2 c_2)] \nonumber \\
&& - T (2E_\nu - T)(b_1 c_2 + b_2 c_1) - m_e T(b_1 -b_2 + b_1 c_1 - b_2 c_2)], \nonumber \\
\left. \frac{d \sigma (\overset{(-)}{\nu}_\beta e )}{d T} \right|_{\rm int} &= & \frac{G_F g_X^2 x_\ell m_e}{\sqrt{2} \pi E^2_\nu (2m_e T +M^2_{Z^\prime}) } [(2E_\nu^2 -2 E_\nu T - T^2)2(b_1c_1 + b_2 c_1) \pm T(2 E_\nu - T)(b_1 c_2 + b_2 c_1)] \nonumber \\
&& - m_e T(b_1 c_1 - b_2 c_2)],  
\eea
where $c_1 = -1/2 + 2 \sin^2 \theta_W$ and $c_2 = -1/2$. Then we calculate differential cross sections and estimate upper limit on $g_X-M_{Z^\prime}$ plane for:

\noindent
{\bf (i) BOREXINO}: The $\nu_e$-$e$ scattering  is measured by the experiment with $\langle E_\nu \rangle = 862$ keV and $T \simeq [270, 665]$ keV for $^7B_e$ solar neutrino. To obtain the constrains, we require the total cross section should not be more than $8\%$ above that of the SM prediction \cite{Bellini:2011rx}. 

\noindent
{\bf (ii) TEXONO}: The experiment measures $\bar \nu_e$-$e$ scatterings using 187 kg of CsI(Tl) scintillating crystal array with 29882/7369 kg-day of reactor ON/OFF data with electron recoil energy $T \simeq [3, 8]$ MeV. 
We estimate the $\chi^2$ value as 
\begin{equation}
\chi^2 = \sum_{\rm bin} \frac{(R_{\rm data} - R_{\rm th})^2}{\Delta R^2},
\label{chi2-2}
\end{equation}
where $R_{\rm data}$ and $R_{\rm th}$ are the event ratios observed by the experiment and predicted by the cross-section from Eq.~(\ref{eq:nu-e-scattering}), $\Delta R$ is the experimental uncertainty, for each recoil energy bin taken from \cite{TEXONO:2009knm}, and anti-neutrino flux from the same reference is applied to estimate upper limit at $90 \%$ C.L from the $\chi^2$ fit. 

\noindent
{\bf (iii) GEMMA}: The experiment measures $\bar{\nu}_e$-$e$ scattering with 1.5 kg HPGe detector where energy of neutrino is $\langle E_\nu \rangle \sim 1$-$2$ MeV and flux is $2.7 \times 10^{13}$ cm$^{-2}$s$^{-1}$. 
Using Eq.~(\ref{chi2-2}) we estimated the  $\chi^2$ value for the data from \cite{Beda:2010hk} with 13000 ON-hours and 3000 OFF-hours to obtain upper limits with $90 \%$ C.L.

\noindent
{\bf (iv) CHARM-II}: The experiment measures $\nu_\mu (\bar \nu_\mu)$-electron scattering where $2677\pm82$ and $2752\pm88$ events are obtained for $\nu_\mu$ and $\bar \nu_\mu$ cases.
The mean neutrino energy is $\langle E_{\nu_\mu} \rangle = 23.7$ GeV and $\langle E_{\bar \nu_\mu} \rangle = 19.1$ GeV, and the range of measured recoil energy is 3-24 GeV.
We estimated the  $\chi^2$ value 
for the data from \cite{CHARM-II:1994dzw} to obtain upper limits with $90 \%$ C.L.

The limits on $g_X-M_{Z^\prime}$ plane are given in Fig:~\ref{fig:lim0}-\ref{fig:lim3} for general $U(1)_X$. In these figures we show the $U(1)_{q+xu}$ scenario in the lower panel where these existing bounds belong to gray shaded region. Bounds for $L_{e}-L_{\mu, \tau}$ scenarios are shown in Fig.~\ref{fig:lim4} whereas those for $L_\mu-L_\tau$ scenario belong to gray shaded region in Fig.~\ref{fig:lim5}. Limits for the $B-3L_e$ scenario are given in Fig.~\ref{fig:lim6} whereas those for $B-3L_\mu$ and $B-3L_\tau$ scenarios belong to gray shaded area. 
\subsection{Limits from coherent neutrino-nucleus scattering}
We coherent elastic $\nu$-nucleus scattering (CE$\nu$NS) measured by COHERENT experiment with CsI and Ar targets~\cite{COHERENT:2020ybo} to estimate upper limits on $g_X-M_{Z^\prime}$ plane by
rescaling the constrains from B$-$L case \cite{Melas:2023olz} by comparing number of events in B$-$L and other cases. The number of events at COHERENT experiment is estimated following the differential 
cross section for CE$\nu$NS process from \cite{Patton:2012jr} 
\begin{equation}
\frac{d \sigma_{\nu-N}}{d T} (E,T) = \frac{G_F^2 M}{\pi} \left(1 - \frac{M T}{2 E^2} \right) Q_{{\rm SM}+Z^\prime}^2,
\end{equation} 
where $T$ is the recoil energy, $E$ is the neutrino energy,  $M$ is the mass of target nucleus and $Q_{{\rm SM}+Z^\prime}$ is the factor coming from SM+$Z^\prime$ interactions.
In the models $Q_{{\rm SM}+Z^\prime}$ is written as 
\begin{equation}
Q_{{\rm SM}+Z^\prime} = \left( g^p_V(\nu_\ell) + 2 \epsilon^{u V}_{\ell \ell} + \epsilon^{d V}_{\ell \ell} \right) Z F_Z(|{\bf q}^2|) + \left( g^n_V(\nu_\ell) +  \epsilon^{u V}_{\ell \ell} + 2\epsilon^{d V}_{\ell \ell} \right) N F_N(|{\bf q}^2|),
\end{equation}
where $Z(N)$ is the number of proton(neutron) in the target nucleus, $g^{p(n)}_V$ is the neutrino-proton(neutron) coupling in the SM and $F_{Z(N)}(|{\bf q}^2|)$ is the from factor of the proton(neutron) for the target nucleus.
The effective coupling $\epsilon^{q V}_{\ell \ell}$ is explicitly given by 
\begin{equation}
\epsilon^{q V}_{\ell \ell} = \frac{g_X^2 x_\ell x_q}{\sqrt{2} G_F ({\bf q}^2 + m^2_{Z^\prime})}.
\end{equation}
For the $\nu$-proton(neutron) coupling, we adopt the values of $g_V^p (\nu_e) = 0.0401$, $g_V^p=0.0318$ and $g_V^n = - 0.5094$ for the SM~\cite{Cadeddu:2020lky,Erler:2013xha}. 
For the form factors $F_{Z(N)}(|{\bf q}^2|)$, we apply Helm parametrization~\cite{Helm:1956zz} using proton rms radii $\{R_{p} ({\rm Cs}), R_p ({\rm I}), R_p ({\rm Ar}) \}  = \{4.804, 4.749, 3.448\}$ [fm] and 
neutron rms radii $\{R_{n} ({\rm Cs}), R_n ({\rm I}), R_n ({\rm Ar}) \}  = \{5.01, 4.94, 3.55\}$ [fm]~\cite{Fricke:1995zz,Angeli:2013epw,Bender:1999yt}. For the CE$\nu$NS event rate in the COHERENT experiment, we use the neutrino fluxes that depend on the neutrino fluxes produced from the Spallation Neutron Source (SNS) at the Oak Ridge National Laboratories. They are written as
\bea
\frac{d N_{\nu_\mu}}{d E} &=& \eta \delta \left( E - \frac{m_\pi^2 - m_\mu^2}{2 m_\pi} \right), \nonumber \\
\frac{d N_{\nu_{\bar \mu}}}{d E} &=& \eta \frac{64 E^2}{m_\mu^3}   \left( \frac34 - \frac{E}{m_\mu} \right), \nonumber \\
\frac{d N_{\nu_e}}{d E} &=& \eta \frac{192 E^2}{m^3_\mu} \left( \frac12 - \frac{E}{m_\mu} \right), 
\eea
where $\eta = r N_{\rm POT}/(4 \pi L^2)$ with $r$, $N_{\rm POT}$ and $L$ being respectively the number of neutrinos per flavor that are produced for each proton-on-target (POT), 
the number of POT, and the distance between the source and the detector.
For these values, we adopt $r = 9 \times 10^{-2}$, $N_{\rm POT} = 13.7 \times 10^{22}$ and $L = 27.5$ m for Ar and 
$r = 0.08$, $N_{\rm POT} = 17.6 \times 10^{22}$ and $L = 19.5$ m for CsI detectors. The theoretical number of events for each energy bin in the COHERENT experiment is estimated by
\begin{equation}
N_i = N (\mathcal{N}) \int^{T_{i+1}}_{T_i} d T A(T) \int_{E_{\rm min}}^{E_{\rm max}} d E \sum_{\nu = \nu_e, \nu_{\mu}, \nu_{\bar \mu}} \frac{d N_\nu}{d E} \frac{d \sigma_{\nu-N}}{d T} (E, T),
\end{equation}
where $i$ distinguishes recoil energy bin,  $E_{\rm min(max)} = \sqrt{M T/2} (m_\mu/2)$ and $A(T)$ is the energy-dependent reconstruction efficiency. We estimate the upper bound of coupling $g_X$ for each mass in our models by rescaling that of B$-$L case from \cite{Cadeddu:2020nbr} comparing the number of events for the upper bound on B$-$L with those in each model to find upper bounds on $g_X$. Corresponding limits are shown in the upper panels of Fig.~\ref{fig:lim0}-\ref{fig:lim3} for $U(1)_X$ case whereas these limits for $U(1)_{q+xu}$ belong to gray shaded region as shown in the lower panels of respective figures. Similar bounds for $B-3L_e$ are shown in Fig.~\ref{fig:lim6}.  
\subsection{Constraints from beam-dump experiments}
\noindent
{\bf (i) Proton beam-dump:} For proton beam-dump experiments (LSND, PS191, NOMAD and CHARM), we estimate bound curves in our models by rescaling those of B$-$L case \cite{Bauer:2018onh,Ilten:2018crw}. 
Approximately, the curves for the upper bound on $g_X-M_{Z^\prime}$ plane are derived applying the scaling formula using $\tau_{Z^\prime}^{}(g_{B-L}^{\rm max}) \sim \tau_{Z^\prime}^{}(g_X^{\rm max}, x_H^{}, x_\Phi^{})$ 
where $g_{B-L}^{}$ indicates the gauge coupling in the B$-$L model and $\tau_{Z^\prime}^{}$ is the lifetime of $Z^\prime$. The curves for the lower bound are scaled using 
\begin{equation}
\label{eq:lowerP}
g_X^{\rm low} \sim g_{B-L}^{\rm low} \sqrt{ \frac{{\rm BR}(\pi^0(\eta) \to Z^\prime_{B-L} \gamma)~ {\rm BR}(Z^\prime_{B-L} \to e^+ e^-) \tilde \tau_{Z^\prime}^{} }{{\rm BR}(\pi^0(\eta) \to Z^\prime \gamma)~ {\rm BR}(Z^\prime \to e^+ e^-) \tilde \tau_{Z^\prime_{B-L} }} }~,
\end{equation}
where $\tilde \tau$ is lifetime with gauge coupling taken to be unity and $Z^\prime$ is produced from $\pi^0(\eta)$ decay for LSND, PS191 and NOMAD (CHARM) following meson decay branching ratio from \cite{Ilten:2018crw}. For proton beam dump experiment $\nu$-cal, the $Z^\prime$ is dominantly produced via bremsstrahlung process studied in \cite{Asai:2022zxw} from which we collect the constraints. Following this line we estimate prospective bounds for the proton beam-dump experiments in DUNE (DUNE-BD), FASER and FASER2 for $U(1)_{q+xu}$ case utilizing the B$-$L as a class of $U(1)_X$ scenarios from \cite{Asai:2022zxw}. 

\noindent
{\bf (ii) Electron/ Positron beam-dump:} For $e^-$ beam-dump experiments (NA64, E774, Orsay and KEK), the limit curves are derived by rescaling the bounds of B$-$L in the line of proton beam dump case. We approximately estimate the constraints for the upper region on $g_X-M_{Z^\prime}$ plane by rescaling $\tau_{Z^\prime}^{}(g_{B-L}^{\rm max}) \sim \tau_{Z^\prime}^{}(g_X^{\rm max}, x_H^{}, x_\Phi^{})$. The constraint for the lower region is calculated by
\begin{equation}
\label{eq:bremsstrahlung}
g_X^{\rm low} \sim g_{B-L}^{\rm low} \sqrt{ \frac{2 {\rm BR}(Z^\prime_{B-L} \to e^+ e^-) \tilde \tau_{Z^\prime}^{} }{(x_\ell^2 + x_e^2) {\rm BR}(Z^\prime \to e^+ e^-) \tilde \tau_{Z^\prime_{B-L}}^{}} }~,
\end{equation}
where $Z^\prime$ is produced from bremsstrahlung process. For $e^-$ beam dump experiments E137 and E141, we adopt the results from \cite{Asai:2022zxw} to estimate limits. Using the similar method we estimate constraints for the $e^-$ beam-dump scenario in ILC (ILC-BD). 

Existing and prospective limits in the proton and electron beam-dump experiments are given in Fig:~\ref{fig:lim0}-\ref{fig:lim3} for $U(1)_X$ cases. We show the prospective bounds (DUNE-BD, FASER, FASER2 and ILC-BD) for the $U(1)_{q+xu}$ scenario in the lower panel of these figures, however, existing bounds belong to gray shaded region. Bounds obtained from $e^\pm$ beam-dump experiments will appear in $L_{e}-L_{\mu, \tau}$ scenarios which are shown in Fig.~\ref{fig:lim4}.  Similarly we estimate limits for $B-3L_e$ scenario which are shown in Fig.~\ref{fig:lim6} whereas those for $B-3L_\mu$ and $B-3L_\tau$ scenarios belong to gray shaded area.  
\subsection{Limits from $g-2$ of muon and electron}
Due to the fact of leptonic interactions with the $Z^\prime$ following $U(1)$ gauge extension of the SM we find the magnetic moment of muon (electron) using one loop $Z^\prime$ mediated processes as 
\bea
\Delta {a_\ell}= \frac{g_X^{2}}{32\pi^2} \int_{0}^{1}dy \frac{\Big\{2y(1-y)(y-4)-4y^3 \frac{m_\ell^2}{M_{Z^\prime}^2}\Big\} x_a^2+ 2 (1-y) y^2 x_b^2}{(1-y) (1-y \frac{m_\ell^2}{M_{{Z^\prime}^2}})+ y \frac{m_\ell^2}{M_{Z^\prime}^2}} 
\eea
 where $m_\ell$ is muon/ electron mass, $x_a= -\frac{3}{2} x_H- 2 x_\Phi$ and $x_b=-\frac{1}{2} x_H$ for the general $U(1)_X$ scenario. For $U(1)_{q+xu}$ scenario, we have $x_a= -\frac{5+x}{3}$ and $x_b= -\frac{x-1}{3}$ respectively. Similarly, we can calculate these coefficients for the flavored cases using the corresponding charges of the left and right-handed charged leptons. The current experimental ranges of muon $(g-2)$ is given as $\Delta a_\mu = (24.9\pm 4.98)\times 10^{-10}$
 \cite{Muong-2:2023cdq}, where the deviation from the world average of SM prediction is 5.0 $\sigma$ level. The current experimental ranges of electron $(g-2)$ are $\Delta a_e (^{133}\text{Cs}) = -(8.8\pm 3.6)\times 10^{-13}$ \cite{Parker:2018vye} and $\Delta a_e (^{87}\text{Rb}) = (4.8 \pm 3.0)\times 10^{-13}$ \cite{Morel:2020dww} where the deviations from the SM prediction are, respectively, $2.4\sigma$ and $1.6\sigma$. Since positive $\Delta a_e$ is obtained from $Z^\prime$ interaction we only show the parameter region that satisfies the second range of $\Delta a_e$. Limits obtained from $(g-2)_{e, \mu}$ are given in upper (lower) panels of Figs.~\ref{fig:lim1}-\ref{fig:lim3} for general $U(1)_X$ $(U(1)_{q+xu})$ scenarios. We show the $(g-2)_e$ limits for $L_{e}-L_{\mu, \tau}$ scenarios in Fig.~\ref{fig:lim4} and those for $L_\mu-L_\tau$ scenario in Fig.~\ref{fig:lim5}. We show the limits for $B-3L_e$ $(B-3L_\mu)$ in upper(lower) panel of  Fig.~\ref{fig:lim6}. 
\subsection{Limits from GRB221009A}
Estimating enhancement in energy deposition rates from GRB221009A, we calculate bounds on $g_X-M_{Z^\prime}$ plane from bounds on $M_{Z^\prime}/g_X$ given in Tab.~\ref{tab:lim} where HT  case provides stronger limits than other metrics. However, from our analysis we find that bounds on $M_{Z^\prime}/g_X$ are also very close, so that they overlap with each other on $g_X-M_{Z^\prime}$ plane. We estimate constraints with $10\%$ and a prospective $1\%$ precisions where later one could be stronger. As we have considered, $v_\Phi >> v$ therefore $x_H=-1$ in $U(1)_X$ case cannot be constrained from GRB221009A. $Z^\prime$ has no direct coupling with $e^\pm$ in $L_{\mu}-L_{\tau}$ and $B-3L_{\mu, \tau}$ scenarios. Therefore $\nu \bar{\nu} \to e^- e^+$ process from GRB cannot constrain these models. In the following we discuss about the strongest bounds from HT (with $\frac{R}{M}=3,~\frac{J}{M^2}=0.37)$ scenario:  

Limits on $U(1)_X$ scenarios from GRB221009A are given in Fig.~\ref{fig:lim1} for $x_H=0$ which is the B$-$L case and in the upper panels of Figs.~\ref{fig:lim2} and \ref{fig:lim3} for other values of $x_H$ respectively. The colored regions are already ruled out by existing experimental data. We find that strongest limits are coming from the HT case for $10^{-3}$ GeV $\leq M_Z^\prime \leq 10$ TeV, the results obtained with a prospective $1\%$ precession will be about a factor $4$ times stronger than the current limits with $10\%$ precision. We find that bounds with $10\%$ precision is comparable with the existing bounds obtained from TEXONO, BOREXINO and COHERENT experiments within $10^{-3}$ GeV $\leq M_{Z^\prime} \leq 2.0$ GeV depending on the type of experiments and $U(1)$ charges, however, limits from $1\%$ precision can be stronger than these and cross the prospective FASER, FASER2 and ILC-BD curves within $0.035$ GeV $\leq M_{Z^\prime} \leq 0.055$ GeV where bounds on $g_X$ could reach up to $\mathcal{O}(10^{-5})$. In addition to that, this could provide a strong bound on $g_X$ for $60$ GeV $\leq M_{Z^\prime} \leq 80$ GeV and $92.0$ GeV $\leq M_{Z^\prime} \leq 140$ GeV whereas limits within the $U(1)$ coupling varies between $0.01 \leq g_X \leq 0.1$. 

Limits on the $U(1)_{q+xu}$ scenarios from GRB221009A using HT case are shown in the lower panels Figs.~\ref{fig:lim0} to \ref{fig:lim3} for different $x$ and for $x=1$ (B$-$L case) in Figs.~\ref{fig:lim2} respectively. In these figures we only show the strongest boundary of the existing bounds and the shaded region in gray is already ruled out by existing experiments. We find that Fig.~\ref{fig:lim0} in case of $x=-1$, bounds with $10\%$ precision is weak and already ruled out whereas those for $1\%$ precision is overlapping the boundary of the strongest existing bounds on the gauge coupling $2\times 10^{-5} \leq g_X \leq $. However bounds on $g_X$ is around $\mathcal{O}(0.01)$ for $92.0$ GeV $\leq M_{Z^\prime} \leq 140$ GeV $x=-1$. For $x=1$, the bounds are same as $x_H=0$ of $U(1)_X$ scenario and are shown in Fig.~\ref{fig:lim1} being the B$-$L case. For $x=0$ and $2$ from the lower panels of Figs.~\ref{fig:lim2} and \ref{fig:lim3} we notice that the bounds with $10\%$ precision is already ruled out, however, the prospective bounds with $1\%$ precision is around $3.5$ times stronger than the bounds obtained from $10\%$ precision. For example the prospective bounds on the coupling could reach within $10^{-5} \leq g_X \leq 10^{-4}$ for $0.05$ GeV $\leq M_{Z^\prime} \leq 0.2$ GeV whereas for $92.0$ GeV $\leq M_{Z^\prime} \leq 150$ GeV prospective bounds on $g_X$ is around $\mathcal{O}(0.01)$. GRB bounds for $U(1)_{q+xu}$ scenario for $M_{Z^\prime} < 0.05$ GeV stay in the gray shaded region which is already ruled out. Prospective bounds from beam dump searches at DUNE, FASER, FASER2 are also weak compared to the existing bounds from $\nu-$Cal for $M_{Z^\prime} \leq 0.08$ GeV. 

We estimate bounds from GRB221009A using HT case in $L_e-L_\mu$ and $L_e-L_\tau$ scenarios in Fig.~\ref{fig:lim4}. These bounds are the same because $Z^\prime$ coupling with first generation lepton is same in both cases. We find that bounds obtained at $10\%$ precision overlap with the strongest bounds for TEXONO and BOREXINO. Prospective bounds from $1\%$ precision could be around $2.86$ times stronger than those from $10\%$ precision. We find that prospective bounds on coupling could reach around $10^{-5} \leq g_X \leq 7\times 10^{-4}$ for $0.025$ GeV $\leq M_{Z^\prime} \leq 2$ GeV. For heavy $Z^\prime$, prospective bound on coupling could reach around $\mathcal{O}(0.01)$ for $40$ GeV $\leq M_{Z^\prime} \leq 250$ GeV except for $Z-$pole where dilepton limit from LEP-II is strong. For $M_{Z^\prime} > \sqrt{s}_{\rm LEP-II}$, LEP-II bounds are stronger than those obtained from GRB. However, prospective bounds form ILC could be stronger. We also find that the prospective GRB bounds from $1\%$ precision could be probed if beam-dump set-up could be established in ILC experiment. Both the prospective contours cross each other at $\{M_{Z^\prime},~g_X\}= \{ 0.045~\rm{GeV},~2\times 10^{-5}\}$ whereas bounds from GRB is ruled out for $M_{Z^\prime} \leq 0.025$ GeV by the existing limits of Orsay and E137 experiments respectively. 
 
Bounds estimated from the GRB221009A using HT case in $B-3L_e$ case which can be found in Fig.~\ref{fig:lim6} where $Z^\prime$ couple with the first generation of leptons. The limit obtained with $10\%$ precision is comparable with the results obtained from the TEXONO experiment, however, the bounds obtained from a prospective $1\%$ precision is around $2.67$ times stronger than those from $10\%$ precision. We find that in $B-3L_e$ case, prospective bounds could reach within $3\times 10^{-6} \leq g_X \leq 2\times 10^{-4}$ for $0.025$ GeV $\leq M_{Z^\prime} \leq 1.5$ GeV, $10^{-3} \leq g_X \leq 10^{-2}$ for $10$ GeV $\leq M_{Z^\prime} \leq 89$ GeV and the bound on the coupling is around $\mathcal{O}(0.01)$ for $92$ GeV $\leq M_{Z^\prime}\leq 150$ GeV respectively. We find that limits from GRB with prospective $1\%$ precision can cross the prospective bounds from FASER, FASER2 and ILC-BD which could be probed in future. However, GRB bounds with prospective $1\%$ precision for $M_{Z^\prime} \leq 0.03$ GeV is ruled out by Orsay and $\nu-$Cal experiments. 

\subsection{Limits from neutrino-DM scattering}

\noindent
{\bf (i) Cosmic blazar TXS0506+056 and active galaxy NGC1068:} Neutrino-DM scattering depends on neutrino energy, $M_{Z^\prime}$, $g_X$, $U(1)$ charges of neutrino and DM candidate. To estimate constrains on $g_X-M_{Z^\prime}$ plane we solve the cascade equation given in Eq.~(\ref{eq:cascade})  which can be  modified as  
\begin{equation}
\label{eq:cascade-1}
    \frac{d\Phi}{dx} = -\sigma(g_X, M_{Z^\prime}) \frac{\Sigma(r)}{m_{\rm DM}}\;\Phi + \frac{\Sigma(r)}{m_{\rm DM}} \int_{E^\prime}^{\infty} dE \; \frac{d\sigma (g_X, M_{Z^\prime})}{dE^\prime} \Phi(E) 
\end{equation}
where we have $E_{\nu}^2+{E_{\nu}^\prime}^2 >> m_{\rm DM} (E_{\nu}^\prime- E_{\nu})$ in the energy range under consideration. In order to solve Eq.~(\ref{eq:cascade-1}) using $m_{\rm{DM}} = 3 M_{Z^\prime}$, we find that the second term is proportional to the differential scattering cross section which can be parametrized using two dimensionless quantities $\alpha_1 = \frac{g_{X}^4 \; \Sigma}{8\pi M_{Z^\prime}^4} (1\;\text{TeV}) , \;\; \alpha_2 = \frac{m_{DM}}{M_{Z^\prime}^2} (1\;\text{TeV})$. To constrain $g_X-M_{Z^\prime}$ plane using the cosmic blazar (B) and active galaxy (A) events we apply the respective initial fluxes from Eqs.~(\ref{flux-1}) and (\ref{flux-2}) respectively to compute final attenuated flux $(\Phi_{\tilde{a}})$ by solving the cascade equation for different values of $\alpha_{1,2}$ being equivalent to different values of $\{g_X,~M_{Z^\prime}\}$. Using Eq.~(\ref{eq:eventseqn}) and $\Phi_{\tilde{a}}$ for TXS0506+056 and NGC1068 we calculate corresponding expected number of events at IceCube. Finally we require these estimated number of events be at least $10\%$ of the total observed events with no $\nu$-DM scattering (i.e, 0.1 events in case of TXS0506+056 and 3.1 events in case of NGC1068) to estimate limits on $g_X-M_{Z^\prime}$ plane at $90\%$ C.L. for different $U(1)$ scenarios. Due to the observation of $\nu_\mu$ we can not estimate limits on $L_e-L_{\tau}$ and $B-3L_{e,\tau}$ scenarios because here the second generation of leptons do not couple with $Z^\prime$ directly. In this analysis we have considered Dirac (D1) and complex scalar (CS) type DM candidates with $U(1)$ charge $Q_{\chi}=1000$ (under perturbative limit $g_X|Q_{\chi}| \leq 4 \pi$) whereas Majorana (M2) type DM candidate has $U(1)$ charge $-5$. Hence the limits from cosmic blazar (active galaxy) are denoted as B-D1(A-D1), B-CS(A-CS) and B-M2(A-M2) in Figs.~\ref{fig:lim0}- \ref{fig:lim7} for different types of DM candidates

Bounds estimated from TXS0506+056 and NGC1068 are weak compared to the existing bounds calculated from $\nu-e$ and $\nu-$nucleon scattering experiments lying in the colored(shaded) region for $U(1)_X$($U(1)_{q+xu}$) scenarios shown in the upper(lower) panels of Figs.~\ref{fig:lim0}- \ref{fig:lim3} for different charges whereas B$-$L scenario with $x_H=0$ and $x=1$ is shown in Fig.~\ref{fig:lim1}. The B-D1 (A-D1) and B-CS (A-CS) derive stronger limits overlapping with each other because $E_{\nu}^2+{E_{\nu}^\prime}^2 >> m_{\rm DM} (E_{\nu}^\prime- E_{\nu})$ with same $U(1)$ charges, however, limits obtained from Majorana DM are comparatively weak for different $U(1)$ charge. Among these cases we find that NGC1068 provides stronger limits than cosmic blazar. Comparing with existing bounds we find that limits obtained from TXS0506+056 and NGC1068 are weak with respect to those from COHERENT, CHARM-II, GEMMA, BOREXINO, TEXONO and dark photon searches at BaBar(vis and invis), LHCb and CMS experiments, respectively.  The existing bounds from different beam-dump scenarios like NA64, E774 and E141 cover the region for $M_{Z^\prime} \leq 0.02$ GeV ruling out the limits obtained from D1 and CS cases. Similarly existing bounds from KEK, Orsay, NOMAD, CHARM, $\nu$-cal and E137 also provide strong bounds compared to TXS0506+056 and NGC1068.  Prospective search reach for the beam-dump scenario at DUNE (DUNE-BD), FASER, FASER and ILC-BD will also be stronger than these limits. We find that bounds obtained from GRB are roughly $\mathcal{O}(2)-\mathcal{O}(3)$ of magnitude stronger than the bounds coming from TXS0506+056 and NGC1068 depending on $M_{Z^\prime}$ and general $U(1)$ charges.  

Along the same line we find that limits in $L_{e}-L_{\mu}$ scenario from TXS0506+056 and NGC1068  are weak compared to those from different scattering and beam-dump experiments which can be found in the upper panel of Fig.~\ref{fig:lim4}.   Limits from TEXONO, BOREXINO, CCFR  \cite{CCFR:1991lpl}, BaBar 4$\mu$ \cite{Bauer:2018onh} and $4\mu$ search from CMS \cite{CMS:2018yxg} and ATLAS \cite{ATLAS:2023vxg}, respectively. Remaining part of the  boundary of the gray shaded region consists of CCFR and 4$\mu$ search from BaBar respectively. We find that $4\mu$ search from the CMS and ATLAS can constrain the parameter region of $L_{e}-L_{\mu}$ scenario providing bounds for $4$ GeV $\leq M_{Z^\prime} \leq 70$ GeV where limits on $g_X$ could stay within $1.5 \times 10^{-3}\leq g_X \leq 0.01$. 

Limits from TXS0506+056 and NGC1068 in case of $L_\mu-L_\tau$ scenario can be found in Fig.~\ref{fig:lim5}. Bounds from NA62 experiment using $K\to \mu+E_T^{\rm miss}$ mode \cite{Ibe:2016dir,Martellotti:2015kna,Krnjaic:2019rsv} are stronger compared to those from the $(g-2)_\mu$ experiment. Whereas bounds from NA62 using $K\to 3\mu+E_T^{\rm miss}$ is comparable to those from the $4\mu$ search at BaBar experiment belonging to the gray shaded region. We find that a prospective limit from the Belle-II experiment at 50 $\rm{ab}^{-1}$ luminosity can vary within $3.2\times 10^{-4} \leq g_X \leq 1.6 \times 10^{-3}$ for $0.007$ GeV $\leq M_{Z^\prime} \leq 5$ GeV \cite{Belle-II:2022cgf}. We point out that limits obtained analyzing NGC1068 (D1 and CS) data comparing $\nu-$DM scattering could provide strong bound on $g_X$ for $0.007$ GeV $\leq M_{Z^\prime} \leq 0.02$ GeV where the coupling could vary within $3\times 10^{-4} \leq g_X \leq 8\times 10^{-4}$. Recent observations from NA64$\mu$ \cite{Andreev:2024sgn} constrain exactly this parameter region from missing energy search which could be complemented by prospective bounds from COHERENT \cite{Bauer:2018onh} experiment in future. $4\mu$ search from CMS \cite{CMS:2018yxg} and ATLAS \cite{ATLAS:2023vxg} provide strong upper bounds  within $4$ GeV $\leq M_{Z^\prime} \leq 70$ GeV which partly overlap with the disfavored region by $b\to s \mu^+ \mu^-$ anomalies \cite{Tuckler:2022fkz}. 

Constraints from TXS0506+056 and NGC1068 scenarios are weak in $B-3L_{\mu}$ case compared to different existing results. Limits from the recent $(g-2)_\mu$ data are also weak compared to the existing bounds. These results are shown in the upper panel of Fig.~\ref{fig:lim7} where the gray shaded region represents the strongest limits coming from neutrino Non-Standard Interaction (NSI) \cite{Bauer:2020itv}. In addition to that, we show the limits from dark photon searches in LHCb, CMS and $4\mu$ search in CMS and ATLAS experiments, respectively.

\noindent
{\bf (ii) CMB and Lyman-$\alpha$:} We study limits on $g_X-M_{Z^\prime}$ plane from $\nu$-DM scattering which affect the Cosmic Microwave Background (CMB) spectrum parametrizing a dimensionless parameter $u_{\nu \rm DM} \equiv \frac{\sigma_{\nu \rm DM}}{\sigma_{\rm Th}} \left(\frac{m_{\rm DM}}{100 {\rm GeV}}\right)^{-1}$ where $\sigma_{\rm Th}$ is the Thomson scattering cross section and $\sigma_{\nu \rm DM}$ is $\nu-$DM scattering cross section as we have already calculated. We can obtain the upper bounds of the total cross-section for $\nu-$DM scattering from the Lyman-$\alpha$ data \cite{Hooper:2021rjc} following $\sigma_{\nu \rm DM} \left(\frac{m_{\rm DM}}{100 {\rm GeV}}\right)^{-1}=\sigma_{\rm Th} u_{\nu \rm DM} < 3.66\times 10^{-30} ~{\rm cm^2}$. In this context we mention that $\sigma_{\rm Th} u_{\nu \rm DM} <2.56\times 10^{-28}  {\rm cm^2} ({\rm CMB + BAO})$ form the Planck CMB and Baryon Acoustic Oscillation (BAO) data \cite{Mosbech:2020ahp} which is weaker than the Lyman-$\alpha$. Therefore we use the Lyman-$\alpha$ data for the analysis. Comparing the theoretically estimated cross sections with respect to the observed upper limit on $\nu-$DM scattering cross section from the Lyman-$\alpha$ line, we estimate limits on the $g_X$ with respect to $M_{Z^\prime}$ for different $U(1)$ scenarios. We find that in the chiral cases for different $x_H$ or $x$, Lyman-$\alpha$ line with Dirac (Lyman-D) and complex scalar (Lyman-S) DM candidates match with each other due to large value of $Q_\chi$ and $E_\nu^2+{E_\nu^\prime}^2 >> m_{\rm{DM}}(E_\nu^\prime-E_\nu)$ under the perturbative limit $g_X|Q_\chi| < \sqrt{4 \pi}$. The Majorana DM (Lyman-M) case is weak compared to the Lyman-D and Lyman-S cases. Throughout the chiral cases Lyman-$\alpha$ (Lyman-M, Lyman-D and Lyman-S) lines are weak compared to the limits obtained from the scattering experiment, beam-dump experiment and limits obtained from GRB scenarios with current (prospective) precision $10(1)\%$ and the limits are shown in Figs.~\ref{fig:lim0}-\ref{fig:lim3}. 

We find the above behavior of the Lyman lines in the case of $L_e-L_{\mu}$ and $L_e-L_{\tau}$ scenarios shown in the upper and lower panels of Fig.~\ref{fig:lim4}. Significant difference occurs in the $L_{\mu}-L_{\tau}$ scenario as shown in Fig.~\ref{fig:lim5}. Under the perturbative constraints we find that limits obtained from the Lyman-D and Lyman-S scenarios could provide the strongest bound on the $U(1)$ gauge coupling which could go down to $1.75 \times 10^{-4} \leq g_X \leq 5\times 10^{-4}$ for $0.004$ GeV $\leq M_{Z^\prime} \leq 0.025$ GeV, however, Lyman-M is lying in the gray shaded region as it is weak compared to existing limits. Limits from Lyman-D and Lyman-S could provide stronger bounds on $g_X$ than NA62 experiment within $0.004$ GeV $\leq M_{Z^\prime} \leq 0.025$ GeV considering $K\to \mu+E_T^{\rm miss}$, BOREXINO, recent data from $(g-2)_{\mu}$ and NA64$\mu$. Prospective bounds from COHERENT and Belle-II which could be sensitive for this region in future. 

Estimated bounds from $B-3L_i$ scenarios are shown in Figs.~\ref{fig:lim6} and \ref{fig:lim7}. Like the chiral cases, in this flavored scenario, limits obtained from Lyman-M are weaker than Lyman-D and Lyman-S, however, all the lines belong to the gray shaded region. 
\section{Conclusions}
\label{conc}
In this paper we consider chiral and flavored scenarios where $Z^\prime-\nu$ interactions depend on general $U(1)$ charges. We incorporate cosmic bursts such as GRB 221009A, which is the brightest GRB of all time, cosmic blazar TXS0506+056 and active galaxy NGC1068 to probe $Z^\prime-\nu$ interactions using different models. In the context of GRB, we study $\nu \overline{\nu} \to e^-e^+$ process involving SM and $Z^\prime$ gauge boson. Estimating energy deposition rates of the SM and BSM process and involving Sc, HT cases we estimate bounds on $M_{Z^\prime}/g_X$. We find that limits obtained from GRB in $U(1)_X$ scenarios (except $x_H=-1$) with $10\%$ precision overlaps with existing bounds obtained from different $\nu$ scattering experiments like TEXONO, BOREXINO, etc. Similar behavior is observed for $L_{e}-L_{\mu, \tau}$ and $B-3L_e$ scenarios. In these cases prospective limits with $1\%$ precision could provide stronger bounds which could be probed by beam-dump experiments like DUNE, FASER, FASER2 and ILC-BD in future depending on the gauge interaction. However, in case of $U(1)_{q+xu}$ scenario we find that limits with $10\%$ precision are already ruled out and bounds with $1\%$ precision being stronger by some factors could be probed by DUNE, FASER, FASER2 and ILC-BD in future. Studying $\nu$-DM scattering for $U(1)_X$, $U(1)_{q+xu}$, $L_i-L_j$ and $B-3L_i$ scenarios from cosmic blazar and active galaxy we find that considerable limits from active galaxy (D1 and CS) can be obtained in $L_{\mu}-L_{\tau}$ scenario for $0.007$ GeV $\leq M_{Z^\prime} \leq 0.02$ GeV where the coupling goes to $3\times 10^{-4} \leq g_X \leq 8\times 10^{-4}$ which is competitive with the recent NA64$\mu$ bounds and could be probed by Belle-II at 50 ab$^{-1}$ luminosity and COHERENT experiments in future. In that region recent limits from $(g-2)_\mu$ manifest complementarity with active galaxy (D1 and CS) bounds. Recent experimental results from NA64$_\mu$ provide strong bounds within this mass range which is almost comparable with the current sensitivity of $(g-2)_\mu$ and  future sensitivity of Belle-II. Furthermore, in future NA62 and COHERENT could provide considerable bounds in this region. Limits obtained from the Lyman-$\alpha$ line studying $\nu-$DM scattering are weak compared to different $\nu$ scattering experiments for the chiral scenarios, $L_e-L{\mu, \tau}$ and $B-3L_{i}$ scenarios. However, in case of $L_\mu-L_\tau$ scenario Lyman-D and Lyman-S limits provide strongest sensitivity on $g_X-M_{Z^\prime}$ plane for $0.004$ GeV $\leq M_{Z^\prime} \leq 0.025$ GeV where the gauge coupling could reach around $1.75 \times 10^{-4} \leq g_X \leq 5\times 10^{-4}$ manifesting complementarity with the current bounds from recent $(g-2)_\mu$ and NA64$_\mu$ data which could be probed by COHERENT, NA62 and Belle-II in future. 
\begin{acknowledgments}
 SKA thanks the Department of Ministry of Education, Culture, Sports, Science and Technology of Japan for MEXT fellowship to study in Japan. This work is supported by INFN and MIUR [GL] and the Fundamental Research Funds for the Central Universities [TN].  
\end{acknowledgments}

\bibliographystyle{utphys}
\bibliography{bibliography}
\end{document}